\pdfoutput=1
\documentclass[11pt]{article}
\usepackage{epsfig}
\usepackage{amsfonts}
\usepackage{slashed}
\usepackage{amsmath,amssymb}
\usepackage{bbm,bm}
\usepackage{cite}
\allowdisplaybreaks[1]

\usepackage{pifont}
\newcommand{\xmark}{\ding{55}}

\usepackage{tikz}
\usetikzlibrary{shapes,arrows,shadows,automata,positioning}
\newdimen\nodeDist
\usepackage{comment}
\usepackage[normalem]{ulem} 

\usepackage[most]{tcolorbox}
\newtcolorbox{probox}[1]{
  left=0.6em,
  right=0.6em,
  enhanced,
  sharp corners,
  attach boxed title to top right,
  colback=black!5,
  colframe=black,
  fonttitle=\bfseries,
  boxed title style={
    sharp corners,
    size=small,
    colback=black,
    colframe=black,
  },
  title={#1}
}

\usepackage[
  colorlinks=true,
  linkcolor={red!50!black},
  citecolor={blue!50!black},
  filecolor=black,
  urlcolor=black,
  breaklinks=true
  ]{hyperref}
\usepackage{orcidlink}

\newcommand{\nn}{\nonumber \\}
\renewcommand{\vec}[1]{{\bf #1}}
\newcommand{\rmi}[1]{{\mbox{\scriptsize #1}}}
\newcommand{\rmii}[1]{{\mbox{\tiny\rm{#1}}}}
\newcommand{\clog}{\Big(c+\ln\frac{3T}{\Lamd}\Big)}
\newcommand{\bmu}{\bar{\mu}}
\newcommand{\Lamd}{\bmu_{\rmi{3d}}}
\newcommand{\LamD}{\bmu}
\newcommand{\gammaE}{{\gamma_\rmii{E}}}
\newcommand{\Veff}{V_{\rmi{eff}}}
\newcommand{\phiB}{\phi_{\rmii{B}}}
\newcommand{\varphiB}{\varphi_{\rmii{B}}}

\newcommand{\nG}{n_{\rmii{G}}}

\newcommand{\Tc}{T_{\rm c}}
\newcommand{\TcDS}{T_{{\rm c},\rmii{DS}}}
\newcommand{\TpDS}{T_{{\rm p},\rmii{DS}}}
\newcommand{\TcSM}{T_{{\rm c},\rmii{SM}}}
\newcommand{\Tfo}{T_{\rm fo}}
\newcommand{\yc}{y_{\rm c}}
\newcommand{\yp}{y_{\rm p}}

\newcommand{\Tp}{T_{\rm p}}
\newcommand{\vw}{v_w}
\newcommand{\cs}{c_s}

\newcommand{\YS}{Y_s}
\newcommand{\YX}{Y_\rmii{$X$}}
\newcommand{\epsS}{\epsilon_s}
\newcommand{\epsV}{\epsilon_\rmii{$V$}}
\newcommand{\thetaS}{\theta_s}
\newcommand{\thetaV}{\theta_\rmii{$V$}}

\newcommand{\gd}{g_d}
\newcommand{\gX}{g_\rmii{$X$}}
\newcommand{\mD}{m_\rmii{D}}
\newcommand{\yD}{y_\rmii{D}}

\newcommand{\ms}{m_s}
\newcommand{\mh}{m_\rmii{$h$}}

\newcommand{\mV}{m_\rmii{$V$}}

\newcommand{\mX}{m_\rmii{$X$}}
\newcommand{\mZ}{m_\rmii{$Z$}}

\newcommand{\mpl}{m_\rmii{Pl}}

\newcommand{\vrel}{v_\rmi{rel}}

\newcommand{\muX}{\mu_\rmii{$X$}}

\newcommand{\Lb}{L_b}
\newcommand{\Lf}{L_f}

\newcommand{\re}{\mathop{\mbox{Re}}}

\newcommand{\T}{\rmii{$T$}}
\newcommand{\Tstar}{T_\star}
\newcommand{\Hstar}{H_\star}
\newcommand{\Tds}{T_{\rmii{DS}}}
\newcommand{\Treh}{T_{\rmi{reh}}}

\newcommand{\bsl}[1]{\,\slash\!\!\!\!{#1}\,}
\newcommand{\msl}[1]{\,\slash\!\!\!{#1}\,}
\newcommand\MSbar{$\overline{\rm MS}$}

\newcommand{\geff}{g_\rmi{eff}}

\newcommand{\Ogw}{\Omega_\rmi{GW}}

\newcommand{\fsw}{f_\rmi{sw}}

\newcommand{\thetaw}{\theta_\rmi{w}}
\newcommand{\cw}{c_\rmi{w}}
\newcommand{\sw}{s_\rmi{w}}

\newcommand{\nF}{n_\rmii{F}}

\newcommand{\Tint}[1]{{\hbox{$\sum$}\!\!\!\!\!\!\!\int\,}_{\!\!\!\!\raise-0.9ex\hbox{$\scriptstyle{#1}$}}}
\newcommand{\Tinti}[1]{{{\Sigma}\!\!\!\!\raise0.3ex\hbox{$\int$}_\rmii{${#1}$}}}
\newcommand{\Tintip}[1]{{{\Sigma'}\!\!\!\!\!\raise0.3ex\hbox{$\int$}_\rmii{${#1}$}}}

\newcommand{\vev}{VEV}
\newcommand{\threeDEFT}{3dEFT}

\makeatletter \@addtoreset{equation}{section} \makeatother
\renewcommand{\theequation}{\arabic{section}.\arabic{equation}}
\makeatletter
\renewcommand\section{\@startsection{section}{1}{\z@}%
  {-5.5ex \@plus -1ex \@minus -.2ex}
  {2.3ex \@plus.2ex}%
  {\normalfont\large\bfseries}}
\renewcommand\subsection{\@startsection{subsection}{2}{\z@}%
  {-3.25ex\@plus -1ex \@minus -.2ex}%
  {1.5ex \@plus .2ex}%
  {\normalfont\normalsize\bfseries}}
\renewcommand\thesection{\@arabic\c@section}
\renewcommand\thesubsection{\thesection.\@arabic\c@subsection}
\renewcommand{\@seccntformat}[1]{%
  \csname the#1\endcsname.\hspace{1.0em}}
\makeatother

\begin{document}

\flushbottom

\begin{titlepage}

\begin{flushright}
  July 2026
\end{flushright}
\begin{centering}

\vfill



{\Large{\bf
A critical look at low-scale cosmological phase transitions
\\
in the PTA era}
}
%
\vspace{0.8cm}

\renewcommand{\thefootnote}{\fnsymbol{footnote}}
Simone Biondini%
\orcidlink{0000-0003-2533-9003}%
\,$^{\rm a,}$%
\footnote{simone.biondini@physik.uni-freiburg.de}
and
Philipp Schicho%
\orcidlink{0000-0001-5869-7611}%
\,$^{\rm b,}$%
\footnote{philipp.schicho@unige.ch}

\vspace{0.8cm}

$^\rmi{a}$%
{\em
Institute of Physics, University of Freiburg, \\
Hermann-Herder-Straße 3, 79014 Freiburg, Germany}
\vspace{0.3cm}

$^\rmi{b}$%
{\em
  D\'epartement de Physique Th\'eorique, Universit\'e de Gen\`eve,\\
  24 quai Ernest Ansermet, CH-1211 Gen\`eve 4, Switzerland
}

\vspace*{0.8cm}

\mbox{\bf Abstract}

\end{centering}

\vspace*{0.3cm}

\noindent
Motivated by the recent evidence for a stochastic gravitational-wave (GW) background
reported by pulsar timing array (PTA) collaborations,
we perform a precision study of low-scale phase transitions in
a dark Abelian Higgs sector,
a minimal gauge theory
of spontaneous symmetry breaking relevant for cosmological phase transitions.
Using dimensionally reduced high-temperature effective field theory,
we quantify the impact of thermal resummation, higher-order matching corrections,
and higher-dimensional operators on the phase-transition thermodynamics and the
resulting GW signal.
We find that the parameter region favored by current PTA observations lies close to the
boundary of validity of the effective field theory,
where higher-dimensional operators become increasingly important.
Even within this controlled region, the predicted signal
remains disfavored by the PTA data, despite the substantial shifts induced by
higher-order thermal corrections.
We further delineate parameter regions where the dark and visible sectors
are thermally and hydrodynamically coupled or decoupled, and
revisit the dark matter phenomenology, identifying asymmetric freeze-out as
naturally compatible with both the observed relic abundance and the gauge
couplings favored by strong phase transitions.
Our results underscore the importance of systematically controlled
finite-temperature calculations for reliable GW predictions from
low-scale cosmological phase transitions.

\noindent

\vfill
\end{titlepage}

{\hypersetup{hidelinks}
\tableofcontents
}
\clearpage

\renewcommand{\thefootnote}{\arabic{footnote}}
\setcounter{footnote}{0}

%
\section{Introduction}
\label{sec:intro}

Despite the remarkable success of
the Standard Model of particle physics (SM) in describing
a vast range of phenomena with
few fundamental particles and interactions,
several observations remain unexplained.
On the particle-physics front, the origin of neutrino masses
still motivates extensive theoretical and
experimental efforts~\cite{Strumia:2006db,Cai:2017jrq,SajjadAthar:2021prg}.
On the cosmological side,
the existence of dark matter
(DM)~\cite{Rubin:1980zd,Bosma:1981zz,Planck:2018vyg,Bertone:2016nfn},
the dominant matter component of the Universe, and
the observed baryon asymmetry~\cite{WMAP:2012fli,Planck:2018vyg}
both call for physics beyond the SM (BSM).
Despite decades of dedicated searches, no conclusive evidence for
a weakly interacting massive particle (WIMP) near the electroweak scale
has emerged, and increasingly stringent constraints have excluded
large parts of the simplest scenarios~\cite{Bertone:2016nfn,Arcadi:2017kky}.
While this does not rule out the WIMP paradigm,
it has motivated a broader exploration of dark-sector frameworks
with sub-electroweak-scale particles and alternative signatures.

Gravitational-wave (GW) astronomy has opened
a new observational window on such hidden sectors,
since a stochastic GW background (SGWB) may
hold information about phase transitions and other dynamical processes in the early Universe.
Gravitational-wave production associated with
an electroweak-scale phase transition typically requires
new particles with masses of
$\mathcal{O}(100)$~GeV to $\mathcal{O}(1)$~TeV
and sizable couplings to the Higgs sector.
In such scenarios, the resulting SGWB is expected to peak in
the mHz frequency range,
making it a prime target for future space-based interferometers such
as LISA~\cite{Caprini:2015zlo,LISA:2017pwj} or
Taiji~\cite{Ruan:2018tsw}.
Gravitational-wave observations
thus provide a powerful complement to collider searches for
new physics~\cite{Ramsey-Musolf:2019lsf,Friedrich:2022cak,Ramsey-Musolf:2024ykk}.

A dark sector with its own gauge interactions and symmetry-breaking pattern
may undergo a first-order phase transition that sources such a background.
Recent measurements by several
pulsar timing arrays (PTAs),
such as
NANOGrav~\cite{NANOGrav:2020bcs,NANOGrav:2023gor},
EPTA~\cite{EPTA:2021crs,EPTA:2023fyk,EPTA:2023xxk},
PPTA~\cite{Goncharov:2021oub,Reardon:2023gzh},
CPTA~\cite{Xu:2023wog}, and
MPTA~\cite{Miles:2024seg}
have provided strong evidence for a SGWB in the nHz frequency range,
reporting the quadrupolar Hellings-Downs correlations~\cite{Hellings:1983fr}
expected to be of GW origin.
While the signal may ultimately be explained by a population of merging
supermassive black-hole binaries~\cite{%
  Middleton:2020asl,Izquierdo-Villalba:2021prf,Curylo:2021pvf},
reproducing its amplitude and spectral shape is non-trivial, not least
because of the unresolved final-parsec problem~\cite{Milosavljevic:2002ht}.
Hence, the possibility that the signal originates from BSM physics
remains an intriguing alternative.

Several cosmological mechanisms have been proposed to account for the SGWB
reported by PTA collaborations, most notably cosmic strings, primordial black holes, and
first-order phase transitions~\cite{NANOGrav:2023hvm,EPTA:2023xxk,Buchmuller:2020lbh}.
In this work, we focus on first-order phase transitions in a dark sector.
For the resulting GW spectrum to peak in the nHz frequency range,
the transition must occur well below the electroweak scale,
typically at temperatures of $\mathcal{O}(1)$--$\mathcal{O}(100)$~MeV.%
\footnote{%
  A delayed electroweak phase transition could in principle also produce nHz gravitational waves,
  but such scenarios have been ruled out~\cite{Athron:2023mer}.
}
Such low-scale transitions point to new-physics states with masses well below
the electroweak scale, giving rise to a rich phenomenology
that connects GW observations with collider and fixed-target experiments
as well as cosmological probes such as
Big Bang Nucleosynthesis (BBN) and
$\Delta N_{\rm eff}$.

The many particle-physics realizations of dark-sector phase transitions
range from
classically conformal models~\cite{%
  Hambye:2013dgv,Carone:2013wla,
  Kierkla:2022odc,Kierkla:2023von,Kierkla:2025qyz,Kierkla:2025vwp,
  Madge:2023dxc,
  Balan:2025uke,Goncalves:2025uwh,
  Christiansen:2025xhv}
to
(multi-)scalar extensions~\cite{%
  Dorsch:2013wja,Hashino:2016xoj,Benincasa:2023vyp}.
Among them,
the Abelian Higgs model
is the simplest Higgs-gauge theory of spontaneous symmetry breaking
and represents a broader class of models in which a radiatively
generated, gauge-boson-induced cubic barrier renders the transition
first order.
This makes the model a long-standing benchmark exposed to finite-temperature methods,
from thermal resummation to lattice simulations~\cite{Kajantie:1997hn}.
A dark Abelian Higgs sector therefore provides a minimal realization of
a first-order phase transition capable of generating a PTA-scale SGWB
if the dark scalar and gauge boson masses lie well below
the electroweak scale, often in the MeV range.%
\footnote{%
During radiation domination, the Hubble rate scales as
$H \propto T^2/M_{\rm Pl}$.
Since the observed GW frequency today satisfies
$f_0=(a_*/a_0)f_*$ with
$a_*/a_0 \propto T_0/T_*$,
one finds the parametric scaling
$f_0 \sim T_*T_0/M_{\rm Pl}$,
where $T_0$ is the present CMB temperature.
A phase transition occurring at
$T_* \sim 1~{\rm MeV}$
therefore naturally produces a signal in the nHz frequency range.
}
Its proximity to BBN, however,
introduces significant cosmological constraints.
Relativistic dark-sector particles present during BBN contribute to $N_{\rm eff}$,
whereas unstable ones must decay early enough to preserve
predictions of BBN~\cite{Yeh:2022heq,Kawasaki:2017bqm,Bringmann:2023opz}.

A large body of recent work has explored the possibility that MeV-scale phase transitions
could account for the PTA signal and
is divided into two complementary directions:
\begin{itemize}
  \item[(i)]
    model-independent descriptions via macroscopic quantities such as
    the transition temperature, latent heat, and
    inverse duration~\cite{Bringmann:2023opz,Addazi:2023jvg,Ghosh:2023aum,Winkler:2024olr,Bringmann:2026xcx};
  \item[(ii)] 
    specific particle-physics realizations that directly relate
    masses and couplings to
    the GW signal~\cite{%
      Addazi:2023jvg,Han:2023olf,DiBari:2023upq,Banik:2024zwj,
      Feng:2026iqo,Costa:2025csj,Bringmann:2026xcx}.
\end{itemize}
Our work focuses on the second category.
Whereas previous studies identified
parameter regions compatible with PTA observations,
whether genuinely or only through tuning,
our primary objective is to provide
a state-of-the-art thermodynamic analysis
that tests the
theoretical robustness of those predictions.
To this end, we perform a precision thermodynamics analysis using
high-temperature three-dimensional (3d) effective field theory (EFT)~\cite{%
  Ginsparg:1980ef,Appelquist:1981vg,Farakos:1994kx,Kajantie:1995dw,Braaten:1995cm,Kajantie:1996mn},
which systematically incorporates thermal corrections and
infrared (IR) effects in a gauge-invariant manner~\cite{%
  Gould:2021ccf,Lofgren:2021ogg,Hirvonen:2021zej},
allowing for controlled thermal resummation together with
renormalization-group improvement~\cite{Gould:2021oba}.
We quantify theoretical uncertainties in
the critical and percolation temperatures and compare with
non-perturbative results~\cite{Kajantie:1997hn}.
We find that the PTA-favored region is particularly sensitive to
higher-dimensional operators and lies close to
the limit of validity of
the high-temperature expansion and
perturbativity.

As emphasized in~\cite{Bringmann:2023opz},
a \textit{decaying} dark sector is favored if a MeV-scale phase transition
underlies the PTA signal, requiring portal interactions that
allow the dark scalar and gauge boson to decay before BBN and
thereby avoiding stringent constraints
on the
effective number of relativistic species, $\Delta N_\text{eff}$.
Motivated by this, we extend the minimal Abelian Higgs model with
a fermionic dark matter candidate charged under
the dark gauge group~\cite{Han:2023olf,Banik:2024zwj},
rendering it stable.
This setup allows us to study both
the GW phenomenology of the phase transition and
the dark matter implications of the model.
In particular, we analyze symmetric and asymmetric freeze-out scenarios,
including Sommerfeld enhancement and bound-state formation, and
assess the complementarity between couplings favored by
the observed dark matter abundance and those required for
a strong first-order phase transition.

In addition, we revisit the thermalization between the dark and visible sectors.
Because cosmological constraints require small portal couplings,
thermal equilibrium is not guaranteed {\em a priori}.
We therefore compute the relevant interaction rates and identify
the dominant processes that maintain thermal contact between
the two sectors during the epoch relevant for the phase transition.
Beyond thermal equilibrium,
we examine hydrodynamic equilibrium between the two sectors during the phase transition.
We identify parameter regions where the SM plasma decouples from
the bubble dynamics.
In this regime,
the hydrodynamic evolution on scales of the bubble size and below
is governed only by
the dark sector, while the total radiation density still determines
the expansion history and GW amplitude.

The structure of the paper is as follows.
Section~\ref{sec:model}
introduces the dark-sector model and discusses its phenomenology,
including the thermalization conditions between the dark and visible sectors.
The thermodynamic framework for the phase transition is outlined
in sec.~\ref{sec:PT_and_GW},
where we derive the thermal EFT.
The dark matter phenomenology is discussed in sec.~\ref{sec:dark_matter},
while sec.~\ref{sec:pta}
addresses the GW spectra and their reconciliation with PTAs.
Finally, we present our conclusions
in sec.~\ref{sec:conclusions},
while additional technical details are collected in the appendices.

%
\section{Model setup}
\label{sec:model}

The primary goal of this work is to investigate the thermodynamics of
a phase transition at 
\emph{low energy scales},
specifically in the range $\mathcal{O}(1\text{--}100)$~MeV, and
to determine the associated SGWB signal.
To this end, we consider an Abelian dark Higgs model, 
consisting of
a complex scalar field ($S$),
a dark gauge boson ($V_\mu$), and
a U(1)$_d$ gauge symmetry.
This constitutes a minimal field content,
an archetypal framework,
that may trigger a first-order phase transition~\cite{%
  Weinberg:1992ds,Metaxas:1995ab,Kajantie:1997hn,Garny:2012cg,
  Hirvonen:2021zej,Lofgren:2021ogg,Ekstedt:2024etx,Bernardo:2025vkz}
in a dark sector~\cite{Croon:2018erz,Bringmann:2023opz}.
We introduce portal interactions with the SM through renormalizable operators.
The Higgs portal~\cite{Patt:2006fw,Pospelov:2007mp,March-Russell:2008lng} for the dark scalar and
kinetic mixing~\cite{Galison:1983pa,Holdom:1985ag} for the dark gauge boson $V_\mu$.
These couplings allow for efficient decays into the SM,
ensuring that both dark particles decay away before BBN.

To remain compatible with Big Bang Nucleosynthesis (BBN),
these decays must occur at lifetimes below
$\tau_{\rmii{BBN}} \approx 0.1$~s
to avoid disrupting light element abundances~\cite{Depta:2020zbh}.
Furthermore, late-time energy injection from light particles can affect
the Cosmic Microwave Background (CMB) through
changes in the effective number of relativistic degrees of freedom
$N_{\rmi{eff}}$~\cite{EscuderoAbenza:2020cmq},
spectral distortions from black-body radiation~\cite{Fixsen:1996nj,Chluba:2011hw}, and
the ionization history~\cite{Chen:2003gz}.
Details regarding the portal interactions and constraints on the corresponding couplings
are presented in sec.~\ref{sec:model_portal}.

Since we also aim to inspect the interplay between the dark phase transition and
a stable relic DM particle, which may account for
the observed DM energy density
$\Omega_{\rmii{DM}} h^2 =0.1200 \pm 0.0012$~\cite{Planck:2018nkj},
we consider a next-to-minimal dark sector that comprises
a dark fermion ($X$).
The dark fermion is taken to be a SM gauge singlet,
whereas it is charged under U(1)$_d$.
As a result, we end up in a class of DM models that have been extensively studied~\cite{%
  Pospelov:2007mp,Feng:2008ya,Feng:2008mu,Arkani-Hamed:2008hhe,
  Bell:2016fqf,Duerr:2016tmh,Evans:2017kti}.
Here, the DM relic abundance
is mainly determined by interactions internal to the dark sector with little or no
involvement of the SM degrees of freedom.%
\footnote{%
  There are various possibilities for naming such a model class,
  such as \emph{two-mediator} models~\cite{Bell:2016fqf,Duerr:2016tmh,Evans:2017kti} or
  hidden/secluded sectors \cite{Pospelov:2007mp,Feng:2008ya}.
  The model realization that we consider in this paper features a Dirac fermion dark matter,
  similarly to~\cite{Evans:1993qg}.
} 
For a thermalized dark sector and a dark fermion heavier than the bosonic degrees of freedom,
the DM relic density is fixed via fermion-antifermion annihilations
that drive a freeze-out dynamics.

We take the dark fermion to be of Dirac type,
with its bare mass parameter $\muX$ a free input of the model,
fixed at the input scale to the physical fermion mass, $\muX(\LamD_0) = \mX$
(cf.\ sec.~\ref{sec:MSbar:physical}).
This is the most convenient choice for our purpose of
addressing the phase transition within a viable dark matter model, and
it allows a direct comparison with~\cite{Han:2023olf,Banik:2024zwj},
where the same model was invoked to explain the PTA data.
Alternative realizations comprise
e.g.\ a Majorana dark fermion~\cite{Bell:2016fqf,Duerr:2016tmh} or
chiral fermions with different charge assignments~\cite{Arnold:1992rz,Kajantie:1997hn,Banik:2024zwj},
whose mass is generated only after the spontaneous breaking of the U(1)$_d$ symmetry.%
\footnote{%
  The latter case introduces no additional parameters,
  since a Yukawa coupling replaces the bare mass,
  but the fermion then enters the phase-transition thermodynamics directly,
  which makes the freeze-out and phase-transition dynamics harder to disentangle;
  we leave this option for future work.
}

The phase transition of the model has been studied on the lattice,
both without fermions and with Yukawa-coupled fermions,
and the corresponding phase diagram was obtained in~\cite{Kajantie:1997hn}.
It features a tricritical endpoint that separates
first-order (type~I superconductor) from second-order (type~II superconductor)
transitions, located non-perturbatively in~\cite{Mo:2001fi} and
perturbatively in~\cite{Kleinert:1982dz,Herbut:1996ut}.
In the case with Yukawa-coupled fermions,
the fermions are not simulated directly but
integrated out in the dimensional reduction,
so that their effect enters only through
the matching relations (cf.\ sec.~\ref{sec:dimensional_reduction_details}).

The four-dimensional~(4d) model Lagrangian (cf.\ e.g.~\cite{Evans:2017kti,Kim:2016kxt})
in Minkowski space-time is
\begin{align}
\label{eq:lag:4d}
  \mathcal{L} &=
    \bar{X} (i\bsl{D}-\muX) X
      - \frac{1}{4} V_{\mu \nu} V^{\mu \nu}
      + (D^\mu S)^{*} (D_\mu S)
      - V(S^{*} S)
      + \mathcal{L}_{\rmi{portal}}
  \,, \nn[2mm]
  V(S^{*} S) &=
      - \mu_s^2 S^{*} S
      + \lambda_s^{ } (S^{*} S)^2
  \,,
\end{align}
where
$D_\mu = \partial_\mu - i\gd Y_{i} V_\mu$ is the covariant derivative,
$\gd$ is the dark gauge coupling,
$Y_{i}$ the corresponding hypercharge of the fermion and scalar field, and
$V_\mu$ the gauge field of the dark sector with the field strength tensor $V_{\mu \nu}$.
Since the DM is a Dirac fermion,
a trilinear coupling of the form $\bar{X} X S$ is absent,
unless either the fermion or the scalar is a U(1)$_d$ singlet
(namely $\YS=0$ or $\YX=0$).
We assign equal U(1)$_d$ charges to the fermion and the scalar,
taking $\YX = \YS = 1$.%
\footnote{%
  In the appendix, we keep a generic hypercharge
  in the matching equations for the 3d theory.
}
The portal interactions, which induce a mixing of the dark and SM states, are addressed
in sec.~\ref{sec:model_portal}.

The scalar potential contains the parameters
$\mu_s^{2} > 0$ and
$\lambda_s^{ } > 0$.
We parametrize the complex scalar field by its vacuum expectation value (\vev{}),
$v_s$,
and two real degrees of freedom, $s$ and $\chi$.
Here,
$\chi$ is the Goldstone boson associated with the symmetry breaking,
\begin{align}
\label{eq:param:S}
    S &= \frac{1}{\sqrt{2}}(v_s+s + i \chi)
    \,,&
    v_s &= \sqrt{\frac{\mu_s^2}{ \lambda_s}}
    \,,
\end{align}
when promoting
$v_s$ to a background field, we later use $v_s = \phi$ in sec.~\ref{sec:PT_and_GW}.
After spontaneous symmetry breaking of the dark gauge group,
we find the following Lagrangian
\begin{align}
\label{lag_model_broken}
 \mathcal{L}   &=
  \mathcal{L}_\rmii{$V$}
 + \mathcal{L}_s
 + \mathcal{L}_\chi
 + \mathcal{L}_\rmii{$X$}
 \nn[2mm] &
  + \frac{\gd^2}{2} V_\mu V^\mu s^2
  + \frac{\gd^2}{2} V_\mu V^\mu \chi^2
  + \gd \,  \mV  V_\mu V^\mu  s
  + \gd V^\mu ( \chi \, \partial_\mu s- s \,   \partial_\mu \chi)
  \nn[1mm] &
  - \frac{\lambda_s}{4} s^4
  - \frac{\lambda_s}{4} \chi^4
  - \frac{\lambda_s}{2} s^2 \chi^2 
  -  \ms \sqrt{\frac{\lambda_s}{2}} \, s \chi^2
  -  \ms \sqrt{\frac{\lambda_s}{2}}\, s^3 \,
  + \gd \, \bar{X} \msl{V} X
  + \mathcal{L}_{\textrm{portal}}
  \, .
\end{align}
The first line contains
the free Lagrangian terms for each field.
The second line encodes
the interactions between the scalar fields $s$ and $\chi$ and the gauge boson;
the third line includes scalar self-interactions, and
the fermion–gauge boson interaction, analogous to QED,
as well as the portal Lagrangian.
The resulting particle masses,
including those of the Goldstone boson and the ghost field $c$,
are%
\footnote{%
  \label{ft:ghost}
  The ghost Lagrangian, including its interaction with the scalar $s$,
  reads  
  $\mathcal{L}_{c} = \bar{c} \bigl[ -\partial^2 - \xi \mV^2 \bigl( 1 + \frac{s}{v_s} \bigr) \bigr] c$.
}
\begin{align}
  \label{masses_minimum}
  \ms^2 &= 2 v_s^2 \lambda_s^{ }
  \, , &
  \mV^2 &= (\gd v_s)^2
  \, , &
  \quad m_\chi^2 &= m_c^2 = \xi \mV^2
  \,.
\end{align}
Following the generalized gauge fixing choices of~\cite{Martin:2018emo},
all computations are performed in a general $R_\xi$ (or Fermi) gauge.
The corresponding gauge fixing functional
that enters the gauge fixing Lagrangian
$\mathcal{L}_\rmii{GF} = \frac{1}{2\xi} \bigl[ F(v_s)\bigr]^2$
is
$F(v_s)= \partial_\mu V^\mu + \xi \gd  v_s \chi$.
Since gauge invariance is manifest
in all our computations,
we display results in Landau gauge ($\xi \to 0$) throughout the paper
when dealing with the thermodynamics of the phase transitions.
For the dark matter annihilation cross-sections, on the other hand,
we work in Feynman gauge ($\xi \to  1$).

%
\subsection{Portal interactions}
\label{sec:model_portal}

Portal interactions can be naturally incorporated via renormalizable operators such as
the Higgs portal~\cite{Patt:2006fw,Pospelov:2007mp,March-Russell:2008lng} and
kinetic mixing~\cite{Galison:1983pa,Holdom:1985ag}.
The corresponding portal Lagrangian is~\cite{Duerr:2016tmh,Evans:2017kti}
\begin{eqnarray}
  \label{eq:portal_lagrangian}
    \mathcal{L}_{\textrm{portal}} =
        - \frac{\epsV}{2 \cw} V^{\mu \nu} B_{\mu \nu}
        - \epsS S^*S H^\dagger H
    \,,
\end{eqnarray}
where $\epsV$ and $\epsS$ parametrize the couplings of the visible-to-dark sector interactions;
$\thetaw$ is the Weinberg angle with
$\cos \thetaw \equiv \cw$ and
$\sin \thetaw \equiv \sw$;
$B_{\mu \nu}$ is the field strength tensor of the SM U(1)$_\rmii{Y}$ gauge group;
and $H$ is the SM Higgs doublet.
The SM gauge coupling of
SU(2) is denoted by $g_2$ and for 
U(1)$_\rmii{Y}$ by $g_1$.
The portal couplings,
assumed to be
$\epsV,\epsS \lesssim 1$, are
constrained by numerous experimental bounds, which we summarize in the following. 

The presence of portal interactions induces mixing between the dark and SM states.
After the spontaneous symmetry breaking of both sectors,
the dark Higgs mixes with the SM Higgs boson, and
the dark photon mixes with both the SM photon and the $Z$ boson
(see e.g.~\cite{Evans:2017kti}).
This induces tree-level couplings
of the dark scalar to SM fermions and gauge bosons, and
of the dark photon to SM fermions.
Consequently, decay channels
$s \to \mathrm{SM} \, \mathrm{SM}$ and
$V \to \mathrm{SM} \, \mathrm{SM}$
are generically allowed whenever kinematically accessible.
These portal interactions not only enable experimental searches for dark states, but also
govern the thermal contact between the visible and dark sectors in
the early universe;
see sec.~\ref{sec:thermal_contact}.

We focus on dark-sector mass scales below $\mathcal{O}(100)$~MeV,
where both the dark scalar and dark photon are much lighter than their SM counterparts:
$\ms \ll \mh$ and $\mV \ll \mZ$.
In this limit of small mixing and hierarchical masses,
the portal couplings relate directly to the physical mixing angles.
To leading order (LO),
one finds (see, e.g.,~\cite{Evans:2017kti})
\begin{align}
  \sin\thetaS &\simeq \epsS \frac{v_h v_s}{m_h^2}
  \,, &
  \sin\thetaV &\simeq \epsV \tan\thetaw
  \, .
  \label{portal_mixing_couplings}
\end{align}

\subsubsection{Dark scalar portal}
\label{sec:dark:scalar:portal}

The model is subject to a variety of laboratory, astrophysical, and cosmological constraints. 
Even independently of the dark fermion $X$, stringent bounds apply to both
the dark Higgs $s$ and the dark photon $V_\mu$. 
The portal interaction
$\epsS S^* S H^\dagger H$
induces a mixing between the dark scalar and the SM Higgs,
allowing for direct searches in rare meson decays and fixed-target experiments. 
In the range of phase-transition energy scales relevant for this work,
$\mathcal{O}(1)$--$\mathcal{O}(100)\,{\rm MeV}$,
accelerator constraints are complemented by bounds from stellar cooling
as well as from BBN and the CMB~\cite{%
  Berger:2016vxi,Fradette:2014sza,Fradette:2018hhl,Li:2020roy,Ibe:2021fed}.

For the dark scalar, the most stringent laboratory constraints in the sub-GeV mass range arise from rare kaon decays. 
In particular, NA62 and E949 constrain the branching ratio of $K^+ \to \pi^+ s$,
implying an approximate upper bound
$\sin\thetaS \lesssim 2\times 10^{-4}$ for
$\ms \lesssim 100$~MeV~\cite{NA62:2021zjw,BNL-E949:2009dza}. 
In addition, light scalars with masses below
$\sim 150$~MeV are constrained by observations of core-collapse supernovae,
excluding the mixing-angle window
$3.9\times 10^{-7} \lesssim \sin\thetaS \lesssim 7.0\times 10^{-6}$~\cite{Dev:2020eam}. 
Cosmological bounds from BBN and the CMB depend sensitively on the scalar lifetime and its visible branching fractions. 
Unstable light particles can also significantly affect the thermal history of the universe. Their impact on BBN and the CMB depends sensitively on their lifetime.
Dark scalars with lifetimes $\tau \lesssim 0.1$~s decay sufficiently early
to evade cosmological bounds.
Conversely, longer lifetimes, as favored by current laboratory constraints,
can inject energetic particles into the primordial plasma,
thereby modifying light-element abundances through electromagnetic or hadronic processes, and
distorting the CMB anisotropy spectrum.
A comprehensive analysis of these effects for
Higgs-mixed scalars can be found in~\cite{%
  Berger:2016vxi,Fradette:2014sza,Fradette:2018hhl,Ibe:2021fed}. 
Adopting the most recent analysis of~\cite{Ibe:2021fed},
we therefore focus on the representative parameter region%
\footnote{%
  The same range is adopted in~\cite{Bringmann:2026xcx}.
  The analyses of~\cite{Berger:2016vxi,Ibe:2021fed}
  consider one mediator at a time and do not explicitly incorporate
  an underlying U(1)$_d$ gauge structure.
  In the model~\eqref{eq:lag:4d},
  additional decay channels such as $s\to VV$ could in principle open up
  at $\gd \sim \mathcal{O}(1)$ and
  modify the scalar lifetime and branching fractions.
  The barrier-induced phase transitions in sec.~\ref{sec:PT_and_GW},
  however, require $\mV \gg \ms$ (cf.\ sec.~\ref{sec:PT_and_GW}),
  which also holds at $T=0$,
  so that $s\to VV$ remains kinematically closed.
}
\begin{align}
  \label{mixing_mass_limits}
  \ms &\in [3.8,100]~{\rm MeV}
  \,,& 
  \sin\thetaS &\in \bigl[2\times 10^{-5},2\times 10^{-4}\bigr]
  \,.
\end{align}
These bounds on $\ms$ and $\sin\thetaS$
can be reinterpreted as constraints on the dark scalar self-coupling
$\lambda_s$.
Using the relation
\begin{align}
\label{Dark_higgs_parameters}
\lambda_s &=
  \biggl( \frac{\epsS}{\sin\thetaS} \biggr)^2
  \frac{1}{4 \lambda_h}
  \biggl( \frac{\ms}{m_h} \biggr)^2
  \,, 
\end{align}
and the measured SM Higgs parameters,
we infer the allowed range of $\lambda_s$ for representative choices of the portal coupling $\epsS$.
Since experimental constraints are most directly expressed through the physical mass
$\ms$ and mixing angle $\sin\thetaS$, we trade the portal coupling $\epsS$
for the parameter set $(\ms,\lambda_s,\sin\thetaS)$.
For each scan point, the value of $\epsS$ is fixed by
eq.~\eqref{Dark_higgs_parameters},
which follows from the scalar mass matrix in the small-mixing limit.

We close this section with an important phenomenological implication
for the parameter scan in sec.~\ref{sec:pta}.
The relation in eq.~\eqref{Dark_higgs_parameters}
imposes a constraint on the dimensionful ratio $\lambda_s/\ms^2$.
Using the Higgs boson mass, $\mh=125.1$~GeV,
its derived self-coupling,
$\lambda_h \simeq 0.13$~\cite{ParticleDataGroup:2026aaa},
and
the allowed range of the mixing angle in
eq.~\eqref{mixing_mass_limits} for the dark scalar masses considered here,
we obtain
\begin{equation}
  \label{eq:lambda_s_constraint}
  \frac{\epsS^2}{\left( \sin \thetaS^{\rmi{max}} \right)^2}
  \;\le\;
  \frac{\lambda_s}{1.2 \times 10^{-10}} \left( \frac{\text{MeV}}{\ms} \right)^2
  \;\le\;
  \frac{\epsS^2}{\left( \sin \thetaS^{\rmi{min}} \right)^2}
  \, ,
\end{equation}
where the minimal and maximal mixing angles are given by
eq.~\eqref{mixing_mass_limits}.
This relation guides the phenomenologically viable
parameter space of the model and may impose additional restrictions on
the parameter scan presented in
tab.~\ref{tab:pta:scan}
through the requirement $\epsS < 1$.

\subsubsection{Dark photon portal}

Constraints on massive dark photons are typically parametrized by
the kinetic mixing parameter $\epsV$, which controls the coupling of $V_\mu$
to the electromagnetic current after electroweak symmetry breaking. 
Comprehensive bounds are summarized in~\cite{Fabbrichesi:2020wbt}. 
For the dark photon mass range relevant to this work,
$\mV\sim\mathcal{O}(1\text{--}100)$~MeV,
the most stringent constraints arise from astrophysical and cosmological probes
sensitive to production and decay of
weakly-coupled light vectors in supernova cores and the early universe.

In particular,
SN1987A cooling arguments constrain the emission of dark photons from
the supernova core and typically exclude intermediate values of
the kinetic mixing~\cite{Chang:2016ntp}. 
Additional limits follow from the non-observation of $\gamma$-rays associated with
late decays of supernova-produced dark photons~\cite{DeRocco:2019njg},
as well as from envelope and trapping effects in core-collapse supernovae~\cite{Sung:2019xie}. 
At the same time, BBN bounds derived from photo-dissociation and neutron-proton
conversion processes exclude scenarios in which long-lived dark photons
inject electromagnetic energy during or after nucleosynthesis~\cite{Fradette:2014sza,Berger:2016vxi}. 
Together, these considerations imply that for
$\mV\sim\mathcal{O}(1\text{--}100)$~MeV,
the viable parameter space typically lies at very small kinetic mixing,
often $\epsV \lesssim 10^{-9}\text{--}10^{-10}$,
with details depending on the precise mass range and decay channels.

If the dark photon also couples to the dark fermion $X$,
additional constraints arise from direct-detection searches mediated by $V_\mu$.
For representative dark matter masses $10~{\rm GeV}\lesssim \mX\lesssim 100$~GeV,
the PandaX-II experiment excludes a broad region of parameter space,
roughly
$10^{-11}\lesssim \epsV \lesssim 5\times 10^{-9}$~\cite{PandaX-II:2021lap}.
Complementary constraints on $\epsV$ in this mass range
arise from fixed-target and beam-dump experiments,
although for the small mixing angles relevant here,
astrophysical and cosmological limits dominate~\cite{Fabbrichesi:2020wbt}.

In summary, thermal contact between the Standard Model and the dark sector
is primarily established through
the scalar portal, namely
the Higgs-dark scalar mixing angle $\sin\thetaS$,
while the kinetic mixing of the dark photon
is negligible in the parameter region considered in this work.
Thermal equilibrium within the dark sector, on the other hand,
is efficiently maintained by the dark gauge interactions for
the values of the gauge coupling $\gd\sim\mathcal{O}(1)$ considered throughout this study.

In the following section,
we present our strategy for determining whether the two sectors remain
in thermal equilibrium and for estimating the temperature at which they eventually decouple.

%
\subsection{Thermal equilibrium between dark and visible sectors}
\label{sec:thermal_contact}

In this section,
we investigate whether
the dark and visible sectors remain in thermal equilibrium throughout
the epoch relevant for the dark-sector phase transition.
Throughout this analysis,
we assume that the hidden-sector particles,
the dark matter fermion $X$ and
the dark bosons $S$, $V_\mu$,
efficiently thermalize among themselves, such that
the dark sector can be characterized by a single temperature $\Tds$.
This assumption is well justified for the values of the gauge coupling $\gd$ considered in this work, which lie in the range $0.1 \lesssim \gd \lesssim 1$.

In this work, we assume that thermal contact between the dark and visible sectors is
established exclusively through the scalar portal. We therefore neglect processes
involving the dark gauge boson and SM particles induced by kinetic mixing
in eq.~\eqref{eq:portal_lagrangian}.
This approximation is well motivated in the parameter region of interest,
where current constraints require the kinetic mixing parameter to
be significantly smaller than the scalar mixing angle for
the dark scalar and vector boson masses considered in our analysis.

Whenever the two sectors are in thermal equilibrium,
the same temperature also describes the SM plasma, which we denote by $T$.
Determining whether, and down to which temperature,
the two sectors remain thermally coupled is
particularly relevant for the computation of
the GW signal discussed in sec.~\ref{sec:PT_and_GW}. 
In particular, whether the dark and visible sectors share a common
temperature determines which relativistic degrees of freedom
enter the redshift factors of the GW signal, and hence its
peak frequency and amplitude, as well as justifying the use
of a single temperature throughout the computation.

Thermal equilibrium is assessed by comparing the interaction rate with
the Hubble expansion rate,
requiring
the condition
\begin{align}
  \label{eq:thermal:equilibrium:condition}
  C_\T^\rmi{eq}:\qquad
  \Gamma_{\textrm{int}}(T) > H(T)
  \,.
\end{align}
Here, $\Gamma_{\textrm{int}}$ denotes the total interaction rate,
including all relevant $1\to2$ and $2\to2$ processes involving dark-sector and SM particles,
both number-changing and number-conserving.
The value of the Hubble rate
during the radiation-dominated era at fixed temperature is
\begin{align}
  \label{eq:Hubble:rate}
  H^2(T) &= \frac{8\pi}{3} \frac{\rho_r(T)}{\mpl^2}
  \,,&
  \rho_r(T) &= \frac{\pi^2}{30} g_{\rmi{eff},\rmi{tot}}(T) T^4
  \,,&
  \mpl &= 1.22 \times 10^{19}~{\rm GeV}
  \,,
\end{align}
where
$g_{\rmi{eff},\rmi{tot}} =
g_{\rmi{eff},\rmii{SM}}
+
g_{\rmi{eff},\rmii{DS}}$
is the total effective number of relativistic degrees of freedom
of the SM and the dark sector~(DS).
For the SM contribution,
we use the results of~\cite{Laine:2015kra},
supplemented by the relativistic bosonic degrees of freedom of
the dark sector.%
\footnote{%
  Throughout this work we take $\mX \gtrsim 1$~GeV,
  so that the dark fermion is non-relativistic across
  the entire temperature range of interest.
  Rather than explicitly computing thermodynamic quantities
  for the dark-sector states,
  we adopt a phenomenological prescription for their contribution
  to the energy density entering the Hubble rate.
  More precisely, the scalar and vector boson contributions are
  suppressed by Boltzmann factors
  $e^{-\ms/T}$ and $e^{-m_V/T}$, respectively.
}

For dark-sector phase transitions occurring well
below the electroweak crossover temperature,
$\TcSM \simeq 160$~GeV~\cite{DOnofrio:2015gop},
we can restrict our analysis to the regime where electroweak symmetry
is already broken and all SM particles, except neutrinos, are massive.
We therefore focus on the temperature interval $1~{\rm MeV} \lesssim T \lesssim 1~{\rm GeV}$.
To obtain two separate phase transitions,
$(0,0)\rightarrow(v_h,0)\rightarrow(v_h,v_s)$,
we require the intermediate electroweak-broken phase to be stable against fluctuations
in the singlet direction.
This is controlled by the curvature of the finite-temperature
effective potential near the electroweak transition,
\begin{align}
  \frac{\partial^2 V}
  {\partial S\,\partial S^*}
  \bigg|_{S=0}
  =
  -\mu_s^2
  +\frac{\epsS}{2}v_h^2(T)
  +\left(\frac{\lambda_s}{3}+\frac{\gd^2}{4}\right)T^2
  \,,
\end{align}
where the LO thermal mass correction of the dark scalar is given in
eq.~\eqref{eq:m:S:2l}. Following the standard treatment of Higgs-portal models,
the zero-temperature contribution $\epsS v_h^2/2$
induced after electroweak symmetry breaking is absorbed into
a redefinition of the quadratic parameter $\mu_s^2$.
The above stability condition is therefore imposed only for temperatures
near the electroweak transition, where $v_h(T)$ differs
from its zero-temperature value.
Since $\epsS>0$, the Higgs condensate contributes positively to
the curvature in the singlet direction, thereby favouring
the stability of the intermediate electroweak-broken phase.
We have explicitly verified that this condition is satisfied
throughout the parameter space in tab.~\ref{tab:pta:scan}.

The relevant interaction channels depend on the phase of the dark sector.
A key feature of the model is that mixing between
the dark scalar and
the SM Higgs boson arises only
after spontaneous symmetry breaking in the dark sector, i.e.
when $v_s \neq 0$.
The cross term $\sim \epsS v_h v_s$ then induces
the  physical mixing angle $\sin\thetaS$,
opening decay and scattering channels into SM states
that can substantially enhance thermal contact between the two sectors,
allowing thermal equilibrium to be maintained even for
relatively small values of $\sin\thetaS$.

For temperatures above the critical temperature of the dark sector,
$T>\TcDS$,
the only relevant $1\to2$ process is the Higgs decay
$h \to S S^\ast$ and the corresponding inverse decay.
The associated $2\to2$ number-changing processes are
\begin{align}
    hh &\to S S^\ast
    \, ,&
    ZZ &\to S S^\ast
    \, ,&
    W^{+}W^{-} &\to S S^\ast
    \, ,&
    f\bar{f} &\to S S^\ast
    \, ,
\end{align}
where $f$ denotes SM fermions.
For brevity, only processes with dark scalars in the final state are shown.
The corresponding reverse reactions are implicitly included.
Additionally, there are number-conserving scattering processes and their conjugates,
\begin{align}
    S h &\to S h
    \, ,&
    Z S &\to Z S
    \, ,&
    W^{-} S &\to W^{-} S
    \, ,&
    f S &\to f S
    \, .
\end{align}
The first class of $2\to2$ processes coincides with
those encountered in Higgs-portal scalar dark matter models;
see e.g.~\cite{Wu:2016mbe} for the corresponding cross-sections.

After spontaneous symmetry breaking in the dark sector,
the physical dark scalar $s$ acquires couplings to SM fermions
through Higgs mixing.
Consequently, in addition to the processes listed above
(with the replacement $S\to s$),
several new decay and scattering channels become relevant whenever
kinematically allowed.
Examples include the decays $s \to f\bar{f}$ and scattering processes such as
$s \gamma \to f\bar{f}$ and $s f \to \gamma f$,
where $\gamma$ denotes the SM photon.
Similar processes involving QCD gluons may also occur,
whereas those involving the electroweak massive gauge bosons
are strongly suppressed in the temperature range considered here.
Related processes have been investigated previously
in~\cite{Krnjaic:2015mbs,Evans:2017kti,Ibe:2021fed}.
In the present work, we independently compute all relevant rates and
include them in the evaluation of the total interaction rate
$\Gamma_{\rmi{int}}(T)$.

Compared to earlier studies~\cite{Evans:2017kti,Ibe:2021fed},
we additionally include the class of processes induced by
scalar self-interactions in eq.~\eqref{lag_model_broken}.
Throughout, we work at LO in 
the small mixing angle $\sin\thetaS$ and keep only diagrams containing
a single insertion of a mixing-induced vertex.
For the thermal collision integrals associated with the $2\to2$ processes,
we neglect Fermi-Dirac and Bose-Einstein statistical factors,
following the standard approximation commonly adopted in
phenomenological studies, including~\cite{Evans:2017kti}.
This approximation typically induces uncertainties at
the $\mathcal{O}(10\%)$ level~\cite{Davidson:2008bu}.%
\footnote{%
  Since improvements act in several directions simultaneously,
  and go beyond correcting statistical factors alone,
  we leave such an analysis for future work.
  A more refined computation would also require
  an accurate evaluation of dark scalar decays into SM leptons,
  accounting for multiple soft scatterings and
  thermal masses~\cite{Anisimov:2010gy,Ghiglieri:2016xye,Biondini:2020ric}.
}

In the temperature range of interest, $T<1$~GeV,
heavy SM particles are strongly Boltzmann suppressed.
Hence, thermal equilibrium is maintained predominantly
through interactions with the light SM degrees of freedom,
namely leptons and light quarks.
Below the QCD crossover temperature,
$T \lesssim 155$~MeV~\cite{Bazavov:2018mes},
a hadronic description replaces the partonic one.
Following~\cite{Ibe:2021fed}, we implement the effective interactions
between the dark scalar and pions and kaons,
and find that their contribution to the interaction rate is subleading.
In practice, processes involving quarks and hadrons play only
a minor role throughout most of the temperature range considered here,
either due to Boltzmann suppression of the heavier quarks
or to the comparatively small hadronic interaction rates
below the QCD crossover.

A similar hierarchy applies to leptons.
The mixing-induced couplings
to SM fermions scale as $\propto \sin\thetaS \, m_f/v_h$,
suppressing processes involving very light fermions.
At the same time, the tau lepton, with mass $m_\tau \simeq 1.7$~GeV,
becomes rapidly Boltzmann suppressed below $T \lesssim 100$~MeV.
Consequently, among the leptonic channels, muon-induced processes
typically dominate the interaction rate.

\begin{figure}[t]
    \centering
    \includegraphics[width=0.5\linewidth]{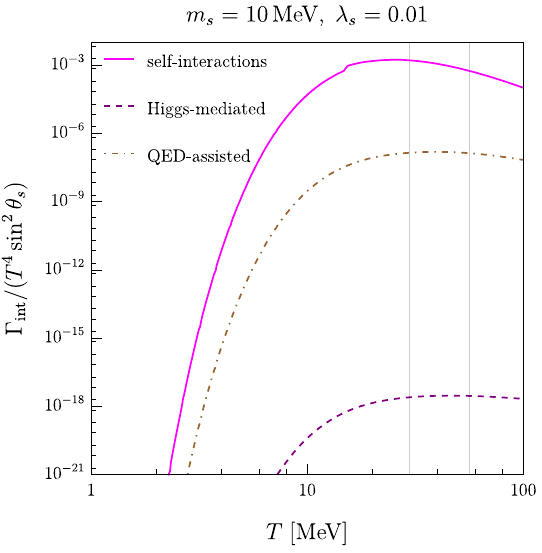}%
    \includegraphics[width=0.5\linewidth]{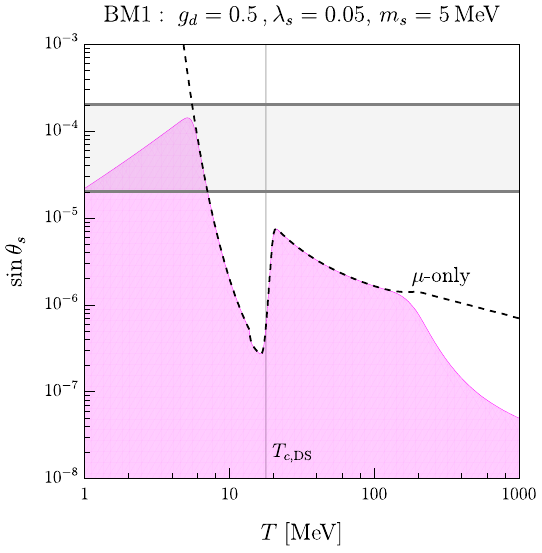}
    \caption{%
    Left:
    Thermal interaction rate normalized by $T^4 \sin^2 \thetaS$ for processes involving muons.
    Three classes of $2 \to 2$ scatterings are shown:
    dark-scalar self-interaction mediated (solid magenta),
    Higgs mediated (dashed purple), and
    QED-assisted processes (dot-dashed brown).
    Right:
    Equilibration condition in the $(T,\sin \thetaS)$
    plane for fixed values of the dark-sector couplings and masses.
    The magenta shaded region corresponds to $\Gamma_{\rmi{int}} < H$,
    where the two sectors lose thermal contact.
    The contribution obtained by including only
    muon-induced processes is also shown (black dashed),
    while the critical temperature $\TcDS$,
    is indicated by the vertical gray line.
  }
  \label{fig:rates_and_eq}
\end{figure}
We now discuss the results of the thermalization analysis.
Figure~\ref{fig:rates_and_eq} (left) shows the interaction rates
in the broken phase of the dark sector, where three classes of
$2 \to 2$ processes involving SM leptons contribute.
Focusing on the muon channel, we find that
processes induced by dark-scalar self-interactions dominate
the interaction rate below the dark-sector phase transition.
The vertical gray lines indicate the critical temperature for
two benchmark choices of $\gd$, with
$\lambda_s=0.01$ and $\ms=10$~MeV held fixed (for $g_d=0.5$ and $g_d=0.8$, corresponding to the leftmost and rightmost vertical lines, respectively.). 
Higgs-mediated processes in the $s$- or $t$-channel
are strongly suppressed by the Higgs mass.
By contrast, the self-interaction-mediated channels are competitive with,
and often dominate over, the QED-assisted processes
$s \mu \to \mu \gamma$ and
$s \gamma \to \mu \bar{\mu}$.
This picture also holds when lowering
$\lambda_s$ and $\ms$ to the lower values relevant to our work.
The enhancement induced by scalar self-interactions is also responsible
for the sharp decrease of the mixing angle required
to maintain thermal equilibrium below the critical temperature
$\TcDS$, as seen in
fig.~\ref{fig:rates_and_eq} (right).

Figure~\ref{fig:rates_and_eq} (right) shows the thermalization
condition $\Gamma_{\rmi{int}} \geq H$ (white region)
for the benchmark point BM1 discussed in sec.~\ref{sec:pta}.
The magenta shaded region marks where thermal equilibrium between
the two sectors is lost.
We also superimpose a gray band indicating the phenomenologically viable
range of $\sin\thetaS$~\eqref{mixing_mass_limits}.

To illustrate the dominant role of muon-induced processes in most of
the temperature range, we additionally display
the contour obtained by including only processes involving muons
(black dashed).
This approximation underestimates the total interaction rate for
temperatures $T \gtrsim 100$~MeV,
where tau leptons and charm quarks still provide sizable contributions due
to their relatively mild Boltzmann suppression.
On the other hand, in the low-temperature regime,
$T \lesssim 5$~MeV, muon-induced processes become exponentially
suppressed and the dominant contribution instead arises from
scalar decays into electrons,
$s \to e \bar{e}$, which constitute
the only kinematically accessible decay channel.
Decays into neutrinos instead remain negligible due to the tiny neutrino masses.

For the parameter scan
in tab.~\ref{tab:pta:scan} and the analysis of
the stochastic GW background in sec.~\ref{sec:pta},
we explicitly verified the thermalization condition using
the interaction rates described in this section.
We find that the two sectors do not always share the same temperature
throughout the parameter space explored in this work.
See the discussion in sec.~\ref{sec:pta}.

%
\section{Gravitational waves from first-order phase transitions}
\label{sec:PT_and_GW}

The model of eq.~\eqref{eq:lag:4d} can undergo
a first-order phase transition in
the early universe~\cite{%
  Weinberg:1992ds,Metaxas:1995ab,Kajantie:1997hn,Garny:2012cg,
  Hirvonen:2021zej,Lofgren:2021ogg,Bernardo:2025vkz}.
We focus on a dark-sector transition
occurring well below the SM electroweak crossover,
$\TcDS \ll \TcSM$.
In this section,
we outline the framework used to determine the corresponding
thermodynamic and GW observables,
starting from the thermal effective action in
sec.~\ref{sec:dim_red},
the phase-transition thermodynamic parameters in sec.~\ref{sec:pt_thermo},
and the resulting GW spectrum in
sec.~\ref{sec:GW_spectrum}.

%
\subsection{Dimensionally reduced effective action}
\label{sec:dim_red}

At high temperatures,
equilibrium thermodynamics
is most efficiently
described via the effective action of
a three-dimensional effective field theory (\threeDEFT{})
obtained via high-temperature dimensional reduction~\cite{%
  Ginsparg:1980ef,Appelquist:1981vg,Nadkarni:1988fh,Landsman:1989be,
  Kajantie:1995dw,Braaten:1995jr,Braaten:1995cm}.
The latter is formulated in the imaginary-time formalism and
in Euclidean space-time.
Since the heat bath singles out a rest frame $u_\mu = (1,\vec{0})$,
Lorentz invariance is no longer explicitly manifest~\cite{Weldon:1982aq} and
additional interactions involving
the temporal component of the gauge field,
the temporal vector $V_0$, can
arise.
For the model of eq.~\eqref{eq:lag:4d},
the field $V_0$
acquires a Debye mass $\mD \sim \mathcal{O}(\gd T)$.

The corresponding
\threeDEFT{} Lagrangian is purely bosonic~\cite{%
  Karjalainen:1996rk,Kajantie:1997vc,Kajantie:1997hn,Andersen:1997ba}
and up to mass dimension four,
its super-renormalizable part is
\begin{align}
\label{eq:lag:3d}
  \mathcal{L}^{3d} &=
      \frac{1}{4} V_{ij}^{2}
    + (D_i S)^* (D_i S)
    + \mu_{s,3}^2 S^* S
    + \lambda_{s,3}^{ } (S^* S)^2
    + \mathcal{L}_{\tau}
    \,,\\[1mm]
\label{eq:lag_3d:temporal}
\mathcal{L}_{\tau} &=
    \frac{1}{2} (\partial_{i}^{ } V_{0}^{ })^2
  + \frac{1}{2} \mD^{2} V_{0}^2
  + h_{3}^{ } V_{0}^2 S^* S
  + \frac{\kappa_{3}^{ }}{4} V_{0}^4
  \;,
\end{align}
where
$\mathcal{L}_{\tau}$ encodes the contributions of the temporal vector.
The spatial covariant derivative
with indices $i = \{1,\dots,d\}$ reads
$D_i = \partial_i - i g_{d,3} V_i$, with $g_{d,3}$ the
three-dimensional effective gauge coupling.
In the broken phase,
the gauge fixing functional takes the same form as
in the fundamental 4d theory of sec.~\ref{sec:model}.

The matching of
the super-renormalizable \threeDEFT{} parameters to
the fundamental 4d theory
is performed automatically using 
{\tt DRalgo}~\cite{Ekstedt:2022bff,Bernardo:2026nyq}.
Since fermions are non-dynamical at high temperatures,
their effects enter only implicitly as ultraviolet (UV) corrections to the effective parameters.
The matching relations up to
next-to-next-to-leading order (NNLO) in the coupling expansion
and dimension four in mass dimension
are listed in appendix~\ref{sec:dimensional_reduction_details}
as well as in~\cite{Hirvonen:2021zej} and
employed in our numerical scans of sec.~\ref{sec:pta}.
Compared to~\cite{Hirvonen:2021zej}, we extend the matching
by including fermionic mass effects at one-loop order.%
\footnote{%
  \label{footnote:fermion_matching}
  Since the dark fermion mass $\mX \sim \mathcal{O}(10\text{--}10^2)$~GeV
  is much larger than the MeV dark-sector phase transition scale,
  fermionic contributions remain Boltzmann suppressed
  for the entire parameter space range relevant to our analysis.
  Our matching relations \eqref{eq:gd:soft}--\eqref{eq:kappa:soft}
  exhibit full fermion-mass dependence and
  are useful when the dark fermion is dynamical during the phase transition.
}

For the strongest transitions,
higher-dimensional operators become relevant~\cite{Bernardo:2025vkz,Chala:2024xll}
and
it is in practice insufficient to consider only the super-renormalizable operators of
eqs.~\eqref{eq:lag:3d} and~\eqref{eq:lag_3d:temporal}. 
The full set of dimension-six operators in
the dimensionally reduced Abelian Higgs model is derived 
in~\cite{Bernardo:2025vkz}
and was recently generalized for generic models in
{\tt DRalgo}~\cite{Bernardo:2026nyq}.
See also~\cite{Fuentes-Martin:2026bhr,Chakrabortty:2026swu,Bandyopadhyay:2026nrv}
for alternative automatic construction of static thermal EFTs.
Including higher-dimensional operators in the thermal EFT can only
extend the validity of the high-temperature expansion over
a limited region of parameter space~\cite{Bernardo:2025vkz}.
If the EFT breaks down at a given point due to the omission of
dimension-six operators, then including those operators merely
postpones the breakdown to the point where dimension-eight operators become
important, and so on.
Hence, we
only indirectly include dimension-six operators in the construction of the EFT
by inspecting the criteria~\eqref{eq:bro:soft} and \eqref{eq:bro:softer}
of~\cite{Bernardo:2025vkz} which
determine if individual parameter points
remain valid under the assumption of high-temperature expansion;
cf.\ sec.~\ref{sec:breakdown_highT}.

The construction of the EFT proceeds in two steps.
First, the hard modes $\sim \pi T$
are integrated out at vanishing scalar background
$\phi = 0$, yielding the
{\em soft-scale} EFT for
the symmetric phase~\cite{Bernardo:2025vkz,Bernardo:2026whs}:

\begin{probox}{Symmetric phase (sym)}
  \phantomsection\label{box:sym}
  The temporal vector $V_0$ with Debye mass $\mD$ are heavy,
  the two scalar degrees of freedom $S$ are light,
  and vector bosons $V_i$ are massless.
  The hierarchy is
  \begin{align}
    m_{\rmii{$V$},3}^{2}
    \ll
    m_{s,3}^{2}
    \ll \mD^{2} \ll \pi T
    \,.
  \end{align}
\end{probox}
Second, allowing for a non-vanishing background
in the soft \threeDEFT{}, $\phi_3$,
and using the scalar-field parametrization~\eqref{eq:param:S}
in three-dimensions and Landau gauge ($\xi=0$),
also the spatial gauge bosons acquire a mass~\cite{Hirvonen:2021zej}
\begin{align}
  m_{s,3}^{2} &=
    \mu_{s,3}^{2}
    + 3\lambda_{s,3}^{ } \phi_{3}^2
  \,,&
  m_{\chi,3}^{2} &=
    \mu_{s,3}^{2}
    + \lambda_{s,3}^{ } \phi_{3}^2
  \,,\nn
  m_{\rmii{$V$},3}^{2} &=
    g_{d,3}^{2}\phi_{3}^2
  \,,&
  m_{\rmii{$V_0$}}^{2} &=
      \mD^{2}
    + h_{3}^{ } \phi_{3}^2
  \,.
\end{align}
At the phase transition,
vector and scalar masses are hierarchically separated
such that
$x \simeq m_{s,3}^2/m_{\rmii{$V$},3}^2$
is the effective expansion parameter~\cite{Ekstedt:2022zro,Lofgren:2021ogg,Gould:2023ovu};
see eq.~\eqref{eq:Veff:bro:LO:dimless}
for a precise definition.
The resulting broken-phase theory
is a 3d Higgs EFT (HEFT) theory~\cite{Kajantie:1997hn}
also obtainable in {\tt DRalgo}~\cite{Ekstedt:2022bff,Bernardo:2026nyq}.
The relevant scale hierarchies are:
\begin{probox}{Broken phase (bro) soft EFT}
  \phantomsection\label{box:bro:soft}
  Spatial $V_i$ and
  temporal vectors $V_0$ acquire a field-dependent mass
  of $\mathcal{O}(g_{d,3} \phi_3)$
  and can be
  integrated out on the same footing.
  The hierarchy and validity range~\cite{Bernardo:2025vkz}
  is\footnotemark
  \begin{align}
    \label{eq:bro:soft}
    m_{s,3}^2 &\ll m_{\rmii{$V$},3}^{2} \sim m_{\rmii{$V_0$}}^{2} \ll \pi T
    \,,&
    0.051\gd^2 &\ll x \ll 0.18
    \,.
    \tag{EFT1}
  \end{align}
\end{probox}%
\footnotetext{%
  Here $m_{s,3}$ denotes the common mass scale of all light scalar
  degrees of freedom, comprising the physical scalar and
  its associated Goldstone boson $m_{\chi,3}$.
  Although in $R_\xi$ gauge the Goldstone mass can reach
  $m_{\chi,3} \sim m_{\rmii{$V$},3}$ individually, the Goldstone and ghost contributions
  combine to give a ultrasoft net correction~\cite{Hirvonen:2022jba}.
}%
In practice,
another hierarchy can emerge in a setup known as the softer EFT setup~\eqref{eq:bro:softer},
where the temporal vector is heavier than the spatial gauge bosons:
\begin{probox}{Broken phase (bro) softer EFT}
  \phantomsection\label{box:bro:softer}
  Spatial $V_i$ and
  temporal vectors $V_0$ acquire a field dependent mass
  of $\mathcal{O}(g_{d,3} \phi_3)$
  and can be
  integrated out successively.
  The hierarchy and validity range~\cite{Bernardo:2025vkz}
  is
  \begin{align}
    \label{eq:bro:softer}
    m_{s,3}^2 &\ll m_{\rmii{$V$},3}^{2} \ll m_{\rmii{$V_0$}}^{2} \ll \pi T
    \,,&
    0.18\gd &\ll x \ll 0.18
    \,.
    \tag{EFT2}
  \end{align}
\end{probox}%
\noindent
In this setup, higher-dimensional operators become
significantly relevant
in the strong transition regime,
which reduces its region of validity.
We will investigate
how such a breakdown manifests in
thermodynamic quantities
in sec.~\ref{sec:breakdown_highT}.

The Abelian Higgs model at finite temperature,
falls into a broader class of models
with a generic radiatively induced cubic barrier.
The LO
broken-phase potential of~\ref{eq:bro:soft} takes the form
\begin{align}
  \label{eq:Veff:bro:LO}
  \Veff^\rmi{bro}(\phi_3) &=
      \frac{1}{2} \mu_{s,3}^2 \phi_3^2
    + \frac{1}{4} \lambda_{s,3} \phi_3^4
    - \eta_{1}^{ }\, \phi_3^3
    - \bigl(\yD^{ } + \eta_2^{2/3} \phi_3^2\bigr)^{3/2}
    \,,
\end{align}
where the barrier is induced by vector and temporal-scalar
UV contributions; see
e.g.~\cite{Bernardo:2025vkz,Kierkla:2025qyz}.
Concretely,
for the Abelian Higgs model,
the barrier is parametrized by
the vector-induced cubic coupling $\eta_1 \sim g_{3,d}^{3}$,
the Debye screening mass $\yD^{ } \sim \mD^2$,
and the coupling between temporal and Lorentz scalars
$\eta_2^{ } \sim h_3^{3/2}$.
In three dimensions, the mass dimensions are
$[\phi_3] = 1/2$,
$[\eta_1] = 3/2$, and
$[\mD] = [\lambda_{\phi,3}] = 1$.

In the limit of a strong transition,
the barrier is dominated by the field-dependent spatial contributions.
Since then $\yD^{ } \ll \eta_2^{2/3} \phi_3^2$,
the temporal vector contribution reduces to a pure cubic term,
and the potential simplifies to the form
\begin{align}
  \label{eq:Veff:bro:LO:strong}
  \Veff^\rmi{bro}(\phi_3) &=
      \frac{1}{2} \mu_{s,3}^2 \phi_3^2
    + \frac{1}{4} \lambda_{s,3} \phi_3^4
    - \eta\, \phi_3^3
    \,,
\end{align}
which has the same structure of
the LO potential of~\eqref{eq:bro:softer}.
Here, however,
$\eta = \eta_1 + \eta_2$ is the effective cubic coupling
where
the first term
is the pure vector contribution from
$\eta_1 = g_{d,3}^3/(6\pi)$,
and the second term is the contribution from the temporal vector.
In the Abelian Higgs model,
the temporal and vector induced radiative contributions
combine to 
\begin{align}
  \label{eq:barrier:shift}
    (d-3)J_{3}(m_\rmii{$V$,3})
  + J_{3}(m_\rmii{$V_0$}) =
  \frac{1}{12\pi}\Bigl(
      2m_\rmii{$V$,3}^3
    + m_\rmii{$V_0$}^3
    \Bigr) =
    \frac{g_{d,3}^3}{6\pi}\Bigl(
        1
      + \frac{1}{2} \tilde{h}_3^{3/2}
    \Bigr) \phi_3^3
  = g_{d,3}^3 \mathcal{E}_{1}^{ } \phi_3^3
  \,,
\end{align}
where
$\tilde{h}_3$ and
$\mathcal{E}_{1}$ are defined in
eqs.~\eqref{eq:temporal:vector:enhancement} and~\eqref{eq:dimless:couplings} and
$J_3(m)$ is defined in eq.~\eqref{eq:1loop-master-3d}.
The modified cubic barrier
amounts to installing an effective gauge coupling
$g_{\rmi{eff}}^2 = g_{d,3}^2\mathcal{E}_{1}^{2/3}$
which in turn
yields an effectively
larger barrier
giving rise to stronger transitions.
Henceforth,
we will display all quantities in units of $\eta \equiv g_{d,3}^3$.

Rescaling $\phi_3 = \eta^{1/3}\varphi$,
with $\varphi$ the dimensionless scalar background,
the LO potential can be written in its dimensionless
form~\cite{%
  Dine:1992vs,Dine:1992wr,Baacke:1993ne,Dunne:2005rt,Ekstedt:2021kyx}
\begin{align}
  \label{eq:Veff:bro:LO:dimless}
  \widetilde{V}_\rmi{eff}^\rmi{bro} \equiv \frac{\Veff^\rmi{bro}(\phi_3)}{\eta^2} =
      \frac{1}{2} y \varphi^{2}
    + \frac{1}{4} x \varphi^{4}
    - \mathcal{E}_{1}^{ }\varphi^{3}
    \,,&&
x &\equiv \frac{\lambda_{s,3}}{\eta^{2/3}}
\,,&
y &\equiv \frac{\mu_{s,3}^2}{\eta^{4/3}}
\,.
\end{align}
In this context, the strongest transitions
occur for $x \lesssim 0.18$~\cite{Bernardo:2025vkz},
with larger values of $\eta$ further enhancing the transition strength.

Both the symmetric- and broken-phase
effective potentials are known to N$^4$LO in this
setup~\cite{Ekstedt:2024etx}.
In our numerical scans,
we modify those results by accounting for temporal vector
effects up to NNLO,
following~\cite{Bernardo:2026whs}
in Landau gauge~\cite{Martin:2018emo},
\begin{align}
  \label{eq:Veff:bro:N1LO}
  \widetilde{V}_{\text{eff}}^\rmi{bro}\big|_{\rmii{NLO}}&=
    - \frac{\tilde{g}_{d,3}^4\varphi^2}{(4\pi)^2}\bigg[
      \mathcal{E}_2
    + \mathcal{E}_3\,\ln\frac{\Lamd^2}{4\mV^2}
    \bigg]
  \,,\\[1mm]
  \label{eq:Veff:bro:NNLO}
  \widetilde{V}_{\text{eff}}^\rmi{bro}\big|_{\rmii{NNLO}}&= -\frac{1}{12\pi}\bigg[
      \bigl(\widetilde{m}_{s,3}^2\bigr)^{\frac{3}{2}}
    + \bigl(\widetilde{m}_{\chi,3}^2\bigr)^{\frac{3}{2}}
    \bigg]
  \,,\\[1mm]
  \label{eq:Veff:sym:N2LO}
  \widetilde{V}_{\text{eff}}^\rmi{sym}\big|_{\rmii{NNLO}}&=
  -\frac{1}{6\pi}y^{\frac{3}{2}}
  \,,
\end{align}
using the temporal vector enhancement factors
\begin{align}
  \label{eq:temporal:vector:enhancement}
    \mathcal{E}_1 &\equiv
      \frac{1}{6\pi}\Bigl(
          1
        + \frac{1}{2} \tilde{h}_3^{3/2}
      \Bigr)
      \,,&
    \mathcal{E}_2 &\equiv
          1
        + \frac{1}{2}\tilde{h}_3^2\bigl(1 - \ln\tilde{h}_3\bigr)
        + 3\tilde{\kappa}_3\tilde{h}_3
    \,,&
    \mathcal{E}_3 &\equiv
          1
        + \frac{1}{2}\tilde{h}_3^2
    \,.
\end{align}
The dimensionless versions of
the effective gauge coupling,
the coupling between temporal and Lorentz scalars, and
the self-coupling of the temporal vectors are defined as
\begin{align}
  \label{eq:dimless:couplings}
  \tilde{g}_{d,3}^2 &= \frac{g_{d,3}^2}{\eta^{2/3}}
  \,,&
  \tilde{h}_3 &= \frac{h_3}{\eta^{2/3}}
  \,,&
  \tilde{\kappa}_3 &= \frac{\kappa_3}{\eta^{2/3}}
  \,.
\end{align}
Setting $\mathcal{E}_1 = 1/(6\pi)$ and
$\mathcal{E}_2 = \mathcal{E}_3 = \tilde{g}_{d,3}^4 = 1$
recovers the results of~\cite{Ekstedt:2024etx,Gould:2023ovu}.
The dimensionless resummed field-dependent scalar masses appearing in
eq.~\eqref{eq:Veff:bro:NNLO} are
\begin{align}
    \partial_\varphi^2 \widetilde{V}_\rmi{eff}^\rmi{bro} 
    = 
    \widetilde{m}_{s,3}^2 &\equiv \frac{m_{s,3}^2}{\eta^{4/3}}=
        y
      + 3x\varphi^2
      - 6\mathcal{E}_{1} \varphi
    \,,\nn
    \varphi^{-1}\partial_\varphi \widetilde{V}_\rmi{eff}^\rmi{bro} 
    =
    \widetilde{m}_{\chi,3}^2 &\equiv \frac{m_{\chi,3}^2}{\eta^{4/3}}=
          y
        + x\varphi^2
        - 3\mathcal{E}_{1} \varphi
    \,.
\end{align}

For the effective action,
rescaling the spatial coordinates and scalar background as
$\vec{x} \to \tilde{\vec{x}}\, y^{-\frac{1}{2}} \eta^{-\frac{2}{3}}$ and
$\varphi \to y/\mathcal{E}_{1} \varphi$,
the LO action takes the form
\begin{align}
  \label{eq:action:LO:dimless}
  S_\rmii{LO} &=
    \int_{\vec{x}}\biggl\{
      \frac{1}{2} (\partial_i \phiB)^2
    + \Veff^\rmii{LO}(\phiB)
    \biggr\}
  \nn &= 
  \kappa
  \int_{\tilde{\vec{x}}}\biggl\{
      \frac{1}{2} (\tilde{\partial}_i \varphiB)^2
    + \frac{1}{2} \varphiB^{2}
    + \frac{\gamma}{4} \varphiB^{4}
    - \varphiB^{3}
    \biggr\}
  \equiv
  \kappa\, \tilde{S}_\rmii{LO}(\gamma)
  \,,
\end{align}
where
$\kappa = y^{3/2}/\mathcal{E}_{1}^{2}$ and
$\gamma = x y/ \mathcal{E}_{1}^{2}$.
The bounce solution $\phiB = \phiB(\vec{x})$ is obtained by solving
the bounce equation of motion and 
$\varphiB$ is the dimensionless bounce solution.
By further shifting
$\mathcal{E}_1 \to 2\mathcal{E}_1^{ }$ and
$x \to 2x$ or 
$\kappa \to \tilde{\kappa} = \kappa/4$ and
$\gamma \to \tilde{\gamma} = \gamma/2$,
we can ensure that
$\tilde\gamma(\yc) = 1$ at the critical temperature $\Tc$;
see eq.~\eqref{eq:yc:NNLO} for the definition of $\yc$.
This way one can expand
the dimensionless action $\tilde{S}_\rmii{LO}(\tilde\gamma)$
around $\tilde\gamma = 1$,
which admits the fit~\cite{%
  Dine:1992wr,Ekstedt:2021kyx,Ekstedt:2022ceo,Matteini:2024xvg,Brdar:2025gyo%
  }
\begin{align}
  \label{eq:action:LO:fit}
  f(\tilde\gamma) &=
      c_1
    + c_2 \tilde\gamma
    + c_3 \tilde\gamma^2
    + c_4 (1-\tilde\gamma)^{-1}
    + c_5 (1-\tilde\gamma)^{-2}
    + c_6 (1-\tilde\gamma)^{-3}
  \,,
\end{align}
provided that $\tilde\gamma < 1$.
Naturally,
$\tilde{S}_\rmii{LO}(\tilde\gamma)$
diverges as $\tilde\gamma \to 1$,
which corresponds to the limit of a vanishing barrier and
tunneling rate.
The bounce solution $\phiB(\vec{x})$ is then obtained
numerically using
{\tt BubbleDet}~\cite{Ekstedt:2023sqc} and
{\tt CosmoTransitions}~\cite{Wainwright:2011kj},
with the fitting coefficients listed in tab.~\ref{tab:fit:action};
we also verified the results using
{\tt FindBounce}~\cite{Guada:2020xnz}.
For a more general fitting approach of bounce actions see~\cite{Bian:2025yfj}.
\begin{table}[t]
  \renewcommand{\arraystretch}{1.1}
  \centering
  \begin{tabular}{|c|rrrrrr|}
    \hline
    \multicolumn{1}{|c|}{Action} &
    \multicolumn{1}{c}{$c_1$} &
    \multicolumn{1}{c}{$c_2$} &
    \multicolumn{1}{c}{$c_3$} &
    \multicolumn{1}{c}{$c_4$} &
    \multicolumn{1}{c}{$c_5$} &
    \multicolumn{1}{c|}{$c_6$} \\
    \hline
    \hline
    $\tilde{S}_\rmii{LO}(\tilde\gamma)$ &
    7.674(7) &
    4.26(3) &
    1.11(5) &
    10.413(2) &
    1.24396(6) &
    0
    \\
    \hline
    $\tilde{S}_\rmii{NLO}^{(a)}(\tilde\gamma)$ &
    55.72(3) &
    15.9(1) &
    $-$16.7(2) &
    53.040(9) &
    5.5867(3) &
    0
    \\
    $\tilde{S}_\rmii{NLO}^{(b)}(\tilde\gamma)$ &
    30.5(2) &
    17(1) &
    36(2) &
    3.17(7) &
    20.182(2) &
    4.95936(2)
    \\
    $\tilde{S}_\rmii{NLO}^{(c)}(\tilde\gamma)$ &
    20.1(3) &
    7(2) &
    75(2) &
    $-$34.9(1) &
    19.019(3) &
    6.87518(2)
    \\
    \hline
    $\tilde{S}_\rmii{NNLO}^{s}(\tilde\gamma)$ &
    1.5(2) &
    2.0(7) &
    $-$5(1) &
    3.40(5) &
    0.548(1) &
    0
    \\
    $\tilde{S}_\rmii{NNLO}^{\chi}(\tilde\gamma)$ &
    1.046(3) &
    0.50(1) &
    $-$1.38(2) &
    1.5133(9) &
    0.46924(3) &
    0.0328801(2)
    \\
    \hline
  \end{tabular}
  \caption{%
    Fitting coefficients for
    the LO dimensionless action
    $\tilde{S}_\rmii{LO}(\gamma)$~\eqref{eq:action:LO:dimless},
    the individual building blocks
    $\tilde{S}_\rmii{NLO}^{(a)-(c)}(\tilde\gamma)$
    of the NLO action~\eqref{eq:action:NLO}, and
    the NNLO scalar and Goldstone fluctuation-determinant
    contributions
    $\tilde{S}_\rmii{NNLO}^{s,\chi}(\tilde\gamma)$.
    Fits are obtained using
    {\tt BubbleDet}~\cite{Ekstedt:2023sqc} and
    {\tt CosmoTransitions}~\cite{Wainwright:2011kj};
    uncertainties stem from the fitting procedure.
    See~\cite{Ekstedt:2021kyx} for a similar construction.
    }
  \label{tab:fit:action}
\end{table}

At NLO
and in derivative expansion,
the effective action receives contributions
from the wave-function renormalization of the scalar field and
from the vector-induced two-loop effective potential~\eqref{eq:Veff:bro:N1LO},
\begin{align}
  \label{eq:action:NLO}
  S_\rmii{NLO} &=
  \int_{\vec{x}}\biggl\{
      \frac{Z(\phiB)}{2} (\partial_i \phiB)^2
    + \Veff^\rmi{bro}\big|_{\rmii{NLO}}(\phiB)
    \biggr\}
  \nn[1mm] &=
    \frac{y^\frac{1}{2}}{\mathcal{E}_{1}}\int_{\tilde{\vec{x}}}\biggl\{
        \frac{\tilde{g}_{d,3}^{ }}{48\pi}
        \frac{\mathcal{E}_\rmii{$Z$}^{ }}{2}\frac{(\tilde\partial_i \varphiB)^2}{\varphiB}
      - \frac{\tilde{g}_{d,3}^{4}\varphiB^2}{(4\pi)^2}\Bigl(
          \frac{\mathcal{E}_{2}}{\mathcal{E}_{1}^{ }}
        + 2\frac{\mathcal{E}_{3}}{\mathcal{E}_{1}^{ }}
        \ln\frac{\tilde{\mu}_\rmii{3d}\mathcal{E}_{1}}{2y\tilde{g}_{d,3}^{ }\varphiB}
      \Bigr)
    \biggr\}
  \,.
\end{align}
The NLO action
is evaluated on the LO bounce solution $\phiB^{ }=\phiB^\rmii{LO}$, and
scales as $\mathcal{O}(y^{1/2})$ compared
to $S_\rmii{LO}$~\eqref{eq:action:LO:dimless}.
The kinetic enhancement factor
is defined in eq.~\eqref{eq:temporal:vector:enhancement:Z}
and is $\mathcal{E}_\rmii{$Z$} = -22$
in the absence of temporal vector contributions.
Here, $\tilde{\mu}_\rmii{3d} = \eta^{-2/3}\Lamd$ is
the dimensionless 3d renormalization scale.

The wave-function renormalization term
$\propto Z(\phiB)$ is obtained from
the NLO term in the derivative expansion
in $\partial_i \sim m_\rmii{$V$,3}, m_\rmii{$V_0$}$ of
the spatial and temporal vector
fluctuation determinant~\cite{Lofgren:2021ogg,Hirvonen:2021zej,Kierkla:2025qyz},
\begin{align}
  \label{eq:wavefunction:renormalization}
  Z(\phi_3) &=
    \frac{1}{48\pi}\biggl[
    - \frac{22g_{d,3}^{ }}{\phi_3}
    + \frac{h_{3}^{2}\phi_3^2}{m_\rmii{$V_0$}^{3}}
    \biggr]
    \stackrel{h_3^{ } \phi_3^2 \gg \mD^2}{=}
    \frac{1}{48\pi}\frac{g_{d,3}^{ }}{\phi_3}
    \mathcal{E}_\rmii{$Z$}
  \,.
\end{align}
The second equality
is obtained in the limit of a strong transition,
where the field-dependent mass contribution dominates over
the Debye mass, $h_3^{ } \phi_3^2 \gg \mD^2$,
(cf.~\eqref{eq:Veff:bro:LO:strong}),
and yields the following enhancement
\begin{align}
  \label{eq:temporal:vector:enhancement:Z}
  \mathcal{E}_\rmii{$Z$} &\equiv
    - 22
    + \tilde{h}_{3}^{\frac{1}{2}}
  \;,
\end{align}
with $\tilde{h}_3 = 1 + \mathcal{O}(\gd^2)$.
In comparison to the temporal vector-induced enhancement of
the cubic barrier~\eqref{eq:temporal:vector:enhancement},
the enhancement of the wave-function renormalization term is
small and of $\mathcal{O}(1\%)$ compared to the pure spatial vector contribution.

While the wave-function renormalization term~\eqref{eq:wavefunction:renormalization}
is often neglected in the PTA literature,
e.g.~\cite{Christiansen:2025xhv,Pascoli:2026tuu},
including the full NLO action is required for
the nucleation rate to be gauge invariant at NLO~\cite{%
  Lofgren:2021ogg,Hirvonen:2021zej}.
In classically conformal models,
where the scalar mass arises entirely from radiative corrections,
the contribution of $Z(\phiB)$ has been found to significantly affect
the nucleation rate~\cite{Kierkla:2023von} and
can even signal the breakdown of
the derivative expansion itself~\cite{Kierkla:2025vwp,Kierkla:2025qyz}
when
$\partial_i \sim m_\rmii{$V$,3}$ vanishes at the bounce tail.

After identifying
the building blocks of the NLO action
in eq.~\eqref{eq:action:NLO},
\begin{align}
  \tilde{S}_\rmii{NLO}^{(a)} &=
    \int_{\tilde{\vec{x}}}
    \frac{(\tilde\partial_i \varphiB(\tilde{\vec{x}}))^2}{\varphiB(\tilde{\vec{x}})}
  \,,&
  \tilde{S}_\rmii{NLO}^{(b)} &=
    \int_{\tilde{\vec{x}}} \varphiB^{2}(\tilde{\vec{x}})
  \,,&
  \tilde{S}_\rmii{NLO}^{(c)} &=
    \int_{\tilde{\vec{x}}} \varphiB^{2}(\tilde{\vec{x}}) \ln \varphiB^{2}(\tilde{\vec{x}})
  \,,
\end{align}
their dimensionless forms are fitted to the same functional
form~\eqref{eq:action:LO:fit},
with coefficients listed in tab.~\ref{tab:fit:action}.

The NNLO contribution to the bounce action arises from
the scalar and Goldstone fluctuation determinants,
$\tilde{S}_\rmii{NNLO}^{s}(\tilde\gamma)$ and
$\tilde{S}_\rmii{NNLO}^{\chi}(\tilde\gamma)$,
again evaluated around the LO bounce solution $\phiB^\rmii{LO}$.
In principle,
these contributions can be approximated via
$A_\rmi{dyn} \times \det_\rmii{$S$} \approx T^4$
in the nucleation rate, which has been shown to be a
robust approximation~\cite{Kierkla:2025qyz}.
Here we instead compute them via their full
determinant~\cite{Ekstedt:2021kyx} using
{\tt BubbleDet}~\cite{Ekstedt:2023sqc},
and fit the results to the functional form~\eqref{eq:action:LO:fit};
the coefficients are listed in tab.~\ref{tab:fit:action}.

%
\subsection{Phase transition thermodynamics}
\label{sec:pt_thermo}

The current state-of-the-art assumption
in computing GW signals from first-order phase transitions 
is that the spectrum depends on
a few microphysical thermodynamic parameters~\cite{Caprini:2019egz}.
One central quantity
is the free energy, or equivalently the pressure $p(T)$,
that encodes the thermodynamic equilibrium behavior.

This section details the computation
of the thermodynamic quantities relevant for
computing the GW spectrum,
including
the critical temperature $\Tc$,
the percolation temperature $\Tp$,
the transition strength $\alpha$,%
\footnote{%
  Not to be confused with the gauge coupling combination $\alpha_d = \gd^2/4\pi$.
}
the inverse duration $\beta/H$,
the symmetric-phase sound speed $\cs$, and
the bubble wall velocity $\vw$.
Below,
all phase-transition quantities are
referenced to the percolation temperature~\cite{Athron:2023rfq}
such that
\begin{align}
  \label{eq:Tstar}
  T_\star = \Tp
  \,.
\end{align}
In turn,
we will encounter the following thermodynamic quantities
consisting of
the pseudotrace anomaly $\bar\theta$~\cite{Giese:2020rtr,Giese:2020znk},
the energy density $e$,
enthalpy density $w$,
entropy density $s$, and
sound speed $\cs$,
\begin{align}
  \label{eq:thermo:quantities}
  \bar\theta &\equiv e - \frac{p}{c_{s,\rmi{bro}}^2}
  \,,&
  e &= T\frac{\partial p}{\partial T} - p 
  \,,&
  w &= T\frac{\partial p}{\partial T}
  \,,&
  s &= \frac{w}{T}
  \,,&
  \cs &= \frac{\partial p}{\partial T} \biggr/ \frac{\partial e}{\partial T}
  \,,
\end{align}
which are valid both in the
symmetric (sym) and broken (bro) phase.
For generic thermodynamic quantities
$X = \{\bar\theta,e,p,w,\Veff,\dots\}$,
we define the symmetric- and broken-phase difference
\begin{equation}
  \Delta X (T) = X_{\rmi{sym}}(T) - X_{\rmi{bro}}(T)
  \,,
\end{equation}
and
the thermal derivatives
$X' = \partial_\T X$.

\begin{figure}[t]
\centering
\includegraphics{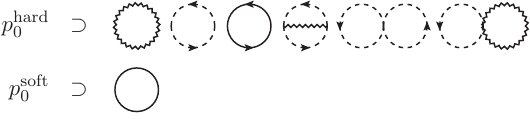}
\caption{%
  One-loop and two-loop vacuum diagrams contributing to
  the dark-sector 
  hard $p_0^\rmi{hard}$~\eqref{eq:p0:hard} and
  soft $p_0^\rmi{soft}$~\eqref{eq:p0:soft}
  unit operator
  in the high-temperature expansion
  computed in the symmetric phase.
  Wiggly lines indicate the vector bosons $V_\mu$,
  dashed directed lines the scalars $S$,
  solid directed lines the fermions $X$, and
  solid undirected lines the temporal vectors $V_0$.
  }
  \label{fig:unit:operator}
\end{figure}

In the high-temperature expansion,
the pressure for
the symmetric and
broken phases
takes the form
\begin{align}
  p_i &= p_0 - T F_{i}
  \,,&
  F_{i} &= \Veff^{i}(\phi_\rmi{min})
  \,,&
  i &\in \{{\rm bro},{\rm sym}\}
  \,,
\end{align}
where
$F_{i}$ is
the free energy, namely
the effective potential of the \threeDEFT{} evaluated at
its minimum for each phase.
The unit operator $p_0$
is the symmetric, field-independent pressure~\cite{%
  Braaten:1995cm,Gynther:2005dj,Tenkanen:2022tly}.
The latter consists of a hard and a soft contribution
$p_0^{ } = p_0^\rmi{hard} + p_0^\rmi{soft}$ as
displayed in fig.~\ref{fig:unit:operator}.
These contributions originate
from thermal corrections to the vacuum.
For the dark sector,
up to NNLO, they take the form 
\begin{eqnarray}
  \label{eq:p0:hard}
  p_0^\rmi{hard} &=&
    T^4\Bigl[
    g_{\rmi{eff},\rmii{DS}}^\rmii{B} \frac{\zeta_4}{\pi^2}
    - \frac{5\gd^2(1 + \nG) + 4\lambda_s}{288}
    + \frac{1}{12} \frac{\mu_{s}^2}{T^2}
    \Bigr]
  \nn[1mm] &&
  - \frac{\muX^{4}}{4(4\pi)^2}\Bigl[
      \ln\frac{\LamD^2}{\muX^{2}}
    + \frac{3}{2}
    \Bigr]
    + \widetilde{J}_{\rmii{$X$}}^{\T}
    + \mathcal{O}(g^3 T^4)
  \nn[1mm] &\stackrel{m_{i} \ll T}{=}&
    T^4\Bigl[
    g_{\rmi{eff},\rmii{DS}} \frac{\zeta_4}{\pi^2}
    - \frac{5\gd^2(1 + \nG) + 4\lambda_s}{288}
    + \frac{1}{12} \frac{\mu_{s}^2 - \muX^2}{T^2}
    + \mathcal{O}\Bigl(g^3 T^4,\frac{\mu_{s}^4}{T^4},\frac{\muX^4}{T^4}\Bigr)
    \Bigr]
    \,,
    \hspace{1cm}\\[2mm]
  \label{eq:p0:soft}
  p_0^\rmi{soft} &=&
    - \frac{T}{12\pi} \mD^{3}
    + \mathcal{O}(g^5 T^4)
    \,,
\end{eqnarray}
where
$g_{\rmi{eff},\rmii{DS}}^\rmii{B}$ denotes
the bosonic relativistic degrees of freedom, and
in the third line we applied the high-temperature expansion.
Here,
$\zeta_4 = \pi^4/90$, and
$\zeta_{s}=\zeta(s)$ for $\re\,(s) > 1$ is the Riemann zeta function.
The divergence at $\mX^4/\epsilon$ in
the hard $\widetilde{J}_{\rmii{$X$}}^{\T}$
in eq.~\eqref{eq:J:highT:fermion}
is cancelled by the potential
4d counterterm~\cite{Croon:2020cgk,Tenkanen:2022tly} and
the thermal master integral is defined in eq.~\eqref{eq:J:T}.
The NNLO contribution to the unit operator,
originating from hard three-loop vacuum diagrams,
can readily be included by using {\tt DRalgo}~\cite{Ekstedt:2022bff}.

In thermal equilibrium,
the relativistic degrees of freedom of the dark sector,
$g_{\rmi{eff},\rmii{DS}}$,
and of the SM,
$g_{\rmi{eff},\rmii{SM}}$~\cite{Laine:2015kra},
add up to the total number
\begin{align}
  \label{eq:geff:total}
  g_{\rmi{eff},\rmi{tot}} &=
  g_{\rmi{eff},\rmii{SM}}
+ g_{\rmi{eff},\rmii{DS}}
  \,, \\[2mm]
\label{eq:geff:DS}
  g_{\rmi{eff},\rmii{DS}} &=
    \underbrace{2}_{S}\,(\text{scalars})
  + \underbrace{2}_{V_\mu}\,(\text{vectors})
  + \frac{7}{8}\Bigl[\underbrace{4\nG}_{\bar{X},X} \Bigr]\,(\text{fermions})
  = \frac{8+7\nG}{2}
  \,,
\end{align}
where each (anti)fermion contributes a factor of $2$ from spin degrees of freedom.
At the transition temperatures considered in this work,
$\Tstar \sim \mathcal{O}(\text{MeV})$,
the dark fermion is non-relativistic for the dark matter masses of interest,
and therefore $g_{\rmi{eff},\rmii{DS}}^{ } = g_{\rmi{eff},\rmii{DS}}^\rmii{B}$.%
\footnote{%
  The precise choice of the dark-sector
  $g_{\rmi{eff},\rmii{DS}}$ is less important,
  since it can always be absorbed by slightly redefining
  the temperature $T_\text{reh}$ at reheating
  of eq.~\eqref{eq:Treh}.
}
If the visible and dark sectors are also hydrodynamically coupled,
$g_{\rmi{eff},\rmi{hyd}} = g_{\rmi{eff},\rmi{tot}}$
(cf.\ sec.~\ref{sec:hydro:decoupling}).

The critical temperature $\Tc$,
or critical mass $\yc$, is determined from the degeneracy condition
of the free energy between the symmetric and broken phase,
\begin{align}
  \label{eq:degeneracy:condition}
  \Delta F(\yc(x),x)  &=
    \Bigl[F_{\rmi{sym}} - F_{\rmi{bro}}\Bigr]\bigl(\yc(x),x\bigr)
    \nn[1mm] &=
    \Bigl[
        \Delta F_\rmii{LO}
      + x^{ } \Delta F_\rmii{NLO}
      + x^{\frac{3}{2}} \Delta F_\rmii{NNLO}
      + \mathcal{O}(x^2)
    \Bigr]_{y = \yc(x)}
    = 0
   \,,
\end{align}
where $x$ and $y$ are the dimensionless variables defined in eq.~\eqref{eq:Veff:bro:LO:dimless}.
From the second line,
this condition can be solved up to N$^4$LO in
a strict perturbative series~\cite{Ekstedt:2022zro,Ekstedt:2024etx}
after expanding
\begin{align}
  \yc^{} &=
      \yc^{\rmii{LO}}
    + x^{ }\yc^{\rmii{NLO}}
    + x^{\frac{3}{2}}\yc^{\rmii{NNLO}}
    + x^{2}\yc^{\rmii{N$^3$LO}}
    + x^{\frac{5}{2}}\yc^{\rmii{N$^4$LO}}
  \,, \\[2mm]
  \phi_\rmi{min}^{ } &=
      \phi_{0}^{ }
    + x^{ }\phi_{1}^{ }
    + x^{\frac{3}{2}}\phi_{2}^{ }
    + \mathcal{O}(x^2)
  \,.
\end{align}
The power-counting parameter $x$ indicates
the suppression of higher-order terms
since the expansion is organized in
powers of $x$ and not by loops.
Focusing on 
the NNLO result, 
all terms of the effective potential
are evaluated at
the minimum,
$\phi_\rmi{min} = \eta^{1/3} \varphi_\rmi{min}$,
at LO~\cite{Kajantie:1997hn}
\begin{align}
  \label{eq:phi_min:LO}
  \varphi_\rmi{min} &=
  \frac{3\mathcal{E}_1+ \sqrt{9\mathcal{E}_1^2- 4xy}}{2x}
  \,,&
  \varphi_{\rmi{min},\rmi{c}} &= \frac{2\mathcal{E}_1}{x}
  \,,
\end{align}
where we
also displayed
the critical value of the minimum $\varphi_{\rmi{min},\rmi{c}}$.

\begin{figure}[t]
  \centering
  \includegraphics[width=0.5\textwidth]{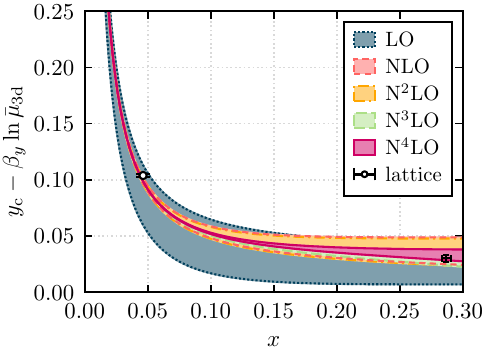}%
  \includegraphics[width=0.5\textwidth]{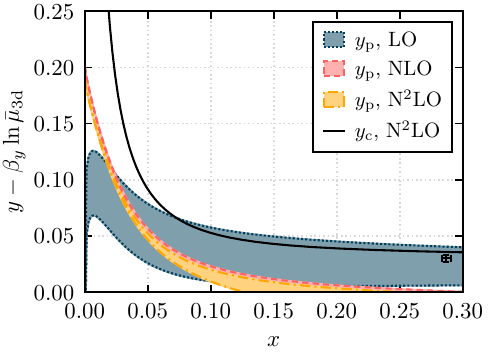}%
  \caption{%
    The renormalization-scale invariant quantity
    $\tilde{y}_{\rm c,p} = y_{\rm c,p}\!-\!\beta_y\ln\Lamd$, where
    $\yc$ is the critical mass (left),
    $\yp$ the percolation mass (right),
    $\Lamd$ the renormalization scale, and
    $\beta_y$ the
    $\beta$-function for
    $y$ from~\cite{Ekstedt:2024etx}.
    Uncertainties for the perturbative calculations are shown as error bands by varying
    $\Lamd/(2\mV) = [10^{-\frac{1}{2}},10^{\frac{1}{2}}]$.
    Left:
    The theoretical uncertainty for
    the lattice data points from~\cite{Kajantie:1997hn,Karjalainen:1996rk,Mo:2001fi}
    are shown as error bars.
    The line of first-order phase transitions ends in
    a second-order transition at $x \sim 0.3$~\cite{Mo:2001fi}.
    The orders NLO and NNLO are almost identical.
    Right:
    Comparison between
    $\yp$ fulfilling
    the condition~\eqref{eq:percolation:criterion}
    and $\yc$ at NNLO.
    }
  \label{fig:yc:yp}
\end{figure}

After solving eq.~\eqref{eq:degeneracy:condition}
order by order in $x$,
the individual orders of the critical mass $\yc$ are given by
\begin{align}
  \Delta F_\rmii{LO}^{ }\Bigr|_{y = \yc^\rmii{LO}}
  &= 0
  \,,&
  \yc^\rmii{NLO}
  &=
  - \frac{\Delta F_\rmii{NLO}^{ }}{\partial_y \Delta F_\rmii{LO}}
    \Bigr|_{y = \yc^\rmii{LO}}
  \,,&
  \yc^\rmii{NNLO}
  &=
  - \frac{ \Delta F_\rmii{NNLO}^{ }}{\partial_{y}^{ } \Delta F_\rmii{LO}^{ }}
    \Bigr|_{y = \yc^\rmii{LO}}
  \,.
\end{align}
The resulting NNLO critical mass $\yc$ with
$\tilde{\mu}_{3} = (x \Lamd/\mathcal{E}_1)\exp\{\frac{\mathcal{E}_{2}}{2\mathcal{E}_3} - 2\ln2\}$
is~\cite{Laine:1999rv,Ekstedt:2022zro,Ekstedt:2024etx}
\begin{align}
  \label{eq:yc:NNLO}
  \yc\bigr|_{\rmii{NNLO}} &=
        \frac{2}{x}\mathcal{E}_{1}^{2}
      + \frac{\mathcal{E}_2}{(2\pi)^2} \ln\tilde\mu_{3}
      - \frac{\mathcal{E}_1}{6\pi}\Bigl(\frac{x}{2}\Bigr)^{\frac{1}{2}}
  \,,
\end{align}
and
gauge independent order by order in $x$~\cite{Gould:2023ovu}.
The temporal vector enhancement factors
$\mathcal{E}_{1,2}$ are
defined in eq.~\eqref{eq:temporal:vector:enhancement}.
The critical mass $\yc$ is
depicted in fig.~\ref{fig:yc:yp}~(left) as a function of $x$
together with lattice data points from~\cite{Kajantie:1997hn,Karjalainen:1996rk,Mo:2001fi} and
the final orders
N$^3$LO and N$^4$LO from~\cite{Ekstedt:2024etx}.
The critical temperature can then be obtained by inverting
the relation $y(\Tc) = \yc$.
By using the LO matching relations
of appendix~\ref{sec:dimensional_reduction_details},
and the leading term of eq.~\eqref{eq:yc:NNLO},
one can recover the LO expression
\begin{align}
  \label{eq:Tc:LO}
  \Tc^{\rmii{LO}} =
  \mu_s \biggr(
      \frac{4\lambda_s + 3 \gd^2}{12}
    - \frac{2\gd^6 \mathcal{E}_1^2}{\lambda_s}
    \biggr)^{-1/2}
  \,,
\end{align}
which restricts the parameter range $(\gd,\lambda_s)$ for which a
first-order transition is realized to a positive argument of the square root.
Larger temporal vector contributions increase $\mathcal{E}_1$ and thus
$\Tc^\rmii{LO}$.

To compute the percolation temperature $\Tp$,
or the percolation mass $\yp$ as a function of $x$,
one can follow a similar strategy~\cite{Ekstedt:2021kyx,Ekstedt:2022ceo}.
After taking the LO action
from eq.~\eqref{eq:action:LO:dimless},
the defining equation,
is the percolation criterion~\cite{Guth:1981uk,Turner:1992tz}
\begin{align}
  \label{eq:percolation}
   S_\rmi{eff} &= \Bigl[
      S_\rmii{LO}
    + x^{ } S_\rmii{NLO}
    + x^{\frac{3}{2}} S_\rmii{NNLO}
    + \mathcal{O}(x^2)
    \Bigr]_{y=\yp(x)}
  = \mathcal{F} [H(\Tp)]
  \,,\\
  \yp &=
      \yp^\rmii{LO}
    + x^{ }\yp^\rmii{NLO}
    + x^{\frac{3}{2}}\yp^\rmii{NNLO}
  \,,
\end{align}
which is also solved
order by order in the expansion parameter $x$.
Up to NNLO,
the individual gauge-independent orders
of $\yp$ are by 
\begin{align}
  S_\rmii{LO}^{ }\Bigr|_{y = \yp^\rmii{LO}}
  &= \mathcal{F}[H(\Tp)]
  \,,&
  \yp^\rmii{NLO}
  &=
  - \frac{S_\rmii{NLO}^{ }}{\partial_y S_\rmii{LO}}
    \Bigr|_{y = \yp^\rmii{LO}}
  \,,&
  \yp^\rmii{NNLO}
  &=
  - \frac{ S_\rmii{NNLO}^{ }}{\partial_{y}^{ } S_\rmii{LO}^{ }}
    \Bigr|_{y = \yp^\rmii{LO}}
  \,,
\end{align}
and depicted in fig.~\ref{fig:yc:yp}~(right) as a function of $x$.
The percolation criterion in eq.~\eqref{eq:percolation}
depends on $\mathcal{F}$ which is a function of
the Hubble parameter $H(T)$;
see eq.~\eqref{eq:Hubble:rate} for its definition during radiation domination.
Here,
we set $\mathcal{F}[H(\Tp)]$ to be a constant~\cite{Enqvist:1991xw},
and impose the standard percolation condition,
for which the probability that a given spatial point remains in the false vacuum,
$P (\Tp) = e^{-I(\Tp)} \simeq 71\%$, corresponds to
$I(\Tp) = 0.34$~\cite{Guth:1981uk,Turner:1992tz}.
This yields
\begin{align}
  \label{eq:percolation:criterion}
  \mathcal{F} [H(\Tp)] &\approx
    \ln I(\Tp)
    + \ln(8\pi)
    + 3\ln \vw
    + \ln \frac{A}{H^4}
    - 4\ln \frac{\beta}{H}
  \\[2mm]
  & =
    181.6
    + \ln I(\Tp)
    + 3\ln \vw
    + \ln \frac{A}{\Tp^4}
    - 4\ln \frac{\beta/\Hstar}{100}
    - 2\ln \frac{g_{\rmi{eff},\rmi{tot}}}{10}
    - 4\ln \frac{\Tp}{1~{\rm MeV}}
  \,,
  \nonumber
\end{align}
where in our numerical studies, we use
$\ln(A/\Tp^4) \simeq -14$~\cite{Carrington:1993ng}, and
the ansatz
$\beta/\Hstar \simeq 100$,
$\vw \simeq 1$,
$g_{\rmi{eff},\rmi{tot}} \simeq 10$, and
$\Tp \simeq 1~{\rm MeV}$.
In practice,
the percolation criterion is not very sensitive
to the precise choice of these parameters but
a fully general computation would require a self-consistent solution of
the percolation criterion together with the thermodynamic parameters.

For strong supercooling
where $\Tp \ll \Tc$ and $\alpha$ is large,
or
very strong transitions where a lot of energy is stored in the form of vacuum
energy that later reheats the universe when the transition ends,
the reheating temperature
can differ significantly from the percolation temperature
$\Tstar = \Tp$~\cite{Ellis:2018mja},
\begin{align}
  \label{eq:Treh}
  T_\rmi{reh} &\approx
    \Tstar (1 + \alpha_\star)^{1/4}
  \,.
\end{align}

The phase-transition strength is
most suitably computed from the pseudotrace
anomaly~\cite{Giese:2020rtr,Giese:2020znk}
defined in eq.~\eqref{eq:thermo:quantities},
\begin{align}
  \label{eq:alpha}
  \alpha &= \frac{\Delta\bar\theta}{3\,\omega_\rmi{sym}}
  =
  \frac{1}{3\, p_\rmi{sym}'}
  \biggl(
    \frac{\Delta F}{c_{s,\rmi{bro}}^2}
    - \Delta \frac{{\rm d}F}{{\rm d}\ln T}
  \biggr)
  \,.
\end{align}
The inverse duration
depends
on the action
$S_3 = S_{\rmii{LO}} + x S_\rmii{NLO} + x^{3/2} S_\rmii{NNLO} + \dots$
fitted in tab.~\ref{tab:fit:action},
and, after applying the chain rule, it factorizes into
\begin{align}
  \beta/H &\simeq
    \frac{{\rm d}\ln \Gamma}{{\rm d}\ln T} 
    =
    \bigl(
        \eta_y \partial_y
      + \eta_x \partial_x
    \bigr)\, S_3
    \,,
\end{align}
with
$\tau = \ln T$ and
$\eta_\kappa = \partial_\tau \kappa$ encoding the
thermal running of the EFT parameters~\cite{Gould:2019qek}.

The subsequent bubble-wall hydrodynamics is controlled by
the sound speeds
in the broken phase $c_{s,\rmi{bro}}$ and
in the symmetric phase $c_{s,\rmi{sym}}$, as well as
the enthalpy ratio across the wall,
\begin{align}
  \label{eq:enthalpyRatio}
  \Psi &= \frac{\omega_\rmi{bro}}{\omega_\rmi{sym}}
  \approx
    1
  - \frac{1}{p_0'}\Bigl(
      F_\rmi{bro}
    + \frac{{\rm d}F_\rmi{bro}}{{\rm d}\ln T}
    \Bigr)
  \,.
\end{align}
The kinetic energy fraction~\cite{Caprini:2024hue,Giese:2020rtr,Giese:2020znk}
is determined by
the kinetic energy density in the fluid $\rho_\rmi{fl}$ and
the total energy density of the plasma $e_\rmi{sym}$~\cite{Huber:2008hg},
\begin{align}
  \label{eq:K}
  K &=
  \frac{\rho_\rmi{fl}}{e_\rmi{sym}}
  =
  \kappa_\rmi{sw}(\alpha_{\star,\rmi{hyd}})\,
  \alpha_{\star,\rmi{tot}}\,
  \frac{3}{4}\Gamma_\rmi{ad}
  \,,
\end{align}
and is related to the efficiency factor
$\kappa_\rmi{sw}(\alpha_{\star,\rmi{hyd}})$.
The latter quantifies the fraction
of the latent heat deposited into bulk fluid motion,
rather than reheating the plasma, and depends on
$\alpha_{\star,\rmi{hyd}}$,
the transition strength normalized
to the energy density of the hydrodynamically active sector.
Here, $\Gamma_\rmi{ad} = \omega_\rmi{sym}/e_\rmi{sym}$ is the adiabatic index
of the symmetric phase, and
$\alpha_{\star,\rmi{tot}}$ is the total transition strength,
both evaluated using the total relativistic degrees of freedom $g_\rmi{eff,tot}$
of eq.~\eqref{eq:geff:total}. The condition for hydrodynamic equilibrium
is discussed in sec.~\ref{sec:hydro:decoupling}.

Imposing local thermal equilibrium (LTE) across the wall,
an upper limit on the terminal wall velocity $\vw$~\cite{Enqvist:1991xw}
follows from the hydrodynamic matching conditions
together with entropy conservation.
The wall velocity $\vw$ and
the sound-wave efficiency factor $\kappa_\rmi{sw}$ are
then fully determined by~\cite{Ai:2023see}%
\footnote{%
  A lower bound on $\vw$
  can in principle be obtained from the ballistic approximation~\cite{Ai:2024btx}.
  It has also been shown that the LTE upper bound cannot always be saturated,
  as entropy production across the wall persists even in
  the zero-friction limit~\cite{Eriksson:2025owh}.
}
\begin{align}
    \label{eq:vw:LTE}
    \vw &= \vw(
        \alpha_{\star,\rmi{hyd}},
        \Psi,
        c_{s,\rmi{sym}},
        c_{s,\rmi{bro}}
    )
    \,,&
    \kappa_\rmi{sw} &= \kappa_\rmi{sw}(
        \alpha_{\star,\rmi{hyd}},
        \Psi,
        c_{s,\rmi{sym}},
        c_{s,\rmi{bro}}
    )
  \,.
\end{align}
Alternatively,
the efficiency factor
$\kappa_\rmi{sw} = \kappa_\rmi{sw}(\alpha,\vw)$ can be taken
from~\cite{Espinosa:2010hh,Jinno:2022mie},
in which case $\vw$ is treated as an input.

For the strongest transitions encountered in this work,
detonations with $\vw \simeq 1$ can occur.
Throughout, we adopt the LTE upper bound on $\vw^\rmii{LTE}$
using the implementation of~\cite{Ai:2023see}.
A general
perturbative determination of
$\vw$ requires including out-of-equilibrium effects~\cite{%
  Laurent:2022jrs,DeCurtis:2022hlx,DeCurtis:2023hil}, as
recently automated in~\cite{DeCurtis:2024hvh,Ekstedt:2024fyq,vandeVis:2025plm}.
In the detonation regime,
where $\vw \simeq 1$,
one has $\kappa_\rmi{sw}(\alpha,1) = \kappa_\rmi{sw}(\alpha)$ and
both the efficiency factor and the kinetic energy fraction
are well approximated by~\cite{Steinhardt:1981ct,Espinosa:2010hh}
\begin{align}
  \label{eq:detonation:approx}
  K &= 0.6\,\kappa_\rmi{sw}(\alpha_{\star,\rmi{hyd}}) \frac{\alpha_{\star,\rmi{tot}}}{1 + \alpha_{\star,\rmi{tot}}}
  \,,&
  \kappa_\rmi{sw}(\alpha) &\approx
  \frac{\alpha}{
        0.73
      + 0.083\,\sqrt{\alpha}
      + \alpha
    }
  \,.
\end{align}
Besides $\kappa_\rmi{sw}$, the wall velocity also sets
the sound-shell thickness $\Delta_{w}$ of eq.~\eqref{eq:soundshell}.

\subsubsection{Hydrodynamic equilibrium and decoupling}
\label{sec:hydro:decoupling}

Besides the thermal equilibration condition~\eqref{eq:thermal:equilibrium:condition},
we also examine the hydrodynamic contact between the SM and dark sectors.
The relevant criterion is whether the mean free path for SM--DS momentum exchange
is shorter than the mean bubble separation $R \sim \vw/\beta$,
the characteristic length scale of the bubble hydrodynamics.
This translates to the condition
\begin{align}
  \label{eq:hydro:equilibrium:condition}
  C_\rmi{hyd}^\rmi{eq}: \qquad
  \Gamma_\rmi{int} \cdot R \gtrsim 1
  \,,
\end{align}
where $\Gamma_\rmi{int}$ is the total SM--DS interaction rate
of eq.~\eqref{eq:thermal:equilibrium:condition}.
Depending on whether this condition is satisfied,
all thermodynamic quantities entering the hydrodynamics are evaluated
either in the dark sector alone or in the full plasma of eq.~\eqref{eq:geff:total},
motivating two distinct transition strengths:
\begin{itemize}
  \item[$\alpha_{\star,\rmi{hyd}}$:]
  normalized to the energy density of the hydrodynamically active sector,
  this governs the bubble-wall dynamics and efficiency factor $\kappa_\rmi{sw}$
  in eq.~\eqref{eq:K}.
  \item[$\alpha_{\star,\rmi{tot}}$:]
  normalized to the full relativistic energy density of eq.~\eqref{eq:geff:total},
  this sets the overall amplitude of the GW signal through eq.~\eqref{eq:K}.
\end{itemize}
The two strengths can differ significantly when the
condition~\eqref{eq:hydro:equilibrium:condition} is not met.

%
\subsection{Gravitational wave spectrum}
\label{sec:GW_spectrum}
A first-order phase transition proceeds through the nucleation and
expansion of bubbles of the broken phase, whose dynamics sources
GWs in the early universe~\cite{Caprini:2015zlo,Caprini:2018mtu}.
Three mechanisms have been identified that may linearly combine
into the total stochastic gravitational-wave background (SGWB),
\begin{align}
\label{eq:Ogw:sum}
  \Ogw h^2 \simeq
    \Omega_\rmi{coll} h^2
  + \Omega_\rmi{sw}   h^2
  + \Omega_\rmi{turb} h^2
  \,,
\end{align}
namely bubble-wall collisions and shocks~\cite{Huber:2008hg},
sound waves expanding into the
plasma~\cite{Hindmarsh:2015qta,Hindmarsh:2017gnf}, and
magnetohydrodynamic (MHD) turbulence in the post-collision
plasma~\cite{Caprini:2009yp}.
An additional source of
feebly interacting particles has been identified recently~\cite{Jinno:2022fom}.
In the parameter range relevant for our analysis
(cf.\ tab.~\ref{tab:pta:scan}),
the bubbles do not run away and the signal is dominated by
the sound-wave contribution.

We write the sound-wave spectrum as a peak amplitude times
a dimensionless spectral shape function,
\begin{align}
  \label{eq:Omega:sw}
    \Omega_\rmi{sw} h^2 &=
    \mathcal{R} h^2\,
    A_\rmi{sw}\,
    K^2\,
    \mathcal{Y}_\rmi{sw}\,
    \bigl( R\Hstar \bigr)\,
    S(f)
    \,,
\end{align}
and take
$H_0 = 100 h\,{\rm km\,s^{-1}\,Mpc^{-1}}$ with
the observed Hubble parameter today given by
$h = 0.6737 \pm 0.0054$~\cite{Planck:2018vyg}.

The overall amplitude structure
of eq.~\eqref{eq:Omega:sw}
is common to multiple spectral models,
while the spectral shape $S(f)$ and its characteristic
frequencies differ.
Here,
$A_\rmi{sw}$ is the simulation-extracted amplitude,
$K$ the kinetic energy fraction~\eqref{eq:K},
$R$ the mean bubble separation, and
the redshift to today reads~\cite{Caprini:2024hue,Bringmann:2026xcx}%
\footnote{%
  The factors $\mathcal{R}$ and $F_{\rmi{gw},0}$ of~\cite{Caprini:2019egz}
  are equivalent.
}
\begin{align}
  \label{eq:redshift}
  \mathcal{R} h^2 &=
    \Omega_{\gamma} h^2\,
    \biggl( \frac{h_0}{h_{\rmi{eff},\rmi{tot}}(\Treh)} \biggr)^{4/3}
    \frac{g_{\rmi{eff},\rmi{tot}}(\Treh)}{g_\gamma}
  \,,
\end{align}
with
$\Omega_\gamma h^2 = 2.473\times10^{-5}$
is the radiation redshift today~\cite{Planck:2018vyg},
$g_\gamma = 2$, and
$h_0 = 3.930$~\cite{Escudero:2025kej}.
Here,
$g_{\rmi{eff},\rmi{tot}}(\Treh)$ and
$h_{\rmi{eff},\rmi{tot}}(\Treh)$ denote
the total number of
relativistic energy and
entropy degrees of freedom
in the thermalised SM and dark sector bath
after reheating at the end of the phase transition
as defined in eq.~\eqref{eq:geff:DS}.
For $\Tstar > 0.1$~MeV,
it is appropriate to take
$g_{\rmi{eff},\rmi{tot}}(\Treh) = h_{\rmi{eff},\rmi{tot}}(\Treh)$~\cite{Kolb:1990vq}.

The finite source lifetime is included through
the shock-formation factor~\cite{%
  Caprini:2019egz,Ellis:2020awk,Guo:2020grp,Balan:2025uke}
\begin{align}
\label{eq:lifetime:SW}
  \mathcal{Y}_\rmi{sw} &=
  \min\{1,\tau_\rmi{sh}\Hstar\}
  \,,&
  \tau_\rmi{sh}\Hstar &\simeq
  \sqrt{\frac{\Gamma_\rmi{ad}}{K}}\, R\Hstar
  \,,
\end{align}
where
$\tau_\rmi{sh}$ is the shock-formation timescale and
$\Gamma_\rmi{ad} = \omega_\rmi{sym}/e_\rmi{sym} \approx 4/3$ is the adiabatic index.
The mean bubble separation is set by the inverse
duration~\cite{Caprini:2019egz}
\begin{align}
\label{eq:RHstar:betaH}
  R \Hstar &\approx
    \biggl( \frac{8\pi}{f_\rmi{p}} \biggr)^{1/3}
    \frac{\Hstar}{\beta}
    \max\{\vw,c_{s,\rmi{bro}}\}
  \,,
\end{align}
with the true-vacuum fraction
$P_\rmi{t}(\Tp) \equiv f_\rmi{p} \simeq 0.28957$
at percolation~\cite{Megevand:2016lpr,Athron:2023xlk,Matuszak:2026xsz}.
The Hubble rate at percolation redshifted to today is
common to
a generic GW spectrum~\cite{Caprini:2024hue}
\begin{align}
  \label{eq:aHstar}
  H_{\star,0} = a_\star\Hstar &=
    11.2 \times 10^{-9}~\textrm{Hz}\,
    \biggl( \frac{\Treh}{100~\textrm{MeV}} \biggr)
    \biggl( \frac{g_{\rmi{eff},\rmi{tot}}(\Treh)}{10} \biggr)^{1/2}
    \biggl( \frac{10}{h_{\rmi{eff},\rmi{tot}}(\Treh)} \biggr)^{1/3}
  \,.
\end{align}

We now focus on two specific
templates that differ in
the spectral shape function and
its characteristic breaks.

\subsubsection{Single broken power law}
\label{sec:single:broken:powerlaw}

The single broken power law extracted from the sound-shell
model~\cite{Hindmarsh:2017gnf,Hindmarsh:2019phv,Caprini:2019egz,Huber:2008hg,Ertas:2021xeh}
peaks at a single
frequency $\fsw^\rmi{peak}$,
\begin{align}
  \label{eq:S:single}
    S_1(f) &=
      \mathcal{N}_1\,
      \biggl( \frac{f}{\fsw^\rmi{peak}} \biggr)^{\!3}
      \biggl[ \frac{7}{4 + 3\bigl(f/\fsw^\rmi{peak}\bigr)^2} \biggr]^{7/2}
    \,,&
    \fsw^\rmi{peak} &=
      1.56
      \,
      \Bigl( \frac{z_p}{10} \Bigr)
      \Bigl(\frac{a_\star\Hstar}{R\Hstar}\Bigr)
    \,,
    \tag{SBPL}
\end{align}
with $\mathcal{N}_{1} = \frac{1080}{343}\sqrt{\frac{3}{7}}$
normalizing the respective peak amplitude
such that
$\int {\rm d}\ln f\, S_1(f) \stackrel{!}{=} 3$~\cite{Hindmarsh:2017gnf}.
The constant factors
$A_\rmi{sw} = 1.2 \times 10^{-2}$ in eq.~\eqref{eq:Omega:sw}, and
$z_p \approx 10$
which accounts for the observed peak value $z_p = (k\Hstar)_\rmi{max}$,
are determined from simulations~\cite{Hindmarsh:2017gnf}.

\subsubsection{Double broken power law}
\label{sec:double:broken:powerlaw}

The template for
the double broken power law can be determined from the more recent Higgsless
simulations~\cite{Jinno:2022mie,Caprini:2024gyk,Caprini:2024hue}.
The spectrum develops a plateau between
two breaks $f_1 < f_2$~\cite{RoperPol:2023bqa,Caprini:2024hue},
\begin{align}
  \label{eq:S:double}
    S_2(f) &=
      \mathcal{N}_2\,
      \biggl( \frac{f}{f_2} \biggr)^{\!3}
      \biggl[ 1 + \biggl( \frac{f}{f_1} \biggr)^{\!2} \biggr]^{-1}
      \biggl[ 1 + \biggl( \frac{f}{f_2} \biggr)^{\!4} \biggr]^{-1}
    \,, &
    f_1 &\simeq 0.2 \Bigl(\frac{a_\star\Hstar}{R\Hstar}\Bigr)
    \,,
    \nn[1mm]
    &&
    f_2 &\simeq \frac{0.5}{\Delta_{w}^{ }} \Bigl(\frac{a_\star\Hstar}{R\Hstar}\Bigr)
    \,,
  \tag{DBPL}
\end{align}
where
the exact position of the peaks $f_1$ and $f_2$ are
determined from simulations~\cite{Caprini:2024gyk}
and are sensitive to the strength of the transition.
The normalization factor
$\mathcal{N}_2 = \frac{2\sqrt{2}}{\pi}\bigl[(1 + f_2^2 / f_1^2) + \sqrt{2} f_2 / f_1\bigr]$,
$A_\rmi{sw} \approx 0.11$, and
the relative sound-shell thickness~\cite{Hindmarsh:2019phv}
\begin{align}
  \label{eq:soundshell}
  \Delta_{w} &=
    \frac{|\vw - c_{s,\rmi{bro}}|}{\max\{\vw, c_{s,\rmi{bro}}\}}
  \,.
\end{align}

Both spectra~\eqref{eq:S:single} and~\eqref{eq:S:double}
share the rising $f^3$ causality tail in the IR~\cite{Caprini:2009fx},
while in the UV the single (double) broken
power law falls off as $f^{-4}$ ($f^{-3}$).
In the strong phase-transition limit, $\alpha \gg 1$, the wall velocity
saturates to $\vw \simeq 1$, so that the sound-shell thickness
$\Delta_{w} \to 1 - c_{s,\rmi{bro}}$ and the two breaks approach
each other, $f_2 \sim f_1$. The double broken power law then collapses
to an effectively single break and the two templates
coincide~\cite{Caprini:2024hue}.
The strongest transitions preferred by the PTA data approach this regime and
the single broken power law~\eqref{eq:S:single}
already provides a reasonable leading description of the signal.
For the finite values of $\alpha$ and $\vw$ realized in sec.~\ref{sec:pta},
the two breaks remain partially resolved,
and we default to the double broken power law~\eqref{eq:S:double}.

In the sound-wave template~\eqref{eq:Omega:sw},
also
the kinetic energy fraction $K$ and
the wall velocity $\vw$ are required as additional inputs.
In our analysis,
we keep both generic albeit choosing
$\vw$ at its LTE value
introduced below eq.~\eqref{eq:vw:LTE},
fixed by the enthalpy
ratio $\Psi$ and the sound speeds~\cite{Ai:2023see}.

%
\section{Dark matter freeze-out}
\label{sec:dark_matter}

%
In this section, we discuss the complementarity between
the phase transition dynamics of
sec.~\ref{sec:PT_and_GW} and dark matter phenomenology.
Although dark matter is not the primary focus,
the model naturally contains a viable dark matter candidate. 

The dark matter candidate is the Dirac fermion $X$ in eq.~\eqref{eq:lag:4d}
with a vector-like mass term,
which therefore constitutes an independent parameter of the model.
We adopt this setup to facilitate comparison with
previous studies~\cite{Han:2023olf},
while leaving the more involved scenario in which
the dark fermion mass is dynamically generated during
the phase transition for future work
(see e.g.~\cite{Kajantie:1997hn,Banik:2024zwj}).

The dark matter fermion undergoes the standard thermal freeze-out mechanism.
For the couplings of $\gd \sim \mathcal{O}(1)$ that are considered throughout this work,
the dark fermions remain in thermal equilibrium with the plasma at
high temperatures.
The parameter region relevant for our analysis,
is dominated by channels into dark scalars and vector bosons.
Annihilations into SM particles are strongly suppressed by
the small portal couplings~\eqref{portal_mixing_couplings}
and become relevant only after the dark-sector phase transition.

In the parameter region considered here,
the dark-matter fermion is significantly heavier than both
the physical dark scalar and vector boson masses.
Since the freeze-out temperature is typically given by
$\Tfo\simeq \mX/25$, one generically finds
\begin{equation}
  \label{eq:freeze_out_condition}
  \Tfo \gg \TcDS, \TpDS
  \, ,
\end{equation}
and the processes driving annihilation in the symmetric phase are
\begin{align}
  \label{eq:ann_channels_symm}
  X \bar{X} &\to VV
  \,,&
  X \bar{X} &\to S S^*
  \,.
\end{align}
We have explicitly verified the condition~\eqref{eq:freeze_out_condition}
by solving the Boltzmann equation~\eqref{Boltzmann_eq_eff}.%
\footnote{%
  A simple estimate can also be obtained from
  the usual freeze-out condition
  $n_X^{\rm eq}\,\langle \sigma_{\rm ann} v_{\rm rel} \rangle \simeq H$,
  where the Hubble rate in the radiation dominated era is given by
  eq.~\eqref{eq:Hubble:rate}.
}
As discussed in sec.~\ref{sec:thermal_contact},
the visible and dark sectors remain in thermal equilibrium down to
temperatures of $\mathcal{O}(\text{MeV})$ for the range of scalar mixing angles
allowed by current constraints;
cf.~\eqref{mixing_mass_limits}.
It is therefore consistent to describe both sectors with
a common temperature
$T = T_\rmii{DS} = T_\rmii{SM}$
during the epoch relevant for dark matter freeze-out.

Dark matter annihilation,
even when proceeding dominantly into dark-sector states rather than
directly into SM particles,
can still be constrained by indirect detection probes.
This is particularly relevant for dark matter masses below
$\mathcal{O}(10^2)$~GeV,
where late-time annihilations may lead to observable signatures.
In particular, annihilations occurring around the time of recombination
inject energy into the primordial plasma and can therefore distort
the CMB anisotropy spectrum.
More precisely, exotic energy injection modifies
the recombination history and alters the optical depth of the CMB.
In this regime, the relevant annihilation processes occur in
the broken phase of the dark sector, and the dominant channels are
\begin{align}
  \label{eq:ann_channels_broken}
  X \bar{X} &\to VV
  \,,&
  X \bar{X} &\to V s
  \,,&
  X \bar{X} &\to V \chi
   \,.
\end{align}

The strongest constraints on exotic energy injection during
recombination are provided by measurements of the CMB anisotropies
by the Planck collaboration~\cite{Planck:2018vyg}.
Following the model-independent treatment of energy deposition
developed in~\cite{Slatyer:2015jla},
one obtains the approximate bound
\begin{equation}
\label{indirect_ann_exp}
\langle \sigma_{\rm ann} v_{\rm rel}\rangle
\lesssim
2.6\times10^{-27}
\left(\frac{\mX}{\rm GeV}\right)
\frac{\rm cm^3}{\rm s}
\,,
\end{equation}
where the efficiency factor $f_{\rm eff}$ parametrizes the absorption efficiency of
injected energy in the intergalactic medium.%
\footnote{%
  We adopt the conservative estimate $f_{\rm eff}=0.137$ for
  all annihilation channels, following~\cite{Slatyer:2015jla}.
}
The thermal average in the CMB bound should be evaluated using
the dark matter velocity distribution at recombination.

More generally, indirect detection constraints are particularly stringent 
in scenarios featuring sizable late-time annihilation
cross-sections~\cite{Galli:2009zc,Slatyer:2009yq,Bringmann:2016din},
especially in the presence of
Sommerfeld enhancement~\cite{Hisano:2006nn,Zavala:2009mi,Beneke:2014hja} and
bound-state formation effects~\cite{Baldes:2020hwx,Biondini:2023ksj}, which
can substantially enhance the annihilation rate at low velocities.
We therefore focus on indirect detection constraints, 
while leaving aside direct detection searches.
The latter are strongly suppressed by the tiny
portal couplings allowed in the mediator mass range relevant for our analysis.
Collider constraints are effectively encoded in the allowed range of
scalar mixing angles~\eqref{mixing_mass_limits}.

The bound on the annihilation cross-section~\eqref{indirect_ann_exp}
is especially severe for the symmetric dark matter scenario,
which is the standard assumption for thermal dark sectors.
In this case, particles and antiparticles remain equally abundant after freeze-out,
and their late-time annihilations in
dense astrophysical environments provide
the basis for indirect detection searches.
Alternatively, asymmetric dark matter scenarios have
attracted considerable interest due to their possible connection with
the matter-antimatter asymmetry of
the visible sector~\cite{Gu:2010ft,Shelton:2010ta,Graesser:2011wi}.
In such scenarios, the dark sector may also contain
a particle-antiparticle asymmetry, resulting in different
relic abundances for dark matter particles and antiparticles.
Consequently, the late-time annihilation rate is suppressed by
the factor~\cite{Graesser:2011wi}
\begin{eqnarray}
    \langle \sigma_{\rmi{ann}} \vrel \rangle  \to
    \langle \sigma_{\rmi{ann}} \vrel \rangle  \frac{4 r_{\infty}}{(1+r_{\infty})^2}
    \, .
\end{eqnarray}
Here,
$r_\infty$ denotes the ratio between the antiparticle and particle abundances at
times well after chemical freeze-out.
See sec.~\ref{sec:asymm_DM} for more details.

In the following, we discuss
the symmetric dark matter scenario in sec.~\ref{sec:symm_DM} and
the asymmetric scenario in sec.~\ref{sec:asymm_DM},
thereby extending the analysis of~\cite{Han:2023olf}
in the context of dark-sector phase transitions
in a PTA-favored GW parameter space.

\subsection{Symmetric dark matter scenario}
\label{sec:symm_DM}
%
In the standard symmetric freeze-out scenario,
the dark matter abundance is determined by solving the Boltzmann equation~\cite{Gondolo:1990dk}
\begin{equation}
\label{Boltzmann_eq_eff}
    (\partial_t + 3H) n = - \frac{1}{2}\langle \sigma_{\textrm{eff}} v_{\textrm{rel}} \rangle \left(n^2-n^2_{\textrm{eq}}\right) \, ,
\end{equation}
where $n$ denotes the total number density of particles and antiparticles.
The quantity $\langle \sigma_{\textrm{eff}} v_{\textrm{rel}} \rangle$ is
the thermally averaged effective annihilation cross-section,
which incorporates non-perturbative effects relevant in
the non-relativistic regime.

One such non-perturbative effect arises from the repeated exchange of
soft vector mediators between the annihilating particle-antiparticle pair.
Such interactions distort the two-body wave function and,
for an attractive potential, enhance the annihilation cross-section.
This phenomenon is known as
Sommerfeld enhancement~\cite{Sommerfeld,Arkani-Hamed:2008hhe}.
In the Coulombic regime, corresponding to the symmetric phase where $\mV=0$,
the Sommerfeld factor admits an analytic expression
(see e.g.~\cite{Iengo:2009ni,Cassel:2009wt})
\begin{align}
\label{Somme_0}
  S_{\hbox{\scriptsize ann}}(\zeta)
  &=
  \frac{2 \pi \zeta}{1-e^{-2 \pi \zeta}}
  \,,&
  \zeta &\equiv \frac{\alpha_d}{v_{\rm rel}}
  \, ,
\end{align}
and multiplies the perturbative annihilation cross-section. Here,
$\alpha_d = \gd^2/(4\pi)$ is the dark fine-structure constant and
$v_{\rm rel}$ is the relative velocity of the annihilating pair.

After the phase transition, the vector mediator acquires a finite mass and
the interaction becomes Yukawa-like.
In this case, no exact analytic expression for the Sommerfeld factor is available,
although approximate analytic results can be obtained using
the Hulth\'en potential~\cite{Cassel:2009wt}.
The Sommerfeld factor must therefore be computed numerically and depends on two parameters,
$S_{\rm ann}(\zeta,\xi)$, where
\begin{equation}
\label{xi_def}
    \xi = \frac{\mX \alpha_d}{2\mV}
    \, .
\end{equation}
This parameter corresponds to the ratio between the would-be
Coulombic Bohr radius and the mediator Compton wavelength, and
therefore quantifies the screening effects induced by the finite mediator mass.
The Coulomb limit is recovered for $\xi \gg 1$.

A second non-perturbative effect originates from
the formation of metastable particle-antiparticle bound states.
Whenever bound states are efficiently formed and not immediately dissociated
by the thermal bath, they provide an additional channel for
dark matter depletion into light degrees of freedom,
as originally pointed out in~\cite{Feng:2008ya,vonHarling:2014kha}.
In this case, the relevant quantities are the bound-state formation cross-section,
$\sigma_{\textrm{bsf}} v_{\rm rel}$,
the bound-state dissociation rate, $\Gamma_{\textrm{bsd}}$, and
the bound-state annihilation width, $\Gamma_{\textrm{ann}}$. 

For the model~\eqref{eq:lag:4d},
the LO bound-state formation process
proceeds through radiative emission of
a dark vector boson~\cite{vonHarling:2014kha,Petraki:2016cnz}.
This process can occur both before and after the phase transition.
In the broken phase, however,
kinematics requires that the energy difference between the
incoming scattering state and the bound state exceeds the dark photon mass.
As in the case of Sommerfeld enhancement,
analytical expressions for bound-state formation, dissociation, and
bound-to-bound transitions are available only in the Coulombic limit
(cf.\ e.g.~\cite{vonHarling:2014kha,Biondini:2023zcz}),
while numerical computations are required once the mediator mass
becomes non-zero (cf.\ e.g.~\cite{Petraki:2016cnz}).

At NLO, several additional processes contribute to
bound-state formation and
dissociation~\cite{%
  Biondini:2018pwp,Binder:2019erp,Binder:2020efn,Biondini:2025jvp}.
However, these corrections significantly affect
the dark matter relic abundance only for comparatively large values of
the coupling. Since the observed relic density in the mass range
$\mX \in [10,100]$~GeV is reproduced for relatively small couplings in our setup,
we neglect these higher-order contributions.%
\footnote{%
  Explicit NLO computations available in
  the literature are typically derived for additional light fermionic species
  coupled to the dark vector,
  rather than for light scalars as in the present model.
}

Neglecting transitions among different bound states,
the Boltzmann equation can be expressed solely in terms of
the density of scattering states through the effective cross-section~\cite{Ellis:2015vaa}
\begin{equation}
\label{Cross_section_eff}
   \langle \sigma_{\textrm{eff}}^{ } v_{\textrm{rel}}^{ } \rangle
   =
   \langle \sigma_{\textrm{ann}}^{ } v_{\textrm{rel}}^{ } \rangle
   + \sum_{n}
   \langle \sigma_{\textrm{bsf}}^{n} v_{\textrm{rel}}^{ } \rangle
   \,
   \frac{\Gamma_{\textrm{ann}}^n}
   {\Gamma_{\textrm{ann}}^n+\Gamma_{\textrm{bsd}}^n}
   \, .
\end{equation}
The first contribution corresponds to direct annihilation from scattering states,
while the second accounts for the formation of
unstable bound states that subsequently decay.
The annihilation cross-section $\sigma_{\textrm{ann}} v_{\textrm{rel}}$
is obtained by multiplying the perturbative cross-section by the Sommerfeld factor.

Since annihilation in the present model is dominated by
the $s$-wave contribution, we keep only
the corresponding Sommerfeld enhancement factor.
In the symmetric phase, the non-relativistic annihilation cross-section,
corresponding to the processes~\eqref{eq:ann_channels_symm},
reads%
\footnote{%
  The non-relativistic annihilation cross-section receives
  two leading contributions, namely
  $X\bar{X}\to VV$ and $X\bar{X}\to SS^\ast$.
  The former yields the well-known result at LO in the relative velocity, 
  $\sigma^{\text{sym}}_{\text{ann}} v_{\text{rel}}(X\bar{X}\to VV)=\pi\alpha_d^2/\mX^2$
  (cf.\ e.g.~\cite{Dirac:1930bga}),
  while the latter gives
  $\sigma^{\text{sym}}_{\text{ann}} v_{\text{rel}}(X\bar{X}\to SS^\ast)=\pi\alpha_d^2/(4\mX^2)$.
  Our result differs from the expression reported in~\cite{Han:2023olf}.
}
\begin{equation}
\label{sigma_ann_symm}
    \sigma_{\rmi{ann}}^{\rmi{sym}} v_{\rm rel}^{ }
    =
    \frac{5\pi\alpha_d^2}{4\mX^2} + \mathcal{O}(\vrel^2)
    \, .
\end{equation}
We refer to the appendix~\ref{app:cross_sections}
for details on the cross-section calculations. 

Including near-threshold effects,
we determine the pairs $(\mX,\alpha_d)$,
or equivalently the coupling $\gd$,
for which the observed dark matter relic abundance,
$\Omega_{\rmii{DM}} h^2 \big|_{\rm obs.}
=
0.1200 \pm 0.0012
$,
is reproduced~\cite{Planck:2018vyg}.
The Boltzmann equation is solved numerically for the yield
$Y_\rmii{$X$}=n_\rmii{$X$}/s$, where
$s$ denotes the total entropy density of the visible and dark sectors.%
\footnote{%
  Strictly speaking, the Boltzmann equation should be integrated down to
  temperatures well below the chemical freeze-out temperature.
  For the standard variable $z=\mX/T$,
  matching the precision of the experimental uncertainty on the relic abundance
  typically requires $z \gtrsim 10^4$.
  We explicitly checked that switching to the broken-phase annihilation
  cross-section at low temperatures has negligible impact on
  the predicted relic abundance.
  Consequently, $\Omega_{\rmii{DM}} h^2$ is effectively independent of
  the scalar and vector masses, or equivalently of the quartic coupling $\lambda_s$.
} 

After including Sommerfeld enhancement and bound-state formation effects,
we find that the corrections to the perturbative freeze-out prediction
remain modest, at the level of approximately $4\%$.
This can be traced back to the relatively small couplings required
to reproduce the observed relic density in the mass range
$\mX \in [10,10^2]$~GeV,
corresponding to $\gd \in [0.06,0.2]$.

Notably, the gauge couplings required to realize
a strong first-order phase transition in
the scanned parameter space of tab.~\ref{tab:pta:scan} are significantly larger than
those compatible with the observed relic abundance via thermal freeze-out.
As a consequence, for couplings relevant to the phase-transition dynamics,
the dark matter relic density is generically several orders of magnitude
below the observed value.

We now turn to late-time annihilations and consider
the leading velocity-independent contribution to the annihilation cross-section
in the broken phase of the dark sector,
denoted by $\sigma^{\rmi{bro}}_{\rmi{ann}} v_{\rm rel}^{ }$.
The explicit expression is reported in appendix~\ref{app:cross_sections}.
The main qualitative difference with respect to the symmetric phase is that
the Sommerfeld factor now depends on the finite mediator mass through eq.~\eqref{xi_def}.
In addition, bound-state formation proceeds via the emission of a massive dark vector boson.

\begin{figure}[t!]
  \centering
   \includegraphics[width=0.5\linewidth]{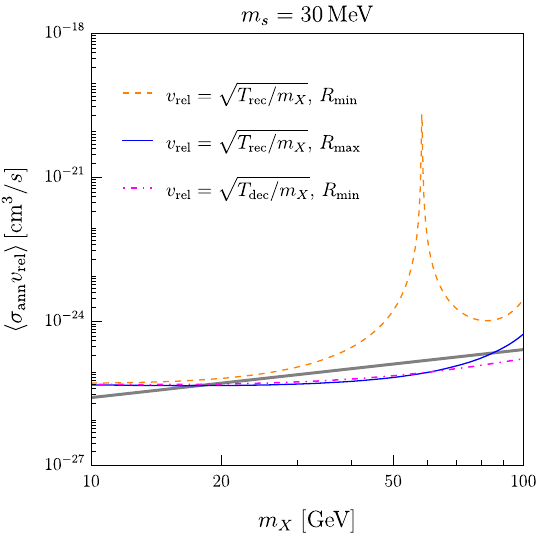}%
  \includegraphics[width=0.5\linewidth]{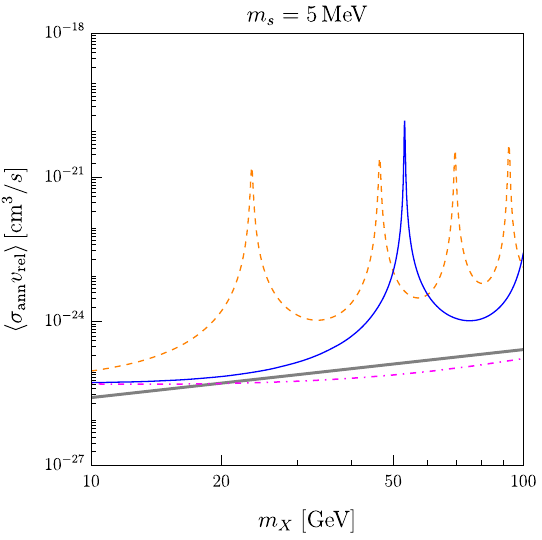}
  \caption{
    Thermally averaged annihilation cross-section for
    two benchmark choices of the dark scalar mass,
    $\ms=30$~MeV (left) and
    $\ms=5$~MeV (right), same legend holds for both panels.
    The solid gray line denotes exclusion limit as given in eq.~\eqref{indirect_ann_exp}.
    Dashed orange and blue curves show the annihilation cross-section
    at relative velocities characteristic of the recombination epoch.
    The magenta dot-dashed
    curve corresponds to larger relative velocities
    expected for an earlier kinetic decoupling at $T\sim \mathcal{O}(1)$~MeV.}
  \label{fig:Planck_symmetric}
\end{figure}
The comparison between the predicted annihilation cross-section and
the \textit{Planck} exclusion bound is shown
in fig.~\ref{fig:Planck_symmetric} for
two benchmark values of the dark scalar mass, namely
$\ms=30$~MeV and
$\ms=5$~MeV.
One clearly observes the resonant structure characteristic of
Sommerfeld enhancement in a Yukawa potential.
We further verified that the contribution from bound-state formation
remains subleading throughout the parameter space of interest.
This is due both to the smallness of the relevant couplings and
to the phase-space suppression associated with the emission of
a massive mediator in the final state.

Overall, a substantial portion of the dark matter mass range considered in this work is strongly constrained.
We analyze two benchmark configurations for the ratio between the vector mediator and dark matter masses,
$R \equiv \frac{\mV}{\mX}$, namely
$R_{\rm min}$ corresponding to $\mV=2\ms$, and
$R_{\rm max}$ corresponding to $\mV=10\ms$.
If one assumes that dark matter particles remain in kinetic equilibrium with the thermal bath until recombination,
their typical relative velocity can be estimated as $v_{\rm rel} \approx \sqrt{T_{\rm rec}/\mX}$.
For the mass range $\mX \in [10,10^2]$~GeV and using $T_{\rm rec}\simeq 1$~eV,
one obtains relative velocities in the range
$v_{\rm rel} \sim 10^{-6}\text{--}10^{-5}$.

In practice, kinetic decoupling is expected to occur well before recombination.
Since the massive dark vector and scalar particles must decay before BBN,
the scattering processes responsible for
maintaining kinetic equilibrium between dark matter and
the thermal bath are expected to become inefficient already at
temperatures of $\mathcal{O}(1)$~MeV (see sec.~\ref{sec:thermal_contact}).
This occurs well before recombination and therefore implies
larger relative velocities,
$v_{\rm rel} \approx 10^{-3}$.
The resulting enhancement is substantially reduced,
and the magenta dot-dashed curves shown in fig.~\ref{fig:Planck_symmetric} consequently
lie much closer to, and in some regions slightly below,
the current experimental exclusion limits.

\subsection{Asymmetric dark matter scenario}
\label{sec:asymm_DM}

A primordial asymmetry may be generated in the dark sector,
in analogy with the baryon asymmetry of the visible sector.
In the latter case, observations indicate
$\eta_{\rmii{B}}=(0.88 \pm 0.021)\times10^{-10}$,
as inferred from CMB measurements~\cite{Planck:2018vyg}.
We do not attempt to address the microscopic origin of
the dark-sector asymmetry
(see e.g.~reviews of the subject~\cite{Petraki:2013wwa,Zurek:2013wia}).
Instead, we assume that such an asymmetry
is generated before chemical decoupling and is
therefore already present during the freeze-out epoch.
In the following, we adopt the notation and formalism
of~\cite{Graesser:2011wi,Gelmini:2013awa}.

We define the asymmetry parameter through the difference between
the particle and antiparticle yields,
\begin{align}
\label{def_asy_quantitites}
  Y_{\rmii{$X$}}-Y_{\rmii{$\bar X$}}
  &=
  \eta_{\rmii{DS}}
  \, ,
  &
  Y_{\rmii{$X$}}
  &=
  \frac{n_\rmii{$X$}}{s}
  \, ,
  &
  Y_{\rmii{$\bar X$}}
  &=
  \frac{n_{\rmii{$\bar X$}}}{s}
  \, ,
  &
  r_{\infty}
  &=
  \frac{Y_{\rmii{$\bar X$}}(z_{\rmi{fin}})}
  {Y_{\rmii{$X$}}(z_{\rmi{fin}})}
  \, ,
\end{align}
where
$\eta_{\rmii{DS}}=\epsilon \eta_{\rmii{B}}$ denotes
the conserved dark-sector asymmetry, normalized to the observed baryon asymmetry.
Without loss of generality, we assume $\eta_{\rmii{DS}}>0$,
corresponding to an excess of dark matter particles over antiparticles.
The present-day dark matter abundance is then determined by
the combined contribution of both components.
The ratio $r_\infty$ parametrizes the residual
antiparticle fraction at late times.%
\footnote{%
  The Boltzmann equations~\eqref{particle_BE_asy} and \eqref{anti_particle_BE_asy},
  are solved up to $z_{\text{fin}}=10^5$.
}

In contrast to the symmetric freeze-out scenario,
the equilibrium number densities involve a non-vanishing chemical potential.
In the non-relativistic regime, the equilibrium density of particles is
approximately given by
\begin{align}
    n_{\rmii{$X$}}^{\textrm{eq}} (\muX)
    &\simeq
    n_{\rmii{$X$}}^{\textrm{eq}}
    \,
    e^{\muX/T}
    \, ,&
     n_{\rmii{$X$}}^{\textrm{eq}}
     =&
    \gX
     \left(
     \frac{\mX T}{2 \pi}
     \right)^{3/2}
     e^{-\mX/T}
    \, ,
\end{align}
where $n_{\rmii{$X$}}^{\textrm{eq}}$ corresponds to the equilibrium density in the symmetric limit and
$\muX$ is the chemical potential of the dark fermion.
The corresponding expression for antiparticles is
obtained by reversing the sign of the chemical potential.%
\footnote{%
  In the limit of vanishing asymmetry,
  particle and antiparticle equilibrium densities coincide
  {\em viz.}
  $n_{\rmii{$X$}}^{\textrm{eq}}=n_{\rmii{$\bar X$}}^{\textrm{eq}}$.
}

By introducing the equilibrium yield in the symmetric case,
\begin{equation}
  Y^{\rmi{eq}} \equiv n^{\textrm{eq}}_{\rmii{$X$}}/s
  \, ,
\end{equation}
one obtains the following relation for the chemical potential
\begin{equation}
    e^{\muX/T}
    =
    \frac{1}{2}
    \Biggl(
      \frac{\eta_{\rmii{DS}}}{Y_{\rmi{eq}}}
    + \sqrt{
      4 +
      \frac{\eta_{\rmii{DS}}^2}{Y_{\rmi{eq}}^2}
    }
  \Biggr)
  \, .
\end{equation}

The Boltzmann equations for the particle and antiparticle yields
can then be written as
\begin{align}
\label{particle_BE_asy}
    \frac{{\rm d} Y_\rmii{$X$}}{{\rm d} z}
    &=
    - \frac{s \xi(T)}{z H}
    \langle \sigma_{\textrm{eff}} v_{\textrm{rel}} \rangle
    \left(
    Y_\rmii{$X$}^2
    - \eta_{\rmii{DS}}^{ } Y_\rmii{$X$}^{ }
    - Y_{\rmi{eq}}^2
    \right)
    \, ,
    \\[2mm]
\label{anti_particle_BE_asy}
     \frac{{\rm d} Y_\rmii{$\bar X$}}{{\rm d} z}
    &=
    - \frac{s \xi(T)}{z H}
    \langle \sigma_{\textrm{eff}} v_{\textrm{rel}} \rangle
    \left(
    Y_\rmii{$\bar X$}^2
    - \eta_{\rmii{DS}}^{ } Y_\rmii{$\bar X$}^{ }
    - Y_{\rmi{eq}}^2
    \right)
    \, ,
\end{align}
where
\begin{equation}
    \xi (T)
    \equiv
    \left(
    1 +
    \frac{d \ln h_{\textrm{eff}}(T)}
    {3 d\ln T }
    \right)
    \, ,
\end{equation}
accounts for the temperature dependence of the entropic degrees of freedom.
The effective annihilation cross-section $\langle \sigma_{\textrm{eff}} v_{\textrm{rel}} \rangle$
is the same quantity as in eq.~\eqref{Cross_section_eff} and
includes Sommerfeld enhancement and bound-state formation effects.

The present-day dark matter abundance is finally expressed as
\begin{eqnarray}
    \Omega_{\rmii{DM}} h^2
    =
    0.2743 \times 10^{9}
     \,
     Y_{\rmii{$X$}}(z_{\textrm{fin}})
     (1+r_\infty)
     \frac{\mX}{\textrm{GeV}}
    \, ,
\end{eqnarray}
using the last relation in~\eqref{def_asy_quantitites},
together with the present-day entropy density and critical density.

Our results are consistent with the general picture established in
previous studies of asymmetric dark matter~\cite{Graesser:2011wi,Gelmini:2013awa}.
The relic abundance of particles is predominantly determined by
the primordial asymmetry $\eta_{\rmii{DS}}$ and is therefore largely insensitive
to the annihilation cross-section.
In contrast, the antiparticle abundance
is strongly controlled by the annihilation efficiency and can be substantially depleted depending on the values of the coupling $\gd$ and the asymmetry parameter $\eta_{\rmii{DS}}$. As expected, larger couplings lead to a stronger suppression of the antiparticle component and therefore to smaller values of $r_\infty$.

\begin{figure}[t!]
     \centering
    \begin{minipage}[t]{0.475\linewidth}
        \vspace{0pt}
        \includegraphics[width=\linewidth]{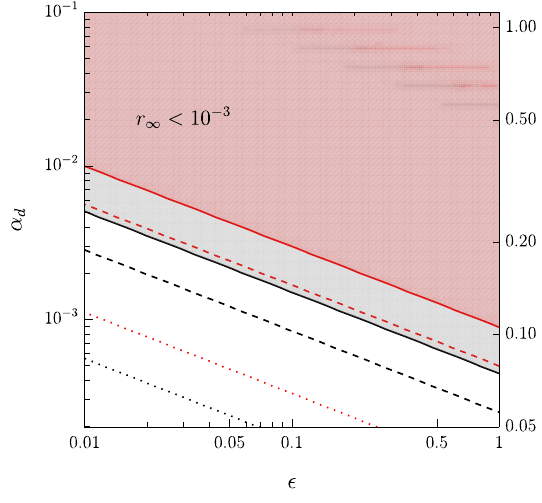}%
    \end{minipage}%
    \begin{minipage}[t]{0.525\linewidth}
        \vspace{-0.35 cm}
        \includegraphics[width=\linewidth]{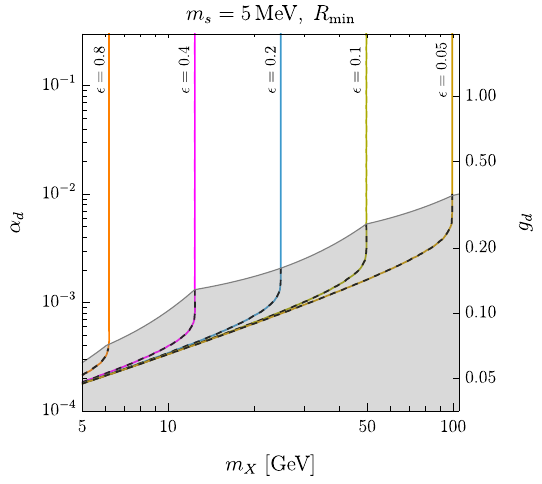}%
    \end{minipage}
    \caption{%
      Left:
      Contour lines of the residual antiparticle fraction $r_\infty$
      in the $(\epsilon,\alpha_d)$ plane for
      $\mX=10$~GeV (black) and
      $\mX=50$~GeV (red).
      Dotted, dashed, and solid lines denote
      $r_\infty=0.9$, $0.1$, and $10^{-3}$, respectively.
      Right:
      Contour lines reproducing the observed dark matter relic abundance
      for different values of the dark-sector asymmetry
      $\eta_{\rmii{DS}}=\epsilon \eta_{\rmii{B}}$.
      Black dotted segments and the gray shaded region indicate
      the parameter space excluded by
      \textit{Planck} limits on late-time dark matter annihilations.
    }
    \label{fig:asy_parameter_space}
\end{figure}
In fig.~\ref{fig:asy_parameter_space} (left),
we show the particle-antiparticle ratio
$r_\infty$ in the $(\epsilon,\alpha_d)$ plane,
together with contours of the gauge coupling $\gd$ for ease of comparison with
the phase transition analysis.
One observes that sizable antiparticle fractions are
realized only for sufficiently small couplings, while $r_\infty$
rapidly decreases to the per-mille level as $\alpha_d$ increases.
Consequently, Sommerfeld enhancement and bound-state effects become relevant only in
the regime where the antiparticle abundance is already strongly suppressed,
implying that their impact on the relic density remains moderate.

We present results for two benchmark dark matter masses, namely
$\mX=10$~GeV and
$\mX=50$~GeV,
shown respectively by the black and red curves.
Their behavior can be understood from the scaling of
the annihilation cross-section,
$\sigma_{\rmi{eff}} \, v_\rmi{rel}\propto \alpha_d^2/\mX^2$.
Larger dark matter masses therefore require larger values of the coupling
to achieve the same level of antiparticle depletion.

Finally,
we turn to the region of parameter space consistent with
the observed dark matter relic abundance.
In fig.~\ref{fig:asy_parameter_space} (right),
we show representative curves in the $(\mX,\alpha_d)$ plane
for different values of the asymmetry parameter $\epsilon$.
These curves indicate the combinations of
the dark matter mass and coupling for which the observed relic abundance,
$\Omega_{\rmii{DM}} h^2 \big|_{\rm obs.}$,
is reproduced.
At small values of $\alpha_d$, the coupling exhibits
a non-trivial dependence on the dark matter mass.
As discussed above, this regime corresponds to sizable residual antiparticle
fractions, such that the relic abundance becomes sensitive to
the annihilation cross-section. This effect becomes increasingly pronounced
for smaller asymmetries. Indeed, decreasing $\epsilon$ reduces
the particle contribution to the relic density, since
$Y_\rmii{$X$}\propto \epsilon \eta_{\rmii{B}}$, and
therefore requires a larger antiparticle component.

In the same fig.~\ref{fig:asy_parameter_space} (right),
we also superimpose the regions excluded by late-time annihilations at
recombination, cf.~eq.~\eqref{indirect_ann_exp}.
The excluded regions correspond predominantly to large values of $r_\infty$,
which are realized for sufficiently small couplings.
In this regime, the antiparticle abundance remains sizable and enhances
the annihilation signal probed by CMB observations.
For the comparison with late-time annihilation constraints,
we also include Sommerfeld enhancement and bound-state formation effects.
We find that these corrections remain relatively mild and
become appreciable only for couplings
$\alpha_d \gtrsim 10^{-3}$.

The accumulation of viable solutions around $\mX \sim 5$~GeV
can be readily understood from the observed relation between
the dark matter and baryonic energy densities.
Since the dark matter abundance is approximately five times
larger than the baryonic one, and the visible matter density is
dominated by baryons with masses of $\mathcal{O}(1)$~GeV,
asymmetric dark matter scenarios naturally favour dark matter masses
in the few-GeV range.

Overall, and in contrast to the symmetric scenario discussed in
sec.~\ref{sec:symm_DM},
we find sizable regions of parameter space
that remain phenomenologically viable,
especially for larger $\gd$.
This feature is particularly appealing in light of the phase transition analysis,
where stronger couplings are generally associated with
a more sizable GW signal.

%
\section{Reconciling first-order phase transitions with PTA data}
\label{sec:pta}

We focus on the dark-sector Abelian Higgs model~\eqref{eq:lag:4d},
as a representative for the model class of
radiatively generated cubic potentials~\eqref{eq:Veff:bro:LO:dimless}.
Using its thermal EFT introduced in sec.~\ref{sec:PT_and_GW},
we assess whether the region of parameter space
that gives rise to a theoretically controlled first-order phase transition is compatible with
the SGWB favored by
the recent PTA data~\cite{NANOGrav:2023gor,EPTA:2023fyk,EPTA:2023xxk,Reardon:2023gzh,Xu:2023wog}.
After establishing the range over which the employed EFTs provide
a controlled description of the phase transition,
we compare the resulting GW predictions with the PTA-preferred
parameter space.
To this end,
we employ
the soft~\ref{eq:bro:soft} and
the softer~\ref{eq:bro:softer}
at the accuracy level outlined in sec.~\ref{sec:PT_and_GW}
which we will refer to as NLO.

The parameter space scan is performed
over the model parameters of the fundamental Lagrangian~\eqref{eq:lag:4d},
spanned by
the dark gauge coupling $\gd$,
the dark scalar quartic coupling $\lambda_s$, and
the dark scalar mass $\ms$,
whose ranges and corresponding
prior distributions are 
informed by
the constraints~\eqref{mixing_mass_limits} and
summarized in tab.~\ref{tab:pta:scan}.
\begin{table}
    \centering
    \begin{tabular}{|l|c|c|c|c|}
        \hline
        & $\gd$ & $\ms$~[MeV] & $\lambda_s$ & $\sin\thetaS$ \\
        \hline
        \hline
        Range &
          $[0.25,1.5]$ &
          $[10^{0},10^{2}]$ &
          $[10^{-4},3\times10^{-1}]$ &
          $[2\times 10^{-5},2\times10^{-4}]$ \\
        Prior & linear & logarithmic & logarithmic & logarithmic \\
        \hline
    \end{tabular}
    \caption{%
      Parameter space scan ranges in the Abelian Higgs model~\eqref{eq:lag:4d},
      focused on a first-order phase transition at the MeV scale
      based on the constraints~\eqref{mixing_mass_limits}.
      Priors are chosen to be
      either linear or logarithmic in the indicated range.
      The parameters are input at the scale of the corresponding dark photon mass
      $\LamD_{0} = \mV$ and run to the thermal scale
      $\LamD_\rmi{ref} = \pi T$~\eqref{eq:LamD:ref}.
    }
    \label{tab:pta:scan}
\end{table}
The dark fermion mass is fixed to $\mX \sim \mathcal{O}(10\text{--}10^2)$~GeV,
and therefore is Boltzmann suppressed at the MeV scale of the phase transition.
The parameters are input at the scale of the corresponding dark photon mass
$\LamD_{0} = \mV$ and run to
the reference scale
\begin{align}
  \label{eq:LamD:ref}
  \LamD_\rmi{ref} &= A \pi T
  \, ,&
  \text{with}\quad
  A &\in [2^{-1},2^{1}]
  \,,
\end{align}
which aligns with the lowest Matsubara mode.
Later,
we explore the impact of varying the reference scale by a factor of two in either direction.
In most of the scans we take the central value $A=1$.
We also relate physical parameters to Lagrangian parameters
at one-loop level in vacuum renormalization,
as detailed in appendix~\ref{sec:MSbar:physical} and
e.g.~\cite{Kajantie:1995dw,Niemi:2021qvp}. 

\begin{figure}[t!]
    \centering
    \includegraphics[width=0.5\linewidth]{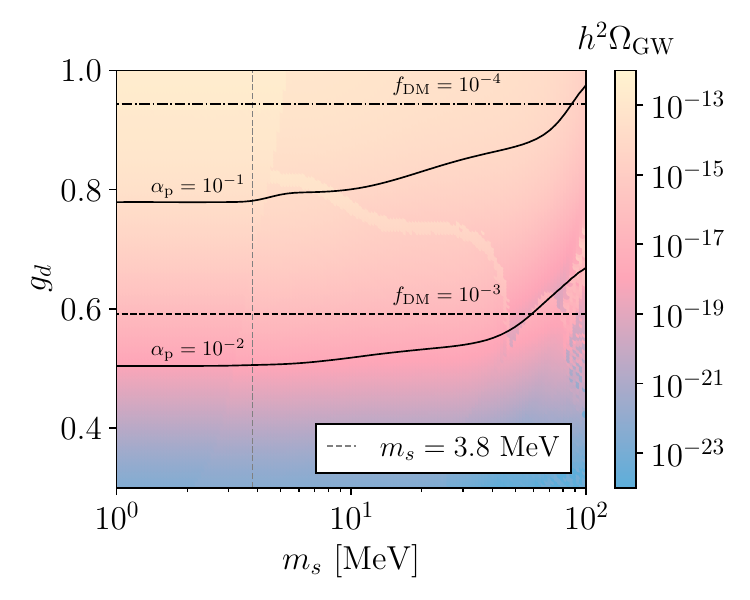}%
    \includegraphics[width=0.5\linewidth]{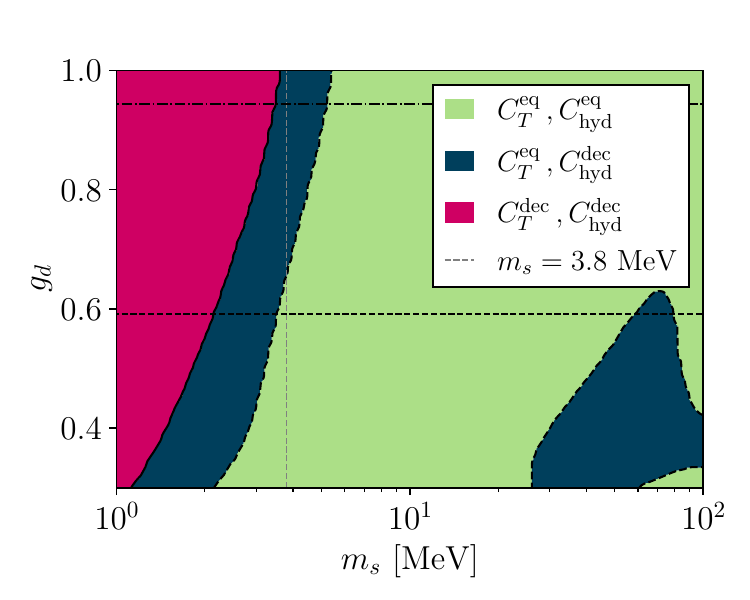}
    \caption{%
      Left: Gravitational-wave amplitude for a first-order phase transition
      in the Abelian Higgs model~\eqref{eq:lag:4d},
      computed in the soft~\eqref{eq:bro:soft}
      in the $(\gd,\ms)$ plane.
      The dark scalar quartic coupling is fixed to $\lambda_s = 10^{-2}$
      and $\sin\thetaS = 2\times10^{-4}$.
      The vertical dashed line at $\ms =3.8$~MeV
      is the lower bound from~\cite{Ibe:2021fed}.
      Black dashed and dot-dashed lines indicate
      $(\gd,\ms)$ values for which symmetric freeze-out yields
      a fraction
      $f_\rmii{DM} \equiv \Omega_\rmii{$X$}/\Omega_\rmii{DM}^{\rm obs}$
      of the observed dark matter abundance,
      $\Omega_\rmii{DM}^{\rm obs} h^2=0.1200(12)$~\cite{Planck:2018vyg}.
      The strongest transitions occur at the largest $\gd$ and smallest $\ms$.
      Towards the upper right corner,
      $\vw^\rmii{LTE} = 1$, as indicated by a faint discontinuous jump
      in $\Omega_\rmii{GW}$.
      Right:
      Thermal equilibrium condition $C_\T^\rmi{eq}$~\eqref{eq:thermal:equilibrium:condition} and
      hydrodynamic equilibrium condition $C_\rmi{hyd}^\rmi{eq}$~\eqref{eq:hydro:equilibrium:condition}
      in the same $(\gd,\ms)$ plane.
      Magenta regions satisfy neither $C_\T^\rmi{eq}$ nor $C_\rmi{hyd}^\rmi{eq}$,
      blue regions satisfy only $C_\rmi{hyd}^\rmi{eq}$, and
      green regions satisfy both $C_\T^\rmi{eq}$ and $C_\rmi{hyd}^\rmi{eq}$.
    }
    \label{fig:pta:scan:soft}
\end{figure}

\begin{figure}[t!]
    \centering
    \includegraphics[width=0.5\linewidth]{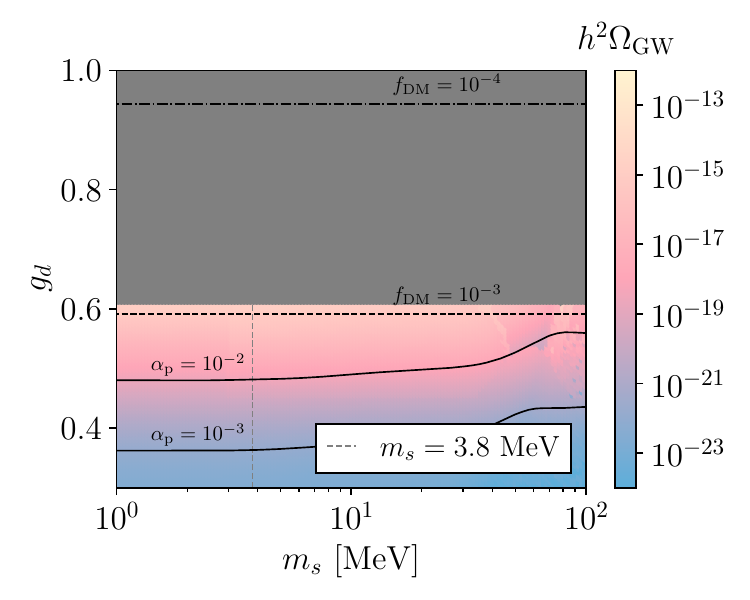}
    \caption{%
      Gravitational-wave amplitude for a first-order phase transition
      in the Abelian Higgs model~\eqref{eq:lag:4d},
      computed in
      the softer~\eqref{eq:bro:softer}
      in the $(\gd,\ms)$ plane.
      Scan ranges and constraint lines the same as in fig.~\ref{fig:pta:scan:soft}.
      Grey shaded regions mark where $x < x_{\rm min}$
      and higher-dimensional operators in the EFT become relevant
      with
      $x_{\rm min} = 0.051\gd^2$
      taken for
      the soft \ref{eq:bro:soft}~\cite{Bernardo:2025vkz}
      which
      coincides with the onset of unphysical values $c^2_{s,\rmi{bro}} < 0$
      for the softer \ref{eq:bro:softer}.
    }
    \label{fig:pta:scan:softer}
\end{figure}

We first focus on a sub-set of the parameter space scan
for fixed
$\lambda_s = 10^{-2}$,
$\sin\thetaS = 2\times10^{-4}$,
up to $\gd = 1.0$ and
compute corresponding the GW amplitude
$\Omega_{\rmii{GW}}h^2$ 
in the single broken power-law template of eq.~\eqref{eq:S:single}
in both
the soft~\ref{eq:bro:soft} in fig.~\ref{fig:pta:scan:soft} and
the softer~\ref{eq:bro:softer} in fig.~\ref{fig:pta:scan:softer}.
Both EFTs possess
only a finite range of validity in their respective
EFT expansion parameter $x$~\cite{Bernardo:2025vkz}
defined in eq.~\eqref{eq:Veff:bro:LO:dimless}.
For \ref{eq:bro:softer} this range is very narrow, and
the breakdown of its validity is
indicated by the gray shaded region in fig.~\ref{fig:pta:scan:softer}.
The breakdown also manifests itself in
the unphysical values $c^2_{s,\rmi{bro}} < 0$
as reported in fig.~\ref{fig:breakdown_diagnostics} (right).
The latter can also be interpreted as a loss of perturbativity,
since
$p_0 \sim T F_\rmi{bro}$ in this regime.
This effect is especially pronounced in our model because
the number of
relativistic degrees of freedom is small, $\geff \sim \mathcal{O}(1)$.%
\footnote{%
  For phase transitions occurring at the electroweak scale,
  the effective number of relativistic degrees of freedom is
  $g_\text{eff} \approx \mathcal{O}(100)$.
  Consequently, the symmetric pressure $p_0$ is significantly larger
  than in the MeV-scale Abelian Higgs model considered here,
  which enhances the pathological behavior of $c_{s,\text{bro}}^2$.
}
By contrast, the less restrictive
approximation in eq.~\eqref{eq:bro:soft} remains valid over a wider range,
so the corresponding gray-shaded region in fig.~\ref{fig:pta:scan:soft} (left)
affects only the large-coupling part
beyond the scan.
The strongest and slowest transitions are realized for
the largest values of $\gd$ and the smallest values of $\ms$.

We assess whether the dark and visible sectors remain
both thermally and hydrodynamically coupled throughout the phase transition,
as established through the criteria of
eqs.~\eqref{eq:thermal:equilibrium:condition} and~\eqref{eq:hydro:equilibrium:condition}.
Thermal equilibrium determines
whether the two sectors share a common temperature
$T_\rmii{DS} = T_\rmii{SM}$.
Hydrodynamic equilibrium determines 
whether both sectors are hydrodynamically coupled
and hence
which relativistic degrees of freedom enter
in the hydrodynamic quantities of sec.~\ref{sec:PT_and_GW}.
In fig.~\ref{fig:pta:scan:soft} (right),
neither criterion holds in the magenta region,
only hydrodynamic equilibrium in the blue region, and
both in the green region.
Notably, the strongest PTA-relevant signals arise at large coupling and
$\ms \gtrsim 3.8$~MeV~\cite{Ibe:2021fed}, precisely the blue region.
There the two sectors still share a common temperature,
yet the SM plasma becomes hydrodynamically decoupled from the dark sector,
so that only the dark-sector degrees of freedom $g_\rmii{DS}$
enter
the hydrodynamic quantities relevant for the GW signal.

The scans shown in
figs.~\ref{fig:pta:scan:soft}
and~\ref{fig:pta:scan:softer}
also illustrate the interplay between the phase-transition dynamics and
the dark matter phenomenology of the model defined in eq.~\eqref{eq:lag:4d}.
As discussed in the previous sections,
we consider dark matter masses in the range $\mathcal{O}(10\text{--}100)$~GeV,
well above the MeV scale associated with the phase transition.
The black dashed and dot-dashed lines indicate the values of
the gauge coupling for which the symmetric freeze-out mechanism reproduces
only a small fraction
$f_\rmii{DM} \equiv \Omega_\rmii{$X$}/\Omega_\rmii{DM}^{\rm obs}$
of the observed dark matter abundance,
$\Omega_\rmii{DM}^{\rm obs}h^2 = 0.1200(12)$~\cite{Planck:2018vyg}.
This is a direct consequence of the relatively large gauge couplings required
to generate a strong first-order phase transition,
which in turn produce annihilation cross-sections significantly
larger than the canonical thermal value,
$\langle \sigma v_{\rm rel}\rangle_{\rm th}\simeq3\times10^{-26} \, \mathrm{cm^3 \, s^{-1}}$~\cite{Steigman:2012nb},
for dark matter masses in the range $\mathcal{O}(10)$~GeV.%
\footnote{%
  This tension is progressively alleviated for heavier dark matter particles,
  since larger gauge couplings are then required to reproduce the observed relic abundance. The DM mass range is also motivated by the comparison with earlier studies \cite{Han:2023olf}.
} 
The curves for $f_{\rmii{DM}}$ are essentially independent of the dark scalar mass,
because freeze-out occurs at temperatures well above the phase-transition temperature;
see~\eqref{sigma_ann_symm} for the corresponding annihilation cross section.
For both lines, the DM mass is set to $\mX = 30 $~GeV.

Comparing
figs.~\ref{fig:pta:scan:soft}
and~\ref{fig:pta:scan:softer} with
fig.~\ref{fig:asy_parameter_space} (right)
shows that the asymmetric dark matter scenario naturally accommodates
the observed relic abundance in the region favored by the phase transition.
In particular, for $\gd \gtrsim 0.3$,
the model reproduces the observed dark matter density while remaining consistent
with current experimental constraints over the range of dark matter masses and
asymmetry parameters $\eta_\rmii{DS}$ considered in this work.
This illustrates the complementarity between the GW and dark matter phenomenology.
The relatively large gauge couplings required to generate a strong first-order phase transition
are naturally compatible with asymmetric freeze-out,
whereas they severely over-deplete the relic abundance in
the conventional symmetric scenario
of sec.~\ref{sec:symm_DM}.

\subsection{Breakdown of high-temperature expansion} 
\label{sec:breakdown_highT}

\begin{figure}[t]
    \centering
    \includegraphics[width=0.5\linewidth]{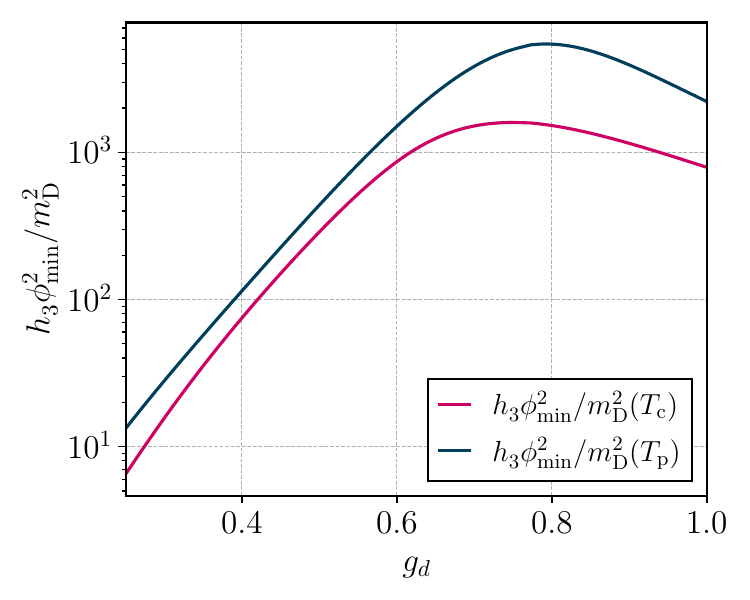}%
    \includegraphics[width=0.5\linewidth]{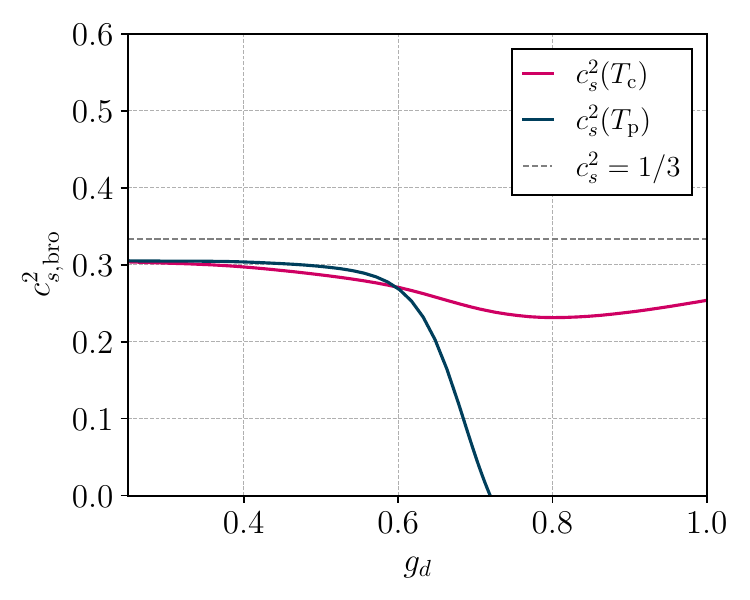}%
    \caption{%
      Tests of
      the validity of the high-temperature and perturbative expansions
      at benchmark point
      {\tt BM7} of tab.~\ref{tab:pta:benchmarks}
      and varying $\gd \in [0.25,1.0]$.
      Left:
      Ratio $h_3\phi_\rmi{min}^2/\mD^2$ across the parameter scan,
      indicating the breakdown of the high-temperature expansion when
      $h_3\phi_\rmi{min}^2/\mD^2 \gg \mathcal{O}(1)$.
      Right:
      Broken-phase speed of sound $c_{s,\rmi{bro}}^2$ across the same scan
      using the softer~\ref{eq:bro:softer},
      indicating the onset of non-perturbative behavior when
      $c_{s,\rmi{bro}}^2 \ll 1/3$.
    }
    \label{fig:breakdown_diagnostics}
\end{figure}

In this section, we scrutinize the limitations of the EFT framework
of sec.~\ref{sec:PT_and_GW} and identify the parameter regimes in which
the super-renormalizable
\ref{eq:bro:soft} and \ref{eq:bro:softer} are invalidated.
A fully consistent extension beyond the high-temperature expansion
as in e.g.~\cite{Navarrete:2025yxy} is beyond the scope of this work.
However, based on extending the EFT operator basis
to dimension-six~\cite{Bernardo:2025vkz,Bernardo:2026whs}, 
we delimit the range of validity of the EFTs
on top of the parameter space scan of tab.~\ref{tab:pta:scan}
using the criteria of eqs.~\eqref{eq:bro:soft} and~\eqref{eq:bro:softer}.
These limits were already indicated in
fig.~\ref{fig:pta:scan:softer}
by the gray shaded regions and
mark the breakdown of the EFT expansion
which coincides, to a good extent, with the unphysical values
of the broken-phase speed of sound $c^2_{s,\rmi{bro}} < 0$
which we display in fig.~\ref{fig:breakdown_diagnostics} (right).

For sufficiently strong transitions, the standard EFT
of eqs.~\eqref{eq:lag:3d} and~\eqref{eq:lag_3d:temporal}
becomes insufficient and higher-dimensional operators have to be
included in the
\begin{itemize}
  \item[(i)]
    \emph{Effective potential.}
    Marginal \threeDEFT{} operators such as $\phi^6$
    induced by hard modes modify the effective potential
    and become quantitatively important when the EFT expansion parameter
    $x = \lambda_{s,3}/\eta^{2/3}$ becomes too small~\cite{Chala:2024xll,Bernardo:2025vkz}.
  \item[(ii)]
    \emph{Nucleation action.}
    Both hard and soft fluctuations
    of $\mathcal{O}(\pi T), \mathcal{O}(gT)$~\cite{Chala:2024xll}, as well as
    fluctuations around the critical bubble of $\mathcal{O}(\mu_\rmi{nucl})$
    contribute to the nucleation action.
    Here, $\mu_\rmi{nucl}$ denotes the inverse length scale of the critical bubble.
    A derivative expansion in these scales, however, is only
    justified for hard and soft fluctuations~\cite{Ekstedt:2021kyx,Kierkla:2025qyz}.
    A consistent treatment requires retaining all dimension-six operators,
    including derivative operators.
    In the present model,
    this corresponds to $\mathcal{O}(10)$ independent operators~\cite{Bernardo:2025vkz,Bernardo:2026nyq}
    which also modify the fluctuation determinant.
\end{itemize}
The absence of these operators
can compromise the characterization of the phase-transition dynamics directly
in certain parameter space regions.
Since for (ii) only the computation of
the bounce in the presence of higher-dimensional operators
is addressed in~\cite{Chala:2024xll},
extending it to fluctuation determinants is beyond the scope of this work.
We therefore restrict to the standard super-renormalizable EFT~\eqref{eq:lag:3d} and,
utilizing the criteria of~\eqref{eq:bro:soft} and~\eqref{eq:bro:softer},
indicate
the regions of parameter space
in which omitting higher-dimensional operators is expected to be most severe.

Since PTA-favored transitions tend to be among the strongest transitions
accessible in the model, they also lie close to the validity limits of the
high-temperature effective theory.
We therefore delineate the theoretical boundaries of our EFT description.
When working in the softer \ref{eq:bro:softer},
after integrating
out the Debye scale in the symmetric phase,
the validity of the EFT can be
assessed by two conditions:
\begin{itemize}
  \item[(A)]
    The validity of the {\em high-temperature expansion},
    of having integrated out the Debye scale, can be monitored by
    the ratio $h_3^{ }\phi_3^2/\mD^2 \ll 1$.
    This ratio controls the hierarchy between the soft and Debye scales.
    When this ratio becomes of order unity,
    the integration over Matsubara zero modes can no longer be performed
    within a strict expansion in temperature scales, and
    the softer-scale~\ref{eq:bro:softer}
    ceases to provide a controlled approximation.

  \item[(B)]
    The validity of the {\em perturbative expansion}
    can be monitored by
    the ratio $h_3/(4\pi \mD) \sim \gd \ll 1$~\cite{Gould:2023ovu,Ekstedt:2024etx,Niemi:2024vzw,Bernardo:2025vkz},
    which controls the corrections to
    the effective expansion parameter within $\bar{\lambda}_{s,3}$
    as discussed in eq.~\eqref{eq:lambdas:softer} and below.
\end{itemize}

Breaking condition (A)
signals the onset of higher-dimensional operators
in the softer~\ref{eq:bro:softer},
as already observed in fig.~\ref{fig:pta:scan:softer}.
The regime where the high-temperature expansion breaks down
coincides with small $x$ (gray region),
which is precisely where higher-dimensional operators
become quantitatively important.
In fig.~\ref{fig:breakdown_diagnostics} (left),
we show the ratio $h_3^{ }\phi_\rmi{min}^{2}/\mD^2$,%
\footnote{%
  For nucleation,
  the relevant field value is the escape point $\phi_\text{esc}$,
  which is slightly larger than $\phi_\text{min}$.
  However, the difference is negligible for the present discussion
  since supercooling is small in the parameter space of interest.
  For stronger supercooling, however, the escape point can become
  hierarchically separated, $\phi_\text{esc} \ll \phi_\text{min}$,
  which can render the high-temperature expansion valid again~\cite{Kierkla:2023von}.
}
across a parameter scan in $\gd$
for fixed $\lambda_s = 10^{-2}$ and
$\sin\thetaS = 2\times10^{-4}$.
For larger values of $\gd$, the ratio approaches $\mathcal{O}(10^3)$,
strongly invalidating the assumption that one can integrate out
the Debye scale $\mD$.

This motivates utilizing the soft-scale \ref{eq:bro:soft},
whose validity range extends to somewhat smaller $x$
than the softer broken-phase setup.
However, statements such as ``in the limit of strongest transition''
refer to parameter points where even this soft EFT
saturates its validity range.
The validity bracket $0.051\gd^2 \ll x \ll 0.18$
is in practice saturated near its lower end
for the parameter points relevant to PTA phenomenology.
A more quantitative inclusion of dimension-six operators in the nucleation action
and fluctuation determinant remains the natural next step for
this class of analyses, and we leave it to future work.
Frameworks that simultaneously incorporate thermal resummation and
remain valid over a broad temperature range are currently being developed.
See e.g.~\cite{Navarrete:2025yxy,Bandyopadhyay:2026nrv} for promising approaches
to go beyond the high-temperature expansion.

Breaking condition~(B) signals a violation of
perturbativity in the fundamental theory.
It can manifest as a breakdown of
the perturbative series of $x$ itself
or of derived quantities, such as $c^2_{s,\rmi{bro}}$ becoming unphysical,
as we show in fig.~\ref{fig:breakdown_diagnostics} (right).
This breakdown is present regardless of
whether the high-temperature expansion is valid,
as can be seen from
the large deviation of $c^2_{s,\rmi{bro}}$ from $1/3$ at both $\Tp$ and $\Tc$,
with the largest deviation at $\Tp$.
Therefore, a violation of perturbativity
can become relevant for the PTA-favored region
already for values of $\gd \gtrsim \mathcal{O}(0.5)$.
As mentioned earlier,
this unphysical behavior of the speed of sound is most pronounced
in phase transitions with a small number of relativistic degrees of freedom,
like the ones addressed in this work and relevant for PTA.
For a larger number of relativistic degrees of freedom,
the breakdown of perturbativity is delayed to larger values of $\gd$.

\subsection{Implications for the PTA-favored region}
\label{sec:pta:compatibility}

While the softer~\ref{eq:bro:softer} setup has the largest range of validity
across the relevant range of couplings,
we employ both the
soft and softer EFTs
to assess the compatibility of
the Abelian Higgs model with the PTA-favored region.
To this end, we employ the {\tt PTArcade}~\cite{Mitridate:2023oar} implementation
of the {\tt ceffyl} likelihood~\cite{Lamb:2023jls}
for the NANOGrav 15~yr data set~\cite{NANOGrav:2023gor},
considering the first 14 Fourier frequencies in the range
$f \in [2,28]$~nHz as recommended by the NANOGrav collaboration~\cite{NANOGrav:2023hvm}.

\begin{figure}[t]
  \centering
  \includegraphics[width=0.5\linewidth]{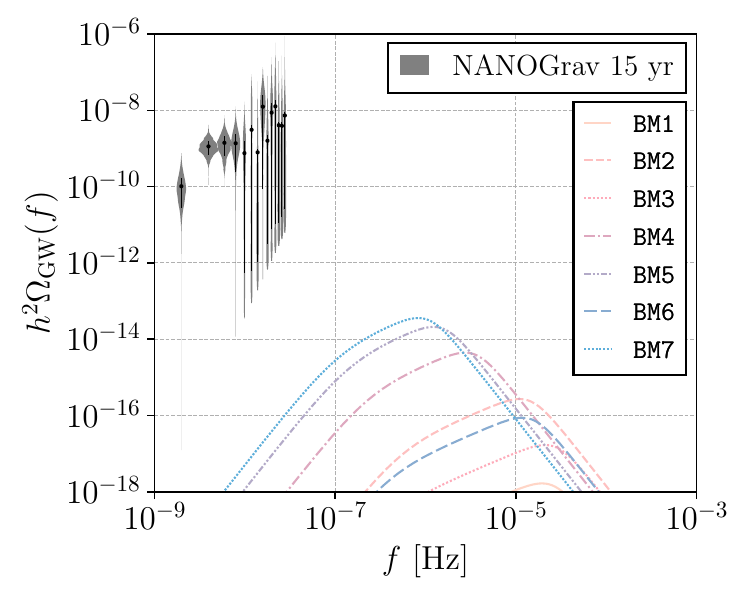}%
  \includegraphics[width=0.5\linewidth]{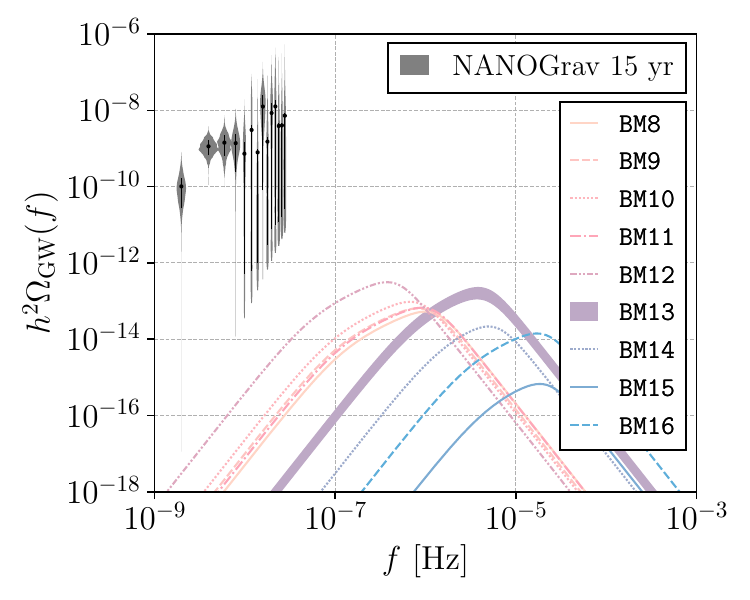}%
  \caption{%
    Left:
    Gravitational-wave spectra for the benchmark points
    {\tt BM1}--{\tt BM7} listed in tab.~\ref{tab:pta:benchmarks},
    computed using the double broken power law template of eq.~\eqref{eq:S:double}.
    The NANOGrav $\mathcal{T}=15$~yr data set~\cite{NANOGrav:2023gor}
    is shown for comparison.
    For {\tt BM13} of tab.~\ref{tab:pta:benchmarks},
    the 4d renormalization scale $\LamD = A \pi T$
    is varied around $A = \{2^{-1},2^{1}\}$ as indicated by
    the band of the predicted GW signal. 
    Right:
    {\tt BM8}--{\tt BM16} of tab.~\ref{tab:pta:benchmarks}.
  }
  \label{fig:pta:scan:spectra} 
\end{figure}

Going beyond the parameter-space scan in
figs.~\ref{fig:pta:scan:soft} and~\ref{fig:pta:scan:softer},
we report a set of benchmark~(BM) points in
tab.~\ref{tab:pta:benchmarks}
at fixed
$\sin\thetaS = 2\times10^{-4}$.
These points
allow for a direct comparison with the PTA signal at nHz frequencies and
illustrate, directly in frequency space in
fig.~\ref{fig:pta:scan:spectra}, the predicted GW signal
and its \textit{incompatibility} with the PTA-favored region.
\begin{table}[t!]
  \centering
  \resizebox{\linewidth}{!}{%
  \begin{tabular}{|l|cllrrcccc|@{\hspace{2pt}}c@{\hspace{2pt}}|@{\hspace{2pt}}c@{\hspace{2pt}}|}
    \hline
    BM & $\ms$~[MeV]
    & \multicolumn{1}{c}{$\gd$}
    & \multicolumn{1}{c}{$\lambda_s$}
    & \multicolumn{1}{c}{$\Tc$\newline[MeV]}
    & \multicolumn{1}{c}{$\Tp$~[MeV]}
    & $c_{s,\rmi{bro}}^2$
    & $\vw^\rmii{LTE}$
    & $\alpha_{\star,\rmi{hyd}}^{ }$
    & $\beta/\Hstar$
    & \hyperref[eq:thermal:equilibrium:condition]{$C_\T^\rmi{eq}$}
    & \hyperref[eq:hydro:equilibrium:condition]{$C_\rmi{hyd}^\rmi{eq}$}
    \\
    \hline
    \hline
    {\tt BM1} & 5.00 & 0.5 & 0.05   & 12.64 & 12.39 & 0.320 & 0.578 & 0.0014 & 1759 & \checkmark & \checkmark \\
    {\tt BM2} & 5.00 & 0.5 & 0.005  & 15.58 & 14.19 & 0.307 & 0.613 & 0.0256 & 4221 & \checkmark & \checkmark \\
    {\tt BM3} & 5.00 & 0.7 & 0.05   & 9.91 & 9.50 & 0.317 & 0.586 & 0.0093 & 4769 & \checkmark & \checkmark \\
    {\tt BM4} & 5.00 & 0.7 & 0.01   & 11.97 & 10.31 & 0.315 & 0.668 & 0.0556 & 2706 & \checkmark & \checkmark \\
    {\tt BM5} & 5.00 & 0.7 & 0.0075 & 12.44 & 10.08 & 0.315 & 0.721 & 0.0841 & 1816 & \checkmark & \checkmark \\
    {\tt BM6} & 5.00 & 0.8 & 0.05   & 9.11 & 8.56 & 0.316 & 0.600 & 0.0187 & 5063 & \checkmark & \checkmark \\
    {\tt BM7} & 5.00 & 0.8 & 0.01   & 11.37 & 8.97 & 0.317 & 0.751 & 0.1029 & 1625 & \checkmark & \checkmark \\
    \hline
    {\tt BM8} & 4.40 & 0.960 & 0.0190 & 8.62 & 6.88 & 0.289 & 0.698 & 0.3894 & 2184 & \checkmark & \xmark \\
    {\tt BM9} & 4.01 & 0.986 & 0.0199 & 7.75 & 6.13 & 0.294 & 0.712 & 0.4061 & 2148 & \xmark & \xmark \\
    {\tt BM10} & 4.10 & 0.921 & 0.0146 & 8.45 & 6.60 & 0.279 & 0.713 & 0.4437 & 1777 & \xmark & \xmark \\
    {\tt BM11} & 4.51 & 0.990 & 0.0200 & 8.71 & 6.87 & 0.294 & 0.714 & 0.4093 & 2138 & \xmark & \xmark \\
    \hline
    {\tt BM12} & 3.39 & 1.2 & 0.03   & 5.94 & 4.38 & 0.349 & 0.817 & 0.5010 & 1801 & \xmark & \xmark \\
    \hline
    {\tt BM13} & \hphantom{0}55.0 & 0.75 & 0.006 & 136.91 & 98.99 & 0.219 & 0.840 & 0.7551 & 1102 & \checkmark & \xmark \\
    {\tt BM14} & 110 & 0.75 & 0.006 & 273.83 & 197.99 & 0.170 & 1.000 & 0.8932 & 1102 & \checkmark & \xmark \\
    {\tt BM15} & 329 & 0.75 & 0.006 & 818.99 & 592.16 & 0.254 & 1.000 & 0.0359 & 1102 & \checkmark & \checkmark \\
    {\tt BM16} & 141 & 0.85 & 0.01 & 313.78 & 238.21 & 0.251 & 0.750 & 0.5463 & 1362 & \checkmark & \xmark \\
    \hline
  \end{tabular}
  }%
  \caption{%
    Benchmark~(BM) points in the parameter space of the Abelian Higgs model~\eqref{eq:lag:4d}
    that yield a first-order phase transition at the MeV scale
    in the soft~\ref{eq:bro:soft}.
    {\tt BM8}--{\tt BM11} are taken from~\cite{Han:2023olf},
    {\tt BM12} is from~\cite{Banik:2024zwj}, and
    {\tt BM13}--{\tt BM16} are from~\cite{Costa:2025csj}.
    The input parameters are
    the dark scalar mass $\ms$,
    the dark gauge coupling $\gd$, and
    the dark scalar quartic coupling $\lambda_s$,
    specified at the input scale $\LamD_{0} = \mV$.
    The portal coupling $\sin\thetaS = 10^{-4}$
    and $\epsS < 1$.
    The remaining columns list
    the critical temperature $\Tc$,
    the percolation temperature $\Tp$,
    the broken-phase sound speed $c_{s,\rmi{bro}}^2(\Tstar)$,
    the wall velocity $\vw^\rmii{LTE}(\Tstar)$ in the LTE approximation of~\cite{Ai:2023see},
    the transition strength of the hydrodynamic sector
    $\alpha_{\rmi{hyd}} (\Tstar)$, and
    the inverse transition timescale $\beta/\Hstar$.
    The last two columns indicate whether
    the dark sector is in thermal equilibrium
    with the SM plasma ($C_\T^\rmi{eq}$ of eq.~\eqref{eq:thermal:equilibrium:condition})
    and whether
    it is in hydrodynamic equilibrium with the SM
    ($C_\rmi{hyd}^\rmi{eq}$ of eq.~\eqref{eq:hydro:equilibrium:condition}).
    All thermodynamic quantities are evaluated at $\Tstar = \Tp$.
  }
  \label{tab:pta:benchmarks}
\end{table}%
Here, we employ the double broken power law template of eq.~\eqref{eq:S:double} and
display the corresponding spectra
for
{\tt BM1}--{\tt BM7} in fig.~\ref{fig:pta:scan:spectra} (left) and
{\tt BM8}--{\tt BM16} in fig.~\ref{fig:pta:scan:spectra} (right).
The latter are taken from earlier studies,
{\tt BM8}--{\tt BM11} from~\cite{Han:2023olf},
{\tt BM12} from~\cite{Banik:2024zwj}, and
{\tt BM13}--{\tt BM16} from~\cite{Costa:2025csj}.
Additionally,
in fig.~\ref{fig:pta:scan:spectra}~(right),
we show the dependence of the predicted GW signal on
the 4d renormalization scale
$\LamD$
for {\tt BM13} of tab.~\ref{tab:pta:benchmarks}.
The band indicates the residual scale dependence
obtained by varying
the reference scale as in eq.~\eqref{eq:LamD:ref}.
This scale dependence estimates the theoretical uncertainty
from missing higher-order perturbative corrections~\cite{Croon:2020cgk},
corresponding to a variation
$\Delta \Omega_\rmii{GW}/ \Omega_\rmii{GW} \sim \mathcal{O}(10)$.
This is typical at NLO and remains much smaller than
the theoretical uncertainty of
LO predictions~\cite{Croon:2020cgk,Gould:2021oba,Gould:2023ovu,Lewicki:2024xan}.
Benchmarks {\tt BM9}--{\tt BM12} are
not in thermal equilibrium with the SM plasma,
failing the condition~\eqref{eq:thermal:equilibrium:condition}.
The number of effective hydrodynamically relevant degrees of freedom
is adapted according to condition~\eqref{eq:hydro:equilibrium:condition}.
In all cases,
the predicted signal is incompatible with the PTA-favored region.

\begin{figure}[t!]
    \centering
    \includegraphics[width=0.5\linewidth]{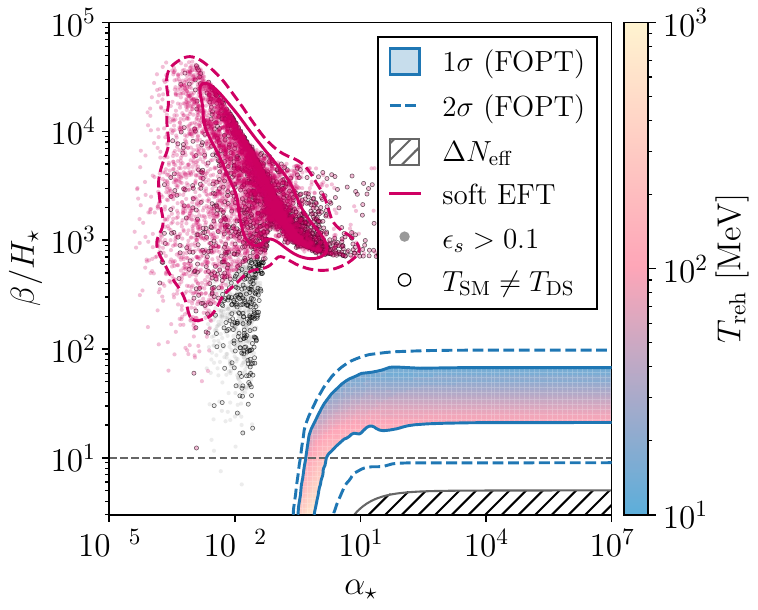}%
    \includegraphics[width=0.5\linewidth]{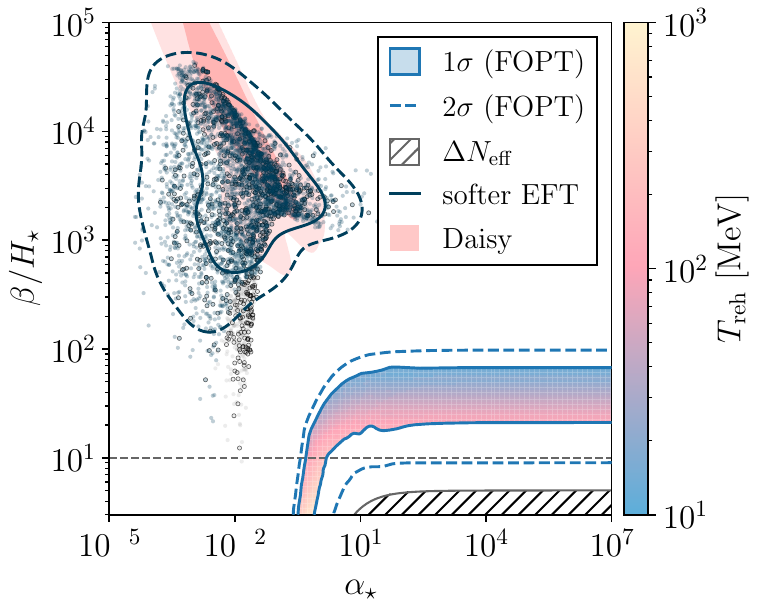}
    \caption{%
      The $1\sigma$ and $2\sigma$ regions favored by current PTA data,
      described by
      the transition strength $\alpha_\star$ and
      the inverse duration $\beta/\Hstar$ of
      a generic dark sector first-order phase transition (light blue), assuming
      sound-wave induced GWs modelled with
      the doubly broken power-law template~\eqref{eq:S:double} and
      the NANOGrav $\mathcal{T}=15$~yr data set~\cite{NANOGrav:2023gor}.
      The colorbar on the right-hand side shows
      the best-fit reheating temperature $\Treh$,
      while the gray hatched region indicates the parameter space excluded by $\Delta N_{\rm eff}$
      constraints from BBN and CMB observations~\cite{Yeh:2022heq}.
      The dashed gray line marks
      values of $\beta/\Hstar$ that are not supported by simulations~\cite{Jinno:2022mie}.
      For
      the soft~\ref{eq:bro:soft}~(left, magenta) and
      the softer~\ref{eq:bro:softer}~(right, dark blue) setups,
      the individual points show the values of
      $\alpha_\star$ and $\beta/\Hstar$
      obtained in the parameter-space scan of tab.~\ref{tab:pta:scan},
      and the contours give the corresponding
      $1\sigma$~(solid) and
      $2\sigma$~(dashed) density regions.
      Gray shaded points correspond to $\epsS > 0.1$.
      Black circled points are out of thermal equilibrium
      with the SM plasma, failing the condition \eqref{eq:thermal:equilibrium:condition}.
      In the right panel,
      we show the corresponding parameter space
      computed \`a la Daisy resummation (orange),
      as in~\cite{Bringmann:2026xcx}.
    }
    \label{fig:pta:alpha:beta}
\end{figure}

Finally, we perform a randomized scan~\cite{AbdusSalam:2020rdj} over
the model parameter ranges and priors in tab.~\ref{tab:pta:scan}
to determine the full distribution of the thermodynamic quantities
$(\alpha_\star,\beta/\Hstar)$ and compare the generic GW template of
eq.~\eqref{eq:S:double} with the PTA data.
The resulting values of $\alpha_\star$ and $\beta/\Hstar$ are shown
in fig.~\ref{fig:pta:alpha:beta},
together with the
$1\sigma$ (solid) and
$2\sigma$ (dashed) density contours of
the scan for
the soft \ref{eq:bro:soft} (left, magenta) and
the softer \ref{eq:bro:softer} (right, dark blue).
While the softer EFT is applicable only over a relatively narrow region
of parameter space, the resulting distributions in the
$(\alpha_\star,\beta/\Hstar)$ plane are robust and agree well with the soft EFT,
cf.~eq.~\eqref{eq:bro:soft}, for small values of the gauge coupling
$\gd$. For larger couplings, however, the soft EFT naturally predicts
denser regions extending towards smaller values of $\beta/\Hstar$ and larger
values of $\alpha_\star$, thereby moving closer to the PTA-favored region.
The soft EFT therefore provides a less restrictive framework for
assessing the compatibility of the model with the PTA observations.

In both scans of fig.~\ref{fig:pta:alpha:beta},
we impose the condition
set by experimentally allowed portal couplings~\eqref{eq:lambda_s_constraint}.
Gray points correspond to $\epsS > 0.1$.
This is a liberal choice, as smaller values of $\epsS$
significantly shrink the available parameter space and
shift the density contours in fig.~\ref{fig:pta:alpha:beta}
towards larger $\beta/\Hstar$,
until all points are excluded once $\epsS < \lambda_s$.
These conclusions extend earlier assessments of the dark Abelian model
in the PTA context, e.g.~\cite{Bringmann:2023opz,Bringmann:2026xcx}.
Black circled points are out of thermal equilibrium (cf.\ eq.~\eqref{eq:thermal:equilibrium:condition}),
so that $T_\rmii{DS} \neq T_\rmii{SM}$.
Treating them consistently requires tracking
the dark-sector temperature $T_\rmii{DS}$ after decoupling,
whose ratio $\xi_\rmii{DS} = T_\rmii{DS}/T_\rmii{SM}$
explicitly enters the hydrodynamic computations of sec.~\ref{sec:PT_and_GW}.
This procedure is beyond the scope of this work,
but has been discussed in~\cite{Breitbach:2018ddu,Bringmann:2023opz}.

The results of the scan shown in
fig.~\ref{fig:pta:alpha:beta} are broadly consistent with the findings
of~\cite{Bringmann:2023opz,Bringmann:2026xcx,Matuszak:2026xsz},
which likewise concluded that the Abelian Higgs model does not reproduce
the PTA-preferred region. In agreement with these studies,
we find that
compatibility with the PTA signal is achieved only
after significant tuning at confidence levels above $2\sigma$ level.

For comparison, we also include in fig.~\ref{fig:pta:alpha:beta} (right, orange)
the corresponding density contours obtained using a Daisy-resummed treatment
which is the softer~\ref{eq:bro:softer} at full LO accuracy.
In comparison, for our NLO computations the populated region extends
towards lower values of $\beta/\Hstar$ and larger values of $\alpha_\star$ by
approximately one order of magnitude.
This shift is a genuine consequence of the EFT construction rather than a mere
{\em thermal detail}.

At NLO,
the soft and softer descriptions
yield compatible results in the regime of
mutual validity for small gauge couplings $\gd$.
For larger couplings, however, the two approaches
lead to quantitatively different phase-transition parameters and
$(\alpha_\star,\beta/\Hstar)$ populated parameter spaces.
Nevertheless, the overall phenomenological
conclusion remains unchanged.
While the analysis of~\cite{Bringmann:2026xcx} is based on a LO thermal
description, our NLO thermodynamics of sec.~\ref{sec:PT_and_GW}
demonstrates that the predicted
phase-transition region remains well separated from the PTA-favored region.
We therefore conclude that the incompatibility of
the Abelian Higgs model with the current PTA data is robust against higher-order
thermal corrections as well as the choice of thermal resummation scheme.

Having established the predictions of our EFT framework, it is instructive to compare them
with previous analyses of the same model also at the level of the predicted spectra.
Since these studies employ different treatments
of the finite-temperature effective potential and thermal resummation, such a comparison
helps isolate the origin of the differing phenomenological conclusions.
Our analysis arrives at different conclusions than~\cite{Han:2023olf,Banik:2024zwj},
where the PTA-favored region was claimed to be compatible with the Abelian Higgs model
as can also be seen in fig.~\ref{fig:pta:scan:spectra}~(right),
which collects the benchmark points
{\tt BM8}--{\tt BM11} of~\cite{Han:2023olf},
{\tt BM12} of~\cite{Banik:2024zwj}, and
{\tt BM13}--{\tt BM16} of~\cite{Costa:2025csj}.

%
\section{Conclusions}
\label{sec:conclusions}

The possibility that the stochastic GW background reported by PTA
collaborations originates from a cosmological first-order phase transition at MeV
temperatures has attracted considerable interest in recent years.
We revisited this scenario
for a model class with a radiatively induced cubic barrier.
As a representative fundamental model,
we focused on a dark Abelian Higgs model, performing a precision
study of the phase-transition thermodynamics using
dimensionally reduced EFTs 
while also addressing
the thermal and hydrodynamic contact between the dark and Standard Model sectors and
the phenomenology of a fermionic dark matter candidate.
Beyond identifying the parameter regions that yield an observable GW signal,
we assessed the theoretical robustness of these predictions by systematically
tracking the impact of thermal resummation,
higher-order matching,
renormalization-scale dependence, and
higher-dimensional operators, using the Abelian
Higgs model as a benchmark for which both perturbative and lattice results are
available.
Currently this is the highest level of precision in thermodynamics
for PTA-era phase transitions.

We compared the resulting GW spectra with the current NANOGrav data using
large parameter-space scans in tab.~\ref{tab:pta:scan} and
individual benchmark points in tab.~\ref{tab:pta:benchmarks}.
Because the region favored by current PTA data lies close to the boundary of
validity of the finite-temperature description of the model,
we first delineated where the predictions remain under theoretical control.
To this end, we have identified three limiting sources of theoretical uncertainty:
\begin{itemize}
  \item
    The high-temperature expansion breaks down, as expected for the strong
    transitions favored by the PTA data.
    The hierarchy between the soft and Debye scales,
    quantified by the ratio $h_3^{ }\phi_\rmi{min}^{2}/\mD^2$,
    exceeds $\mathcal{O}(1)$, so that integrating out the temporal Debye
    scale is no longer parametrically justified.
  \item
    In the same regime, the dark gauge coupling $\gd$ grows large enough to
    challenge a strict perturbative expansion.
  \item
    Where the effective theory remains under control, higher-order corrections
    only shift the predicted parameter space at large gauge coupling, while the
    predictions remain robust at small coupling.
    For the strongest transitions, higher-dimensional operators are expected to
    dominate over loop corrections~\cite{Bernardo:2025vkz,Bernardo:2026nyq}.
\end{itemize}
The strongest phase transitions accessible within the model push against this boundary,
precisely where compatibility with the PTA signal would be maximal.
Within the range where the effective theory remains theoretically controlled,
the predicted phase-transition parameters remain well separated from the
PTA-preferred region, despite
a substantial shift
of the predicted parameter space
induced by higher-order thermal corrections
at large gauge couplings.
Reproducing the PTA-favored signal therefore requires significant tuning.
Quantitative differences with previous studies arise from
the state-of-the-art treatment of thermal resummation
in sec.~\ref{sec:PT_and_GW}, 
higher-order matching corrections in appendix~\ref{sec:dimensional_reduction_details}, and
a systematic assessment of the validity of the effective field theory.

Phenomenological constraints~\eqref{mixing_mass_limits} further shrink the viable parameter space.
Assuming the dark scalar decays exclusively into Standard Model particles through Higgs mixing,
the allowed window for the Higgs-dark scalar mixing angle, combined with the requirement that
it decay before the onset of BBN, strongly disfavors dark scalar masses below $\ms \approx 3$~MeV.
This further restricts the region that can be reconciled with
the PTA observations.

We have also carefully examined the thermal history of the dark sector and the
conditions required to maintain thermal contact with the Standard Model plasma.
Owing to the small portal couplings implied by cosmological and experimental
constraints,
thermal equilibrium between the two sectors is not guaranteed {\em a priori}.
By computing the relevant interaction rates and comparing them with the Hubble
expansion rate, we showed that thermalization is predominantly maintained through the
scalar portal, while the vector portal is typically ineffective in the parameter
region of interest.
We also identified a class of scattering processes mediated by dark-scalar
self-interactions that, to the best of our knowledge, had not been included in
previous analyses.
Overall, we find that the dark and visible sectors remain in thermal equilibrium
throughout the epoch of GW production across most of the parameter space.
This justifies the use of a common temperature and the inclusion of both sectors
in the relativistic degrees of freedom entering the GW redshift.
Also decoupled regions are identified and must be treated accordingly;
see~\cite{Breitbach:2018ddu,Bringmann:2023opz}.

The interaction rates computed in this work also determine whether
the dark and visible sectors remain hydrodynamically coupled during the phase transition.
Comparing
the mean free path associated with momentum transfer between the two sectors
with the mean bubble separation
allows us to assess whether this exchange is efficient and
determine the relativistic degrees of freedom that enter
the hydrodynamic description of the transition.

We further investigated the dark matter phenomenology of the model.
For thermal freeze-out in the symmetric scenario, the gauge couplings required to
obtain a strong first-order phase transition are generally incompatible with
reproducing the observed relic abundance, while late-time annihilation constraints
further restrict the viable parameter space.
In contrast, asymmetric dark matter scenarios naturally accommodate the observed relic
density over sizable regions of parameter space while remaining compatible with the
phase-transition dynamics, illustrating the complementarity between GW
observations and dark matter phenomenology.
Moreover, we showed that the experimentally allowed range of
the Higgs-dark scalar mixing angle significantly constrains
the parameter space explored in our scan (cf.\ fig.~\ref{fig:pta:alpha:beta}),
thereby affecting the compatibility of the model with the PTA-favored region.

Interpreting PTA observations with a particle-physics phase transition requires
both viable phenomenology and quantitatively controlled finite-temperature calculations.
Our analysis provides a precision blueprint for such studies.
The framework in sec.~\ref{sec:PT_and_GW}
applies to a broad class of gauge-Higgs models,
such as non-Abelian gauge theories~\eqref{eq:Veff:bro:LO:dimless}.
There,
gauge self-interactions may enhance the
thermal barrier and yield stronger transitions in a more controlled
perturbative regime.
The framework can also be extended to the full NLO thermodynamics of
classically scale-invariant models,
where significant supercooling can arise.
Natural next steps include
incorporating higher-dimensional operators consistently in
the nucleation dynamics and
adopting frameworks that remain valid beyond the high-temperature regime.

%
\section*{Acknowledgements}
The authors thank
Carlo Tasillo and
Tuomas V.I.\ Tenkanen
for valuable discussions.
We are grateful to
Jorinde van de Vis
for advice on the hydrodynamic equilibration condition and
for many helpful comments during the preparation of this manuscript.
We also thank
Mikko Laine
for comments on the manuscript. 
PS is supported by the Swiss National Science Foundation (SNSF) under
grant \href{https://data.snf.ch/grants/grant/215997}{\tt PZ00P2-215997}.\\

\noindent
{\bf Data availability statement.}
The effective potential expressions used to produce
figs.~\ref{fig:yc:yp}, and
figs.~\ref{fig:pta:scan:soft}--\ref{fig:pta:alpha:beta}
are publicly available via the software {\tt DRalgo}~\cite{Ekstedt:2024etx}.
Feynman diagrams were generated with {\tt Axodraw}~\cite{Collins:2016aya}.

%
\appendix
\renewcommand{\thesection}{\Alph{section}}
\renewcommand{\thesubsection}{\Alph{section}.\arabic{subsection}}
\renewcommand{\theequation}{\Alph{section}.\arabic{equation}}

%
\section{Master integrals}
\label{sec:master_integrals}

This appendix collects the master integrals used
in the effective potential and
the dimensional-reduction matching of
sec.~\ref{sec:dimensional_reduction_details}.
The $d$-dimensional integral measure is
\begin{equation}
\label{eq:measure}
  \int_{\vec{p}} \equiv \int \frac{{\rm d}^d \vec{p}}{(2\pi)^d} =
  \frac{2}{(4\pi)^2\Gamma(\frac{d}{2})} \int_{0}^{\infty} {\rm d}p\,p^{d-1}
  \;.
\end{equation}

\subsection{Zero-temperature master integrals}
\label{sec:master_integrals_zeroT}

At one-loop level and in
dimensional regularization,
the relevant class of master integrals in vacuum is
\begin{equation}
\label{eq:Id:vacuum}
  I_{s}^{d} (m_i) =
  I_{s;i}^{d} \equiv
  \Bigl(\frac{\LamD^{2}e^\gammaE}{4\pi}\Bigr)^{\epsilon}
  \int_{\vec{p}}
  \frac{1}{(P^2+m_{i}^{2})^s} =
  \Bigl(\frac{\LamD^{2}e^\gammaE}{4\pi}\Bigr)^{\epsilon}
  \frac{[m_{i}^{2}]^{\frac{d}{2}-s}}{(4\pi)^\frac{d}{2}}
  \frac{\Gamma\bigl(s-\frac{d}{2}\bigr)}{\Gamma(s)}
  \;.
\end{equation}  

The Euclidean Passarino-Veltman integrals~\cite{tHooft:1978jhc} are
defined as:
\begin{align}
\label{eq:A:B}
  A(m) &= \int_P \frac{1}{P^2+m^2}
  \,,
  \nn
  B_{\{1;\mu;\mu\nu\}}(K;m_1,m_2) &=
    \int_P
    \frac{\bigl\{1; P_\mu; P_\mu P_\nu\bigr\}}{
      [P^2+m_1^2]
      [(K-P)^2+m_2^2]}
    \, ,
\end{align}
where
$\int_P = \int \frac{{\rm d}^D P}{(2 \pi)^D}$
denotes the $D$-dimensional integral measure,
and we define
$B(K;m,m) \equiv B(K;m)$,
and similarly for
the rank-1 and 2 tensor integrals $B_\mu$ and $B_{\mu\nu}$.
For vector self-energies,
the following structures appear,
\begin{align}
\label{eq:Cmu}
  B_\mu(K;m_1,m_2) &=
  \frac{1}{2}\frac{K_\mu}{K^2}
  \Bigl\{
      A(m_1)
    - A(m_2)
    - \bigl[K^2 + m_1^2 - m_2^2\bigr] B(K;m_1,m_2)
  \Bigr\}
\,,
\\[2mm]
\label{eq:Dmunu}
  B_{\mu\nu}(K;m_1,m_2) &=
    \frac{1}{4(D-1)K^2}
    \Bigl(\delta_{\mu\nu} - \frac{D K_{\mu} K_{\nu}}{K^2} \Bigr)
  \nn &
  \hphantom{\frac{1}{4(D-1)K^2}}
  \times
    \Bigl\{
        \bigl[K^2 - m_1^2 + m_2^2\bigr] A(m_1^2)
      + \bigl[K^2 - m_2^2 + m_1^2\bigr] A(m_2^2)
  \nn &
  \hphantom{\frac{1}{4(D-1)K^2}\times\Bigl\{}
    - \bigl[K^2 + (m_1 + m_2)^2\bigr]
      \bigl[K^2 + (m_1 - m_2)^2\bigr] B(K;m_1^2,m_2^2)
    \Bigr\}
  \nn[1mm] &
  + \frac{K_{\mu} K_{\nu}}{K^2}\Bigl\{
      A(m_1^2)
    - m_2^2 B(K;m_1^2,m_2^2)
  \Bigr\}
    \,.
\end{align}
The rank-1 tensor integral
is proportional to the external momentum
$B_\mu(K;m_1,m_2) = K_\mu C(K;m_1,m_2)$, and
for the rank-2 tensor integral,
we only need
its transverse part
$B_{\mu\nu}^{(\T)}$ which is defined through
$B_{\mu\nu} = B_{\mu\nu}^{(\T)} \bigl(
    \delta_{\mu\nu}
  - \frac{K_{\mu} K_{\nu}}{K^2} \bigr)
  + \mathcal{O}(K_\mu K_\nu)$.

In $D = 4-2\epsilon$ dimensions,
the master integrals evaluate to
\begin{eqnarray}
\label{eq:A:master}
  A(m) &=&
    I_{1}^{D} (m)
  \nn[1mm] &\stackrel{D = 4-2\epsilon}{=}&
      - \frac{m^2}{(4\pi)^2}\Bigl(
        \frac{\LamD^{-2\epsilon}}{\epsilon}
        + \Bigl( \ln\frac{\LamD^2}{m^2} + 1\Bigr)
        + \mathcal{O}(\epsilon)
    \Bigr)
  \,,\\[2mm]
\label{eq:B:master}
  B(K,m_1,m_2) &\stackrel{D = 4-2\epsilon}{=}&
    \frac{1}{(4\pi)^2}\biggl(
      \frac{\LamD^{-2\epsilon}}{\epsilon}
      \\ & &
      \hphantom{1/(4\pi)}
       + \biggl[
          \ln\frac{\LamD^2}{m_1 m_2}
        + 1
        - \frac{m_1^2 + m_2^2}{m_1^2 - m_2^2}\ln\frac{m_1}{m_2}
        + F\Bigl(\frac{m_1}{K},\frac{m_2}{K}\Bigr)
      \biggr]
      + \mathcal{O}(\epsilon)
    \biggr)
    \,.
    \hspace{2mm}
    \nonumber
\end{eqnarray}
Here, we defined
\begin{align}
  \re F(r_1,r_2) &=
    1+
    \Bigl(
      \frac{r_1^2 + r_2^2}{r_1^2 - r_2^2}
      +r_1^2 - r_2^2
    \Bigr)\ln\frac{r_1}{r_2}
    \nn &
    - 2\re\sqrt{(r_1 - r_2)^2 + 1}\sqrt{(r_1 + r_2)^2 + 1}
    \;\;
    \mbox{arctanh}
    \frac{\sqrt{(r_1 - r_2)^2 + 1}}{\sqrt{(r_1 + r_2)^2 + 1}}
    \,,
\end{align}
where
$F(r_1,r_1) = F(r_1)$.
Anticipating the analytic continuation to Minkowskian space-time,
via $K \to -i\mathcal{K}$,
we also identify $F(r_1,r_2) = F(i r_1, i r_2)$.
The function $\mathcal{F}(r_1,r_2)$ has
the following special cases
\begin{align}
  \label{eq:F:master}
  \mathcal{F}(1) &=
    2 - \frac{\pi}{\sqrt{3}}
    \, , &
  \mathcal{F}(r) &=
    2
    - 2\sqrt{4r^2-1}
    \arctan\frac{1}{\sqrt{4r^2-1}}
    \,.
\end{align}

\subsection{Thermal master integrals}
\label{sec:master_integrals_finiteT}

The one-loop master integrals for the effective potential
are given by
\begin{align}
\label{eq:1loop-master-d}
J_{d}(x) &\equiv
  \frac{1}{2}
  \Big( \frac{\LamD^2 e^\gammaE}{4\pi} \Big)^\epsilon
  \int_\vec{p} \ln(p^2 + x) =
  -\frac{1}{2}
  \Big( \frac{\LamD^2 e^\gammaE}{4\pi} \Big)^\epsilon
  \frac{x^\frac{d}{2}}{(4\pi)^{\frac{d}{2}}} \frac{\Gamma(-\frac{d}{2})}{\Gamma(1)}
  \;,\\
\label{eq:1loop-master-4d}
J_{4}(x) &=
  - \frac{x^2}{4(4\pi)^2}\Bigl(
      \frac{1}{\epsilon}
    + \ln\frac{\LamD^2}{x}
    +\frac{3}{2}
  + \mathcal{O}(\epsilon)
  \Bigr)
  \;,\\
\label{eq:1loop-master-3d}
J_{3}(x) &=
    - \frac{x^{\frac{3}{2}}}{12 \pi}
    + \mathcal{O}(\epsilon)
  \;.
\end{align}

At one-loop level,
the bosonic and fermionic master integrals decompose into
a vacuum part and a thermal part.
The hard thermal contributions $Z^{\T}$ and $\widetilde{Z}^{\T}$,
for bosons and fermions respectively,
can be evaluated numerically,
\begin{align}
\label{eq:Z:master:boson}
Z_{s;i}^{\alpha} &=
  \Tint{P}' \frac{p_n^\alpha}{[P^2 + m_{i}^{2}]^{s}}
  \;,
  &
Z_{s;i}^{0} =
Z_{s;i}^{ } &=
  \Tint{P}' \frac{1}{[P^2 + m_{i}^{2}]^{s}} =
    I_{s;i}^{4}
  + Z_{s;i}^{\T}
  \;,
  \\
\label{eq:Z:master:fermion}
\widetilde{Z}_{s;i}^{\alpha} &=
  \Tint{\{P\}} \frac{p_n^\alpha}{[P^2 + m_{i}^{2}]^{s}}
  \;,
  &
\widetilde{Z}_{s;i}^{0} =
\widetilde{Z}_{s;i}^{ } &=
  \Tint{\{P\}} \frac{1}{[P^2 + m_{i}^{2}]^{s}} =
    I_{s;i}^{4}
  + \widetilde{Z}_{s;i}^{\T}
  \;,
\end{align}
where $P\equiv(p_n^{ },\vec{p})$ is the Euclidean four-momentum
with Matsubara frequencies
$p_n = 2n\pi T$ for bosons and
$p_n = (2n+1)\pi T$ for fermions ($n\in\mathbb{Z}$).
Curly braces on the sum-integral denote fermionic statistics,
while a prime indicates that the bosonic zero mode is excluded.
Similarly,
the bosonic and fermionic logarithmic master integrals can be written
as a Coleman-Weinberg vacuum part plus a thermal part,
\begin{align}
  J_{i} &=
  \frac{1}{2}\,\Tint{P}'\!\!\ln\bigl(P^2 + m_{i}^{2}\bigr)
  =
  J_{i}^{4} + J_{i}^{\T}
  \,,&
  \widetilde{J}_{i} &=
  \frac{1}{2}\,\Tint{\{P\}}\!\!\ln\bigl(P^2 + m_{i}^{2}\bigr)
  =
  J_{i}^{4} + \widetilde{J}_{i}^{\T}
  \,,
\end{align}
with the corresponding 4d vacuum integral
$J_{i}^{d} = J_d(m_i^2)$ from eq.~\eqref{eq:1loop-master-d}.

The high-temperature expansion ($m_i \ll T$) of
the master integrals in $d=3-2\epsilon$
relevant for the hard scalar and fermionic pressure
in eq.~\eqref{eq:p0:hard}
is%
\footnote{
  Vector bosons carry a dimension-dependent multiplicity,
  $J_{i} \to (D-1) J_{i}$,
  which modifies, e.g., the $\mathcal{O}(\epsilon^{0})$ term
  in eq.~\eqref{eq:1loop-master-4d}.
}
\begin{eqnarray}
\label{eq:J:highT:boson}
  J_{i}^{ } &=&
  J_{i}^{4}
+ J_{i}^{\T}
  \nn &=&
  - \frac{m_{i}^{4}}{4(4\pi)^2}\Bigl(
      \frac{1}{\epsilon}
    + \ln\frac{\LamD^2}{m_{i}^{2}}
    + \frac{3}{2}
    \Bigr)
    + J_{i}^{\T}
    + \mathcal{O}(\epsilon^{ })
  \nn &\stackrel{m_i \ll T}{=}&
  T^4\biggl[
    - \frac{\zeta_{4}}{\pi^2}
    + \frac{1}{24}\frac{m_{i}^2}{T^2}
    - \frac{1}{4(4\pi)^2}\frac{m_{i}^4}{T^4}\Bigl(
        \frac{1}{\epsilon}
      + \Lb
    \Bigr)
    + \mathcal{O}\Bigl(\frac{m_{i}^6}{T^6},\epsilon^{ }\Bigr)
  \biggr]
  \;,
  \hspace{1cm}
  \\[2mm]
\label{eq:J:highT:fermion}
  \widetilde{J}_{i}^{ } &=&
  J_{i}^{4}
+ \widetilde{J}_{i}^{\T}
  \nn &=&
  - \frac{m_{i}^{4}}{4(4\pi)^2}\Bigl(
      \frac{1}{\epsilon}
    + \ln\frac{\LamD^2}{m_{i}^{2}}
    + \frac{3}{2}
    \Bigr)
    + \widetilde{J}_{i}^{\T}
    + \mathcal{O}(\epsilon^{ })
  \nn &\stackrel{m_i \ll T}{=}&
  T^4\biggl[
      \frac{7}{8} \frac{\zeta_{4}}{\pi^2}
    - \frac{1}{48}\frac{m_{i}^2}{T^2}
    - \frac{1}{4(4\pi)^2}\frac{m_{i}^4}{T^4}\Bigl(
        \frac{1}{\epsilon}
      + \Lf
    \Bigr)
    + \mathcal{O}\Bigl(\frac{m_{i}^6}{T^6},\epsilon^{ }\Bigr)
  \biggr]
  \;,
\end{eqnarray}
with $\zeta_4 = \frac{\pi^4}{90}$.
Here,
$\zeta_{s}=\zeta(s)$ for $\re\,(s) > 1$ is the Riemann zeta function, and
$(\ln \zeta_s)'=\zeta'(s)/\zeta(s)$.
To aid compactness,
we defined a shorthand notation
for the thermal logarithms
\begin{align}
\Lb &\equiv
    2 \ln\frac{\mu}{T}
  - 2 \big( \ln(4\pi) - \gammaE \big)
  \;,&
\Lf &\equiv \Lb + 4\ln2
    \;,
\end{align}
where
$\gammaE$ is the Euler-Mascheroni constant.
Using
the vacuum integrals of eq.~\eqref{eq:Id:vacuum}
in $d=4-2\epsilon$ dimensions,
the corresponding contributions to the one-loop matching
relations are
\begin{eqnarray}
  \widetilde{Z}_{1;i}^{ } &=&
  I_{1;i}^{4}
+ \widetilde{Z}_{1;i}^{\T}
  \nn &=&
  - \frac{m_{i}^{2}}{(4\pi)^2}\Bigl(
      \frac{1}{\epsilon}
    + \ln\frac{\LamD^2}{m_{i}^{2}}
    + 1
    \Bigr)
    + \widetilde{Z}_{1;i}^{\T}
    + \mathcal{O}(\epsilon^{ })
  \nn &\stackrel{m_i \ll T}{=}&
  T^2\biggl[
  - \frac{1}{24}
  - \frac{1}{(4\pi)^2}\frac{m_{i}^2}{T^2}\Bigl(\frac{1}{\epsilon} + L_f^{ }\Bigr)
  + \mathcal{O}\Bigl(\frac{m_{i}^4}{T^4},\epsilon^{ }\Bigr)
  \biggr]
  \;,\\[2mm]
  \widetilde{Z}_{2;i}^{ } &=&
  I_{2;i}^{4}
+ \widetilde{Z}_{2;i}^{\T}
  \nn &=&
    \frac{1}{(4\pi)^2}\Bigl(
      \frac{1}{\epsilon}
    + \ln\frac{\LamD^2}{m_{i}^{2}}
    \Bigr)
    + \widetilde{Z}_{2;i}^{\T}
    + \mathcal{O}(\epsilon^{ })
  \nn &\stackrel{m_i \ll T}{=}&
    \frac{1}{(4\pi)^2}\Bigl(\frac{1}{\epsilon} + L_f^{ }\Bigr)
  - 2\frac{14\zeta_{3}}{(4\pi)^4}\frac{m_{i}^2}{T^2}
  + \mathcal{O}\Bigl(\frac{m_{i}^4}{T^4},\epsilon^{ }\Bigr)
  \;,\\[2mm]
  \widetilde{Z}_{3;i}^{ } &=&
  I_{3;i}^{4}
  + \widetilde{Z}_{3;i}^{\T}
  \nn &=&
  \frac{1}{(4\pi)^2} \frac{1}{2m_{i}^{2}}
  + \widetilde{Z}_{3;i}^{\T}
  + \mathcal{O}(\epsilon^{ })
  \nn &\stackrel{m_i \ll T}{=}&
  \frac{1}{T^2}\biggl[
      \frac{14\zeta_{3}}{(4\pi)^4}
    - 3\frac{124\zeta_{5}}{(4\pi)^6} \frac{m_{i}^2}{T^2}
    + \mathcal{O}\Bigl(\frac{m_{i}^4}{T^4},\epsilon^{ }\Bigr)
  \biggr]
  \;,\\[2mm]
  \widetilde{Z}_{4;i}^{ } &=&
  I_{4;i}^{4}
  + \widetilde{Z}_{4;i}^{\T}
  \nn &=&
  \frac{1}{(4\pi)^2} \frac{1}{6 m_{i}^{4}}
  + \widetilde{Z}_{4;i}^{\T}
  + \mathcal{O}(\epsilon^{ })
  \nn &\stackrel{m_i \ll T}{=}&
  \frac{1}{T^4}\biggl[
    \frac{124\zeta_{5}}{(4\pi)^6}
  - 4\frac{1270\zeta_{7}}{(4\pi)^8} \frac{m_{i}^2}{T^2}
  + \mathcal{O}\Bigl(\frac{m_{i}^4}{T^4},\epsilon^{ }\Bigr)
  \biggr]
  \;,
\end{eqnarray}
where
the explicit fermionic thermal integrals up to
$\mathcal{O}(\epsilon^0)$ are
\begin{align}
\label{eq:J:T}
\widetilde{J}_{i}^{\T} &=
  - T\!\int_{\vec{p}} \ln\bigl(1 - \nF(\omega_{p,i})\bigr)
  \;, \\[1mm]
\label{eq:Z1:T}
\widetilde{Z}_{1;i}^{\T} &=
  - \int_{\vec{p}} \frac{\nF(\omega_{p,i})}{\omega_{p,i}}
  \;, \\[1mm]
\widetilde{Z}_{2;i}^{\T} &=
  - \frac{1}{2}\int_{\vec{p}} \frac{\nF(\omega_{p,i})}{p^2 \omega_{p,i}}
  \;, \\[1mm]
\widetilde{Z}_{3;i}^{\T} &=
  - \frac{1}{8}\int_{\vec{p}} \frac{\nF(\omega_{p,i})}{p^2 \omega_{p,i}^2}
    \Bigl[
        \frac{1}{\omega_{p,i}}
      + \frac{1 - \nF(\omega_{p,i})}{T}
    \Bigr]
  \;, \\[1mm]
\widetilde{Z}_{4;i}^{\T} &=
- \frac{1}{16}\int_{\vec{p}} \frac{\nF(\omega_{p,i})}{p^{2} \omega_{p,i}^{3}}
    \Bigl[
      \frac{1}{\omega_{p,i}^{2}}
      + \frac{1 - \nF(\omega_{p,i})}{\omega_{p,i} T}
      + \frac{(1 - \nF(\omega_{p,i}))(1 - 2\nF(\omega_{p,i}))}{3 T^{2}}
    \Bigr]
  \;,
\end{align}
with
$\omega_{p,i} \equiv \sqrt{p^2 + m_{i}^2}$,
$m_{i}$ a general mass, and
the Fermi distribution function
$\nF(\omega) \equiv 1/(\exp(\omega/T) +1)$.

%
\section{Zero-temperature vacuum structure}
\label{sec:vacuum}

\subsection{Renormalization and one-loop $\beta$-functions}
\label{sec:ren}

The renormalization group (RG) equations ($\beta$-functions) for the parameters of the
model in eq.~\eqref{eq:lag:4d} encode their running with respect to
the \MSbar{} renormalization scale $\LamD$.
Defining
\begin{equation}
\label{eq:rge:param}
t \equiv \ln\LamD^2
\;,
\end{equation}
at one-loop level,
we obtain
\begin{align}
\label{eq:beta:gd}
\partial_{t}^{ }\,
\gd^2 &=
  \frac{1}{(4\pi)^2} \frac{\gd^{4}}{3} \Big( \YS^{2} + 4\nG^{ }\YX^{2} \Big)
  \;,
  \\
\partial_{t}^{ }\,
\mu_{s}^{2} &=
  \frac{1}{(4\pi)^2} \mu_{s}^{2}\Big(
      4 \lambda_{s}^{ }
    - 3 \gd^{2}\YS^{2}
  \Big)
  \;,
  \\
\partial_{t}^{ }\,
\muX^{2} &=
- \frac{3}{(4\pi)^2} \muX^{2} \gd^{2}\YX^{2}
  \;, \\
\label{eq:beta:lambda}\,
\partial_{t}^{ }
\lambda_{s}^{ } &=
  \frac{1}{(4\pi)^2} \Big(
      10 \lambda_{s}^{2}
    - 6 \gd^{2}\YS^{2} \lambda_{s}^{ }
    + 3 \gd^{4}\YS^{4}
  \Big)
  \;,
\end{align}
given in terms of
the hypercharge of the scalar and fermion fields, $\YS = \YX = 1$.
For numerical evaluations,
the input scale is chosen as the dark photon mass,
$\LamD_0 = \mV$.
The model parameters are evolved using
eqs.~\eqref{eq:beta:gd}--\eqref{eq:beta:lambda},
with
the number of fermion generations $\nG$
changing at the fermion threshold $\LamD \simeq \mX$.

In the \threeDEFT{},
we employ a different RG scale, denoted by $\Lamd$,
\begin{align}
  t_3 \equiv \ln \Lamd
  \,.
\end{align}
In the soft \threeDEFT{}
both the scalar and Debye masses obtain
a non-zero $\beta$-function~\cite{Farakos:1994kx,Laine:1995np,Hirvonen:2021zej},
\begin{align}
\partial_{t_3}^{ }
\bar{\mu}_{s,3}^2 &=
  \frac{4}{(4\pi)^2} \Bigl(
        g_{d,3}^4
      + \frac{1}{2} h_{3}^2
      - 2 g_{d,3}^2 \lambda_{s,3}^{ }
      + 2 \lambda_{s,3}^{2}
    \Bigr)
    \,,\\
\partial_{t_3}^{ }
\mD^2 &=
  -\frac{4}{(4\pi)^2} \Bigl(
        g_{d,3}^{2} h_{3}^{ }
      - h_{3}^2
      - 24\kappa_{3}^{2}
    \Bigr)
    \,,
\end{align}
where the soft-EFT matching parameters
$g_{d,3}$, $\lambda_{s,3}$ are defined in
eqs.~\eqref{eq:gd:soft}--\eqref{eq:kappa:soft}.
In the softer \threeDEFT{}
only the scalar mass has a non-zero $\beta$-function,
\begin{align}
\partial_{\bar{t}_3}^{ }
\bar{\mu}_{s,3}^2 &=
  \frac{4}{(4\pi)^2} \Bigl(
        \bar{g}_{d,3}^4
      - 2\bar{g}_{d,3}^2 \bar\lambda_{s,3}^{ }
      + 2\bar\lambda_{s,3}^{2}
    \Bigr)
    \,,
\end{align}
where
$\bar{t}_{3} \equiv \ln \Lamd$
and the barred softer-EFT matching parameters
$\bar{g}_{d,3}$, $\bar\lambda_{s,3}$ are defined in
eqs.~\eqref{eq:gd:softer}--\eqref{eq:lambdas:softer}.

\subsection{Relations between \MSbar{}-parameters and physical observables}
\label{sec:MSbar:physical}

The physical input parameters are related
to the \MSbar{} Lagrangian parameters of the model in eq.~\eqref{eq:lag:4d},
\begin{equation}
  \bigl\{ \ms, \alpha_d, \mV, \mX \bigr\} \mapsto
  \bigl\{ \mu_s, \gd, \lambda_s, \muX \bigr\}
  \,,
\end{equation}
at the input scale $\LamD = \LamD_0$ taken as the physical dark photon mass,
\begin{align}
  \label{eq:input:scale}
  \LamD_0 = \mV
  \,.
\end{align}
After inverting the scalar mass eigenvalues~\eqref{masses_minimum},
the \MSbar{}
mass parameter $\mu_s$,
scalar self-coupling, and
fermion mass $\muX$ are related to
the physical masses
$\ms$,
$\mV$,
$\mX$ in the broken phase
using
one-loop vacuum renormalization~\cite{Kajantie:1997hn}
with the relations
\begin{align}
\label{sec:MSbar:relations}
  \mu_s^2(\LamD) &= \frac{\ms^2}{2} \biggl[
      1
    + \frac{\Pi_s(-\ms^2;\LamD)}{\ms^2}
  \biggr]
  \,,\;\;
  \lambda_s(\LamD) = \frac{\gd^2}{2} \frac{\ms^2}{\mV^2}
  \biggl[
    1
    - \frac{\Pi_\rmii{$V$}(-\mV^2;\LamD)}{\mV^2}
    + \frac{\Pi_s(-\ms^2;\LamD)}{\ms^2}
    \biggr]
  \,,
\end{align}
and
$\muX(\LamD) = \mX$
at the input scale
$\LamD = \LamD_0$.
Here,
$\Pi_s$ is the scalar and
$\Pi_\rmii{$V$}$ the vector
one-loop self-energy evaluated at their respective pole mass,
with the explicit expressions given
in eqs.~\eqref{eq:Pi:s} and~\eqref{eq:Pi:A} below
and in~\cite{Kajantie:1997hn,Kim:2016kxt}.
Thus,
both higher-order corrections in the renormalization conditions and
the momentum dependence of the pole masses are consistently
included~\cite{Kajantie:1995dw,Laine:2017hdk,Niemi:2021qvp}.
The expressions in eq.~\eqref{sec:MSbar:relations} hold at
one-loop level and, in general,
receive radiative corrections that can induce a further scale dependence.%
\footnote{%
  In the Coleman-Weinberg limit $\lambda_s \ll \gd^2$,
  the scalar mass is loop-generated and satisfies $\ms \ll \mV$.
  The one-loop relations of eq.~\eqref{sec:MSbar:relations}
  then receive large higher-order corrections,
  and a consistent treatment requires
  resumming the scalar self-energy~\cite{Coleman:1973jx,Laine:1999rv}.
}
Although $\mX$ is an input parameter, the dark fermion is decoupled in
the scans of sec.~\ref{sec:pta} and therefore does not enter
the self-energies below.

The masses $\ms$, $\mV$ are analogs of the Higgs and $Z$-boson
masses of the SM and, together with the fermion mass $\mX$,
could in principle be determined experimentally.
The dark gauge coupling $\alpha_d = \gd^2/(4\pi)$, in contrast, must be fixed by relating $\gd^2(\LamD)$
to a physical observable, such as the analogue of the Fermi constant.
The latter could be extracted from a four-fermion scattering amplitude.
Since the dark model parameters are unknown in any case,
we instead take the \MSbar{} parameter $\gd^2(\LamD)$ and the pole masses $\{\ms,\mV,\mX\}$
as direct inputs at $\mu_0$~\eqref{eq:input:scale}. 

In our parameter scans of sec.~\ref{sec:pta},
we trade the dark photon mass $\mV$ for the scalar quartic $\lambda_s$
via the tree-level relation
$\lambda_s = \frac{\gd^2}{2} \frac{\ms^2}{\mV^2}$ at $\LamD_0$,
and scan over $\{\ms,\lambda_s,\gd\}$.
The transition is insensitive to the dark fermion, since
it remains Boltzmann suppressed for $\mX \gg \mV$ in the dark sector.
All parameters are then evolved to the thermal matching scale
$\LamD_\rmi{ref} = A\pi T$ of eq.~\eqref{eq:LamD:ref} using the one-loop
$\beta$-functions~\eqref{eq:beta:gd}--\eqref{eq:beta:lambda},
where $A$ is varied around unity to quantify the residual scale dependence.

\begin{figure}[t]
\centering
\includegraphics{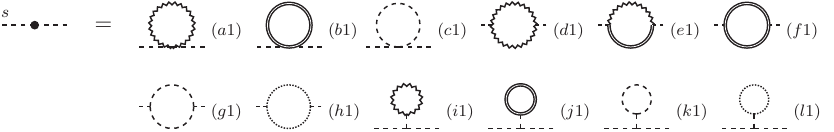}
\caption{%
  One-loop contributions to the vacuum Abelian Higgs model 2-point functions for
  the dark scalar $s$
  with implicit counterterms.
  Dashed lines denote the scalar $s$,
  double lines the Goldstone boson $\chi$,
  wiggly lines dark vector $V_\mu$, and
  dotted lines ghosts $c,\bar{c}$.
  Note that in U(1)$_d$ there is no $(V_\mu\,c\,\bar{c})$ vertex.
  }
  \label{fig:scalar:SE:4D}
\end{figure}

\begin{figure}[t]
\centering
\includegraphics{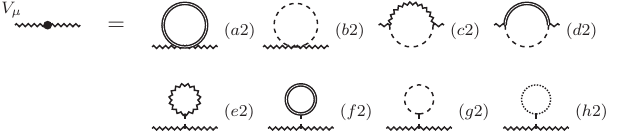}
\caption{%
  One-loop contributions to the vacuum Abelian Higgs model 2-point functions for
  the dark vector $V_\mu$
  with implicit counterterms.
  Line prescription is as denoted in fig.~\ref{fig:scalar:SE:4D}.
  }
  \label{fig:vector:SE:4D}
\end{figure}

Here, we detail the computation of
the self-energies $\Pi(K^2;\LamD)$
that enter the one-loop-corrected propagators
{\em viz.}
\begin{align}
  \bigl\langle s(K) s(-K) \bigr\rangle &=
    \frac{1}{K^2 + \ms^2 - \Pi_{s}(K^2;\LamD)}
  \,,
  \nn[2mm]
  \bigl\langle V_\mu(K) V_\nu(-K) \bigr\rangle &=
    \frac{\delta_{\mu \nu} - \frac{K_\mu K_\nu}{K^2}}{
      K^2 + \mV^2 - \Pi_{\rmii{$V$}}^{(\T)}(K^2;\LamD)}
    + \mathcal{O}\Big(\frac{K_\mu K_\nu}{K^2}\Big)
  \,.
\end{align}
The self-energies are computed
at one-loop level in vacuum
based on the diagrams in
fig.~\ref{fig:scalar:SE:4D} for the dark scalar and
fig.~\ref{fig:vector:SE:4D} for the dark vector.
The result for the dark scalar self-energy
can readily be found in~\cite{Kajantie:1997hn} and
the dark vector self-energy in~\cite{Kim:2016kxt}.
Our strategy is to compute both one-loop self-energies
in dimensional regularization in $D=4-2\epsilon$ Euclidean space-time dimensions
and then perform the analytic continuation to Minkowskian space-time.
We compute the results in general $R_\xi$ gauge,
where the gauge-fixing parameter is denoted by $\xi$
as given in the ghost Lagrangian
(cf.\ footnote~\ref{ft:ghost})
using the particle masses of eq.~\eqref{masses_minimum}.

In Minkowskian space-time (subscript $M$),
$\int_\mathcal{P} = \int\frac{{\rm d}p^0}{2\pi} \int_\vec{p}$
denotes the
$D=(d+1)$-dimensional integral measure with
$\int_\vec{p} = \int\frac{{\rm d}^d \vec{p}}{(2 \pi)^d}$
being its $d$-dimensional spatial part.
The analytic continuation
from Euclidean to Minkowskian space-time is performed through
$K \to -i\mathcal{K}$,
where the propagator in Minkowskian space-time is regulated by
$\Delta \to \Delta + i\epsilon$.

Below, we list the individual diagrammatic contributions from fig.~\ref{fig:scalar:SE:4D}
to the scalar self-energy $\Pi_s(K^2;\LamD)$
for $\xi=1$ (Feynman gauge),
\begin{align}
    \mathcal{D}_{(a1)} &=
      D\,\gd^2 \, A(\mV)
    \, , &
    \mathcal{D}_{(b1)} &=
      \lambda_{s} \, A(\mV)
    \, , \nn[2mm]
    \mathcal{D}_{(c1)} &=
      3 \lambda_s \, A(\ms)
    \, , &
    \mathcal{D}_{(d1)} &= 
      -2D\, \gd^2 \, \mV^2 \,
      B(K,\mV)
    \, , \nn[2mm]
    \mathcal{D}_{(e1)} &=
      -\gd^2 \bigl(
        A(\mV)
      + (2 K^2 - \mV^2) B(K,\mV)
    \bigr)
    \, , &
    \mathcal{D}_{(f1)} &=
      - \ms^2 \lambda_s \,
        B(K,\mV)
    \, , \nn[2mm]
    \mathcal{D}_{(g1)} &=
      - 9 \lambda_s \ms^2 \,
        B(K,\ms)
    \, , &
    \mathcal{D}_{(h1)} &=
        \gd^2 \mV^2 \,
        B(K,\mV)
    \, , \nn[2mm]
    \mathcal{D}_{(i1)} &=
      - 3 \gd^2 \, D \,
      A(\mV)
    \, , &
    \mathcal{D}_{(j1)} &= 
      - \frac{3}{2} \gd^2\frac{\ms^2}{\mV^2}  \,
      A(\mV)
    \, , \nn[1mm]
    \mathcal{D}_{(k1)} &=
      - \frac{9}{2} \gd^2 \frac{\ms^2}{\mV^2} \,
      A(\ms)
    \, , & 
    \mathcal{D}_{(l1)} &= 
      3 \gd^2 \,
      A(\mV)
    \,.
\end{align}
The transverse part of the vector self-energy $\Pi_\rmii{$V$}^{(\T)}(K^2;\LamD)$
from fig.~\ref{fig:vector:SE:4D} reads:
\begin{align}
  \mathcal{D}_{(a2)}^{(\T)} &=
    \gd^2 \, A(\mV)
  \,, &
  \mathcal{D}_{(b2)}^{(\T)} &= 
    \gd^2 \, A(\ms)
  \,, \nn[2mm]
  \mathcal{D}_{(c2)}^{(\T)} &=
    -4 \, \gd^2 \mV^2 \, B(K,\ms,\mV)
  \,, &
  \mathcal{D}_{(d2)}^{(\T)} &=
    - 4\gd^2 
      D_{\mu \nu}^{(\T)}(K;\ms,\mV)
  \,, \nn[2mm]
  \mathcal{D}_{(e2)}^{(\T)} &=
    - 2D\, \gd^{2} \frac{\mV^2}{\ms^2} A(\mV)
  \,, &
  \mathcal{D}_{(f2)}^{(\T)} &=
    - \gd^{2} \, A(\mV)
  \,,
  \nn[1mm]
  \mathcal{D}_{(g2)}^{(\T)} &=
      - 3 \gd^{2} \, A(\ms)
  \,, &
  \mathcal{D}_{(h2)}^{(\T)} &=
      2 \gd^{2} \frac{\mV^2}{\ms^2}\, A(\mV)
  \,.
\end{align}
Here, we used that
eq.~\eqref{masses_minimum} in Feynman gauge implies
$\mV = m_\chi = m_c$ in the minimum.
To simplify, we use the on-shell relation
\begin{align}
  \lambda_s &= \frac{\gd^2}{2}\frac{\ms^2}{\mV^2} = \frac{\gd^2}{2} R_s^2
    \,,&
    R_s &= \frac{\ms}{\mV}
    \, .
\end{align}

Two integral structures arising in the computation
reduce to the basic integrals of eq.~\eqref{eq:A:B}:
\begin{align}
  \mathcal{D}_{(e1)} &\simeq
  \int_{P}
  \frac{(2K-P)^2}{[P^2+m^2]}
  \frac{1}{[(K-P)^2+m^2]}
  =
      A(m^2)
    + (2 K^2 - m^2) B(K^2;m^2)
  \, ,
  \\[3mm]
  \mathcal{D}_{(c2)}
  &\simeq
  \int_{P}
  \frac{
    (2 P_\mu - K_\mu)
    (2 P_\nu - K_\nu)}{
    \left[ P^2+\ms^2 \right]
    \left[ (K-P)^2+\mV^2 \right]
  }
  \nn[2mm] &=
      4B_{\mu\nu}(K;\ms,\mV)
    - K_\mu K_\nu \bigl(
        4 C(K;\ms,\mV)
      - B(K;\ms,\mV)
     \bigr)
  \, ,
\end{align}
where only the first term contributes to the transverse result.
The self-energies are then obtained as
\begin{align}
  \Pi_{s}(K^2;\ms,\mV) &=
    \mathcal{D}_{(a1)}
    + \dots
    + \mathcal{D}_{(l1)}
    \nn[2mm] &=
      - \gd^2 \biggl\{
          \frac{9}{2} R_s^2 \ms^2 B(K;\ms)
        + \frac{(R_s^4+4(D-1))\ms^2 + 4R_s^2 K^2}{2R_s^2} B(K;\mV)
      \nn &
      \hphantom{= - \gd^2 \Bigl\{}
        + (R_s^2 + 2(D-1))A(\mV)
        + 3 R_s^2 A(\ms)
    \biggr\}
    \,,\\
  \Pi_{\rmii{$V$}}^{(\T)}(K^2;\ms,\mV) &=
    \mathcal{D}_{(a2)}^{\rmii{$(\T)$}}
    + \dots
    + \mathcal{D}_{(h2)}^{\rmii{$(\T)$}}
    \nn[2mm] &=
      - 2\gd^2 \biggl\{
          2 B_{\mu\nu}^{(\T)}(K;\ms,\mV)
        + 2\mV^2 B(K;\ms,\mV)
        \nn &
        \hphantom{= - 2\gd^2 \Bigl\{2 D_{\mu\nu}^{(\T)}(K;\ms,\mV)}
        + A(\ms)
        + \frac{D-1}{R_s^2} A(\mV)
    \biggr\}
    \,,
\end{align}
where
$\Pi_\rmii{$V$}^{ } = \Pi_{\mu\nu}^{\rmii{$(\T)$}}$.
The vector self-energy agrees with~\cite{Kim:2016kxt} and
the scalar self-energy with~\cite{Kajantie:1997hn}.

Finally,
after analytically continuing
the self-energies to Minkowskian space-time
via $K \to -i\mathcal{K}$ and
employing the master integrals of eq.~\eqref{eq:F:master},
we find the explicit expressions for
the one-loop self-energies at
the pole,
\begin{align}
\label{eq:Pi:s}
  \Pi_{s}(-\ms^2;\LamD) &=
  \frac{\gd^2}{(4\pi)^2}\ms^2 \Bigl[
    (2R_s^2 - 3)\ln\frac{\LamD^2}{\mV^2}
  - 3 R_s^2 \ln R_s
    - \Bigl(3 - \frac{9}{2} \mathcal{F}(1) \Bigr)R_s^2
  \nn &
  \hphantom{=\frac{\gd^2}{(4\pi)^2}\ms^2 \Bigl[}
    - 1
    - \frac{6}{R_s^2}
    + \frac{R_s^4-4R_s^2+12}{2R_s^2} \mathcal{F}(1/R_s) 
  \Bigr]
  \,, \\[2mm]
\label{eq:Pi:A}
  \Pi_\rmii{$V$}(-\mV^2;\LamD) &=
  \frac{\gd^2}{(4\pi)^2}\frac{\mV^2}{3} \Bigl[
    - \Bigl(9R_s^2 - 10 + \frac{18}{R_s^2}\Bigr)\ln\frac{\LamD^2}{\mV^2}
    + 18R_s^2\frac{R_s^2-2}{R_s^2-1} \ln R_s
  \nn &
  \hphantom{=\frac{\gd^2}{(4\pi)^2}\frac{\mV^2}{3} \Bigl[}
    - 11 R_s^2
    + \frac{26}{3}
    - \frac{6}{R_s^2}
    + (R_s^4-4R_s^2+12) \mathcal{F}(1,R_s) 
  \Bigr]
  \,.
\end{align}

%
\section{Dimensional reduction details}
\label{sec:dimensional_reduction_details}

The effective parameters of the dimensionally reduced theory
setups~\eqref{eq:bro:soft} and~\eqref{eq:bro:softer} are collected below.
Our independent computation here also agrees with
the output of {\tt DRalgo}~\cite{Ekstedt:2022bff,Bernardo:2026nyq} and
the pure Abelian Higgs part of~\cite{Kajantie:1997hn,Hirvonen:2021zej,Bernardo:2025vkz}.
Both references employ the high-temperature expansion.

When constructing
the soft~\ref{eq:bro:soft}, we
follow the reasoning of~\cite{Navarrete:2025yxy} and
apply the high-temperature expansion to
the transitioning scalar field, $\mu_s \sim g T$,
while keeping the inducing field, the Dirac fermion $X$,
over a broader temperature range, $\mX > g T$,
down to $\mX \gg T$, where it becomes Boltzmann suppressed.
While in the broken phase,
vector bosons also act as inducing fields,
we keep them in the high-temperature limit during the matching.
To accommodate fermion masses over a broad temperature
range, including e.g.\ $\mX \sim T$,
the thermal integrals are
kept in general form during matching, as was done
for a Majorana fermion in~\cite{Biondini:2022ggt} and
for EQCD in~\cite{Laine:2019uua} and
were recently automated using the hot Loop-Tree Duality technique~\cite{Navarrete:2025yxy,Capatti}.

The one-loop
bosonic and fermionic master integrals decompose into a vacuum part and a
thermal part,
$Z^{\T}$ for bosons and 
$\widetilde{Z}^{\T}$ for fermions,
that can be evaluated numerically;
see eqs.~\eqref{eq:Z:master:boson}--\eqref{eq:Z:master:fermion}
for the corresponding definitions.
Since only the dark photon couples to the fermion,
the sole fermionic one-loop contribution to
the field normalizations appears in the $\mathcal{O}(p^2)$ term of
the temporal and spatial dark photon correlators
\begin{eqnarray}
  \Pi_{V_0 V_0}' &=&
  \frac{\gd^{2}}{(4\pi)^2}\frac{1}{3}\Bigl[
    \Lb+2
    + 4\nG\YX^2\Bigl(
        \ln\frac{\LamD^2}{\muX^2}
      - 1
      + (4\pi)^2 \bigl(
        \widetilde{Z}_{2;\rmii{$X$}}^{\T}
      + 2\muX^{2}\widetilde{Z}_{3;\rmii{$X$}}^{\T}
      \bigr)
    \Bigr)
  \Bigr]
  \nn[1mm]
  \nn &\stackrel{m_i \ll T}{=}&
  \frac{\gd^{2}}{(4\pi)^2}\frac{1}{3}\Bigl[
    \Lb+2
    + 4\nG\YX^2\bigl(
      \Lf - 1
    \bigr)
  \Bigr]
  \,,
  \nn[2mm]
  \Pi_{V_i V_i}' &=&
  \frac{\gd^{2}}{(4\pi)^2}\frac{1}{3}\Bigl[
    \Lb
    + 4\nG\YX^2\Bigl(
        \ln\frac{\LamD^2}{\muX^2}
      + (4\pi)^2 \widetilde{Z}_{2;\rmii{$X$}}^{\T}
    \Bigr)
  \Bigr]
  \nn[1mm]
  \nn &\stackrel{m_i \ll T}{=}&
  \frac{\gd^{2}}{(4\pi)^2}\frac{1}{3}\Bigl[
      \Lb
    + 4\nG\YX^2\Lf
  \Bigr]
  \,.
\end{eqnarray}
Here, we also displayed the respective high-temperature limits,
$m_i \ll T$.

For $\nG$ fermion generations,
the gauge coupling and Debye mass
one-loop matching relations are
\begin{align}
\label{eq:gd:soft}
g_{d,3}^{2} &=
  \gd^2 T\Big[
    1
    - \frac{\gd^2}{(4\pi)^2}\frac{1}{3}\Big(
        \Lb\YS^{2}
      + 4\nG^{ }\YX^2\Bigl(
          \ln\frac{\LamD^2}{\muX^2}
        + (4\pi)^2 \widetilde{Z}_{2;\rmii{$X$}}^{\T}
      \Bigr)
  \Big)
\Big]
\;,\\
\label{eq:mD}
\mD^{2} &=
    \gd^{2} T^2\Bigl(
      \frac{1}{3}
    - 4\frac{\YS^{2}}{(4\pi)^2}
      \frac{\mu_{s}^{2}}{T^2}
    \Bigr)
    - 8\nG^{ }\YX^2 \gd^{2}\Bigl(
        \widetilde{Z}^{\T}_{1;\rmii{$X$}}
      + \muX^{2}\widetilde{Z}^{\T}_{2;\rmii{$X$}}
      \Bigr)
 \nn &
    - \frac{T^2}{(4\pi)^2} \gd^{2}\Bigl(
        \frac{\Lb - 7}{9} \gd^{2}\YS^{4}
      - \frac{4}{3} \lambda_{s}^{ } \YS^{2}
    \Bigr)
\;,
\end{align}
where the scalar mass effect in the Debye mass
is rooted in the high-temperature expansion.

The remaining matching relations for
the thermal mass of the complex
U(1)$_d$ singlet and
its quartic couplings
take the form
\begin{align}
\label{eq:m:S:2l}
\mu_{s,3}^{2} &=
  - \mu_{s}^2
  + \frac{T^2}{12} \Big(
      4\lambda_{s}
    + 3 \gd^{2}\YS^{2}
  \Big)
  + \frac{\Lb}{(4\pi)^2}
      \mu_{s}^{2}\Big(
          4 \lambda_{s}
        - 3 \gd^{2}\YS^{2}
      \Big)
    \nn &
    - \frac{T^{2}}{(4\pi)^{2}}\frac{\gd^{2}\YS^{2}}{9}\Big(
        2\gd^{2} \YS^{2}
      - 6 \lambda_{s}^{ }
    \Big)
    - \frac{T^{2}}{(4\pi)^{2}}\Lb\Big(
        \frac{13}{12} \gd^{4} \YS^{4}
      - 2 \gd^{2}\YS^{2} \lambda_{s}^{ }
      + \frac{10}{3} \lambda_{s}^{2}
      \Big)
    \nn &
    - \clog\frac{1}{(4\pi)^{2}}\Big(
        4 g_{d,3}^{4} \YS^{4}
      - 8 g_{d,3}^{2} \lambda_{s,3}^{ } \YS^{2}
      + 8 \lambda_{s,3}^{2}
      + \frac{h_{3}^{2}}{2}
    \Big)
    \;, \\[2mm]
\lambda_{s,3} &= T \Big[ \lambda_s + \frac{1}{(4\pi)^2} \Big(
      \big(2- 3 \Lb\big) \gd^{4}\YS^{4}
    + \Lb \big(
          6 \gd^{2}\YS^{2} \lambda_{s}^{ }
        - 10 \lambda_{s}^2
      \big)
    \Big)
    \Big]
  \;, \\[2mm]
h_{3} &= \gd^{2}\YS^{2} T \Big[
  1
  - \frac{1}{(4\pi)^2}\Bigl(
        \frac{\Lb-4}{3}\YS^{2}\gd^{2}
      - 8\lambda_{s}^{ }
  \nn &
  \hphantom{= \gd^{2}\YS^{2} T \Big[
  1
  - \frac{1}{(4\pi)^2}\Big(}
        + \frac{4}{3}\nG\YX^2\gd^{2}\Bigl(
          \ln\frac{\LamD^2}{\muX^2}
        - 1
        + (4\pi)^2 \bigl(
            \widetilde{Z}_{2;\rmii{$X$}}^{\T}
          + 2\muX^{2}\widetilde{Z}_{3;\rmii{$X$}}^{\T}
        \bigr)
      \Bigr)
    \Bigr)
  \Big]
  \;, \\[2mm]
\label{eq:kappa:soft}
\kappa_{3} &=
  T\gd^{4} \Bigl[
    \frac{16\YS^{4}}{(4\pi)^{2}}
    + 192\nG\YX^{4} m_{i}^{4} \widetilde{Z}_{4;\rmii{$X$}}^{\T}
  \Bigr]
\,,
\end{align}
where
$c =
  \frac{1}{2} \bigl(
    \ln \frac{8\pi}{9}
    + (\ln\zeta_{2})'
  - 2 \gammaE \bigr)
$.
For completeness,
we also display the high-temperature-expanded fermionic contributions at
two-loop level
\begin{align}
\label{eq:mD:fermion:highT}
\mD^{2} &\supset
    - \frac{T^2}{(4\pi)^2}\frac{\gd^{4}}{9} \Bigl(
          (\Lb + 4\Lf -2)\YS^{2}\YX^{2}\nG^{ }
        + 18\nG^{ }\YX^{4}
        + 4(\Lf - 1)\YX^{4}\nG^{2}
    \Bigr)
    \;, \\[2mm]
\label{eq:m:S:2l:highT}
\mu_{s,3}^{2} &\supset
    + \frac{T^{2}}{(4\pi)^{2}}\gd^{4}\YS^{2}\YX^{2}\nG\Big(
        \frac{1}{9}
      + \frac{\Lf}{6}
      - \frac{\Lb}{2}
      \Big)
    \;,
\end{align}
where, due to the absence of a tree-level interaction,
fermionic contributions to the scalar mass only start at two-loop level.
In the matching relations used in practice,
we truncate the fermionic effects after one loop order
both in the Debye and scalar mass and
relegate the computation of
the corresponding two-loop thermal integrals,
similar to~\cite{Laine:2019uua,Biondini:2022ggt}, to future work.

Integrating out the temporal vector boson $V_0$ from the soft theory
yields the softer~\ref{eq:bro:softer}.
The corresponding softer-EFT matching parameters read
\begin{align}
\label{eq:gd:softer}
\bar{g}_{d,3}^{2} &=
  g_{d,3}^{2}
  \;,\\[2mm]
\bar{\mu}_{s,3}^{2} &=
    \mu_{s,3}^{2}
  - \frac{h_{3} \mD^{ }}{(4\pi)}
  - \frac{h_{3}}{(4\pi)^2}\Bigl[
      h_{3}\Bigl(1 + \ln\frac{\Lamd^{2}}{4\mD^{2}}\Bigr)
    - \frac{\kappa_{3}^{ }}{24}
  \Bigr]
\;,\\
\label{eq:lambdas:softer}
\bar{\lambda}_{s,3}^{ } &=
    \lambda_{s,3}^{ }
  - \frac{1}{(4\pi)}
    \frac{h_{3}^{2}}{2\mD^{ }}
  + \frac{2}{(4\pi)^2}\frac{h_{3}^{3}}{\mD^2}
\;.
\end{align}
Here, we explicitly included the two-loop contributions
for the softer scalar self-coupling $\bar{\lambda}_{s,3}$,
since they are formally of $\mathcal{O}(\gd^4)$, which corresponds
to NLO accuracy.
The positive sign of this term improves convergence of the softer EFT close to $\Tc$
as indicated in~\cite{Ekstedt:2024etx,Niemi:2024vzw}.
The successive contributions to $\bar{\lambda}_{s,3}^{ }$,
namely $\lambda_{s,3}^{ }$, $h_{3}^{2}/(4\pi \mD^{ })$ and $h_{3}^{3}/(4\pi \mD^{ })^2$,
scale as $\gd^2 T$, $\gd^3 T$ and $\gd^4 T$ under the power counting
$h_{3}^{ }, \lambda_{s,3}^{ } \sim \gd^2 T$ and $\mD^{ } \sim \gd T$.
Scaling out the common factor $\lambda_{s,3}^{ } \sim \gd^2 T$,
the genuine dimensionless expansion parameter is the ratio of
consecutive terms, $h_{3}^{ }/(4\pi \mD^{ }) \sim \gd$, so that the
truncation is justified only for $h_{3}^{ }\ll 4\pi \mD^{ }$;
otherwise one should either use the soft theory directly
or encode temporal effects
to a softer-type EFT as described in sec.~\ref{sec:dim_red}.

%
\section{Annihilation cross-sections across the phase transition}
\label{app:cross_sections}

\begin{figure}[t]
\centering
\includegraphics{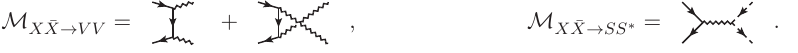}
\caption{%
  Diagrams for the dark matter pair annihilation and co-annihilation with
  the singlet scalar in the unbroken phase.
  The dark fermion $X$ is displayed by an arrowed solid line,
  the complex scalar $S$ by an arrowed dashed line, and
  dark photons by wiggly lines.
}
\label{fig:xx:ub}
\end{figure}

This appendix collects
the annihilation cross-sections relevant for
the dark matter phenomenology discussed in sec.~\ref{sec:dark_matter}.
We distinguish between
the symmetric (sym) and broken (bro) phases of the dark gauge symmetry.

\subsection{Symmetric phase}

In the symmetric phase, the relevant annihilation channels are
\begin{align}
  X\bar{X} &\to VV
  \, ,&
  X\bar{X} &\to SS^*
  \, .
\end{align}
The corresponding tree-level diagrams are shown in fig.~\ref{fig:xx:ub}.
The annihilation cross-sections,
expressed in terms of the Mandelstam variable
$s = (p_\rmii{$X$} + p_\rmii{$\bar{X}$})^2$,
are
\begin{align}
\label{sigma_ann_UNBR_V}
   \sigma^{\rmi{sym}}_{\rmi{ann}}  v_{\textrm{rel}}^{ }(X \bar{X} \to VV)
  &=
    \frac{4 \pi \alpha_d^2}{s}
    \left[
        \vphantom{\frac{\mbox{artanh}\!\left( \sqrt{1-4 \mX^2/s}\right)}
        {\sqrt{1 - 4 \mX^2/s}}}
        - 1
        - \frac{4 \mX^2}{s}
    \right.
    \nn &\left.
    \hphantom{{}=\frac{4 \pi \alpha_d^2}{s}\Biggl[}
        + 2 \left( 1+ \frac{4 \mX^2}{s} - \frac{8 \mX^4}{s^2} \right)
        \frac{\mbox{artanh}\!\left( \sqrt{1-4 \mX^2/s}\right)}
        {\sqrt{1 - 4 \mX^2/s}}
    \right]
    \, ,
\\[2mm]
\label{sigma_ann_UNBR_S}
  \sigma^{\rmi{sym}}_{\rmi{ann}} v_{\textrm{rel}}^{ }(X \bar{X} \to S S^*)
  &=
      \frac{2 \pi \alpha_d^2}{3 s}
    \left( 1+ \frac{2\mX^2}{s}\right)
  \, .
\end{align}

In the non-relativistic regime, relevant for thermal freeze-out and late-time annihilations,
one may expand the cross-sections using $s = 4 \mX^2 /(1-\vrel^2/4)$.
The leading terms in the velocity expansion are
\begin{eqnarray}
    \sigma^{\rmi{sym}}_{\rmi{ann}}  v_{\textrm{rel}}^{ }(X \bar{X} \to VV)
    &=&
    \frac{\pi \alpha_d^2}{\mX^2}
    + \mathcal{O}(\vrel^2)
    \, ,
    \\
     \sigma^{\rmi{sym}}_{\rmi{ann}}  v_{\textrm{rel}}^{ }(X \bar{X} \to S S^*)
    &=&
     \frac{\pi \alpha_d^2}{4 \mX^2}
     + \mathcal{O}(\vrel^2)
     \, ,
\end{eqnarray}
which we have crosschecked also via a non-relativistic matching to dimension-six, velocity independent,
operators~\cite{Bodwin:1994jh,Biondini:2023zcz}.
We find contribution to
${\rm{Im}}[f(^1S_0)] = \pi \alpha^2$ solely from the annihilation into
gauge boson and ${\rm{Im}}[f(^3S_1)] = \pi \alpha^2/12$ for
the annihialtion into scalar pairs, where ${}^S L_J$
is the standard spectroscopic notation for orbital angular momentum, spin and
total angular momentum.  
In this regime, non-perturbative effects become important.
In particular, multiple exchanges of the vector mediator induce Sommerfeld enhancement and
allow for bound-state formation, both of which modify the annihilation dynamics.
In the main text, these effects are incorporated through
the effective cross-section introduced in eq.~\eqref{Cross_section_eff}. 

\subsection{Broken phase}

\begin{figure}[t]
\centering
\centering
\includegraphics{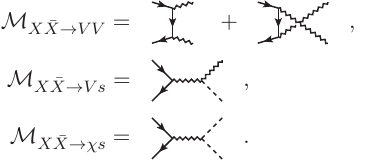}
\caption{%
  Diagrams for the dark matter pair annihilation in the broken phase and in Feynman gauge.
  The dark fermion $X$ is displayed by an arrowed solid line,
  the scalar $s$ and
  the Goldstone $\chi$ by a dashed line, and
  dark photons by wiggly lines.
}
\label{fig:xx:broken}
\end{figure}

After spontaneous symmetry breaking in the dark sector,
additional annihilation channels become kinematically available.
The relevant processes are
\begin{align}
X\bar{X}&\to VV
\, ,&
X\bar{X}&\to Vs
\, ,&
X\bar{X}&\to \chi s
\, ,
\end{align}
where $s$ denotes the physical dark scalar and $\chi$ the Goldstone boson.
Throughout this appendix,
we work in Feynman gauge ($\xi\to 1$).
Accordingly, the Goldstone boson appears as an explicit propagating degree of freedom.
Physical observables are recovered after consistently summing all contributions.
The corresponding diagrams are shown in fig.~\ref{fig:xx:broken}.

The annihilation cross-sections into
dark vectors,
mixed vector-scalar states, and
pure scalar states are given by
\begin{align}
 \label{XX_VV_broken}
  \sigma^{\rmi{bro}}_{\rmi{ann}}  v_{\textrm{rel}}^{ }(X \bar{X}\!\to VV)
  &=
    \frac{4 \pi \alpha_d^2}{s}
    \sqrt{1-\frac{4\mV^2}{s}}
    \left[
      \frac{
        2\Bigl(1
          + \frac{4 \mX^2}{s}
          - \frac{8 \mX^4}{s^2}
          - \frac{4 \mX^2 \mV^2}{s^2}
            \Bigl( 2 - \frac{\mV^2}{\mX^2}\Bigr)
        \Bigr)}{
      \Bigl(1-\frac{2 \mV^2}{s} \Bigr)
      \sqrt{\Bigl(1-\frac{4 \mX^2}{s} \Bigr)\Bigl(1-\frac{4 \mV^2}{s} \Bigr)}}
      \right.
      \nn[2mm] &
      \left.
      \hspace{0.5cm}\times\,
      \mbox{artanh} \left(
        \frac{\sqrt{\bigl( s- 4 \mX^2\bigr)\bigl( s- 4 \mV^2\bigr)}}{s-2 \mV^2}
      \right)
    - \frac{
        1+\frac{4 \mX^2}{s} + \frac{2 \mV^2}{s \mX^2}}{
        1-\frac{4 \mV^2}{s} + \frac{\mV^4}{s \mX^2}}
    \vphantom{\frac{
        \Bigl(\Bigl( 2 - \frac{\mV^2}{\mX^2}\Bigr)\Bigr)}{
        \sqrt{\Bigl(1-\frac{4 \mX^2}{s} \Bigr)\Bigl(1-\frac{4 \mV^2}{s} \Bigr)}}}
    \right]
    \, ,
\\[2mm]
\label{XX_V_s_broken}
    \sigma^{\rmi{bro}}_{\rmi{ann}} v_{\textrm{rel}}^{ }(X \bar{X}\! \to V s)
    &=
    \frac{4 \pi \alpha_d^2 }{s}
    \frac{\mV^2 (s+2 \mX^2)}{(s-\mV^2)^2}
    \left[
      \left( 1\! -\! \frac{(\mV\!+\!\ms)^2}{s}\right)
      \!
      \left( 1\! -\! \frac{(\mV\!-\!\ms)^2}{s}\right)
    \right]^\frac{1}{2}
    \!\!,
\\[3mm]
\label{XX_phi_s_broken}
  \sigma^{\rmi{bro}}_{\rmi{ann}}  v_{\textrm{rel}}^{ }(X\! \bar{X} \to \chi s)
  &=
  \frac{2 \pi \alpha_d^2}{3}
  \frac{(s+2 \mX^2)}{(s-\mV^2)^2}
  \left[
    \left( 1\! -\! \frac{(\mV\!+\!\ms)^2}{s}\right)
      \!
    \left( 1\! -\! \frac{(\mV\!-\!\ms)^2}{s}\right)
  \right]^\frac{3}{2}
  \!\!.
\end{align}
As a consistency check, consider the limit $v_s\to0$,
which restores the ${\rm U}(1)_d$ symmetry.
In this limit,
the scalar,
Goldstone, and
vector masses vanish, cf.\ eq.~\eqref{masses_minimum},
and the expressions above reduce to
the corresponding symmetric-phase cross-sections
in eqs.~\eqref{sigma_ann_UNBR_V} and~\eqref{sigma_ann_UNBR_S}.
Assuming $\mX \gg \mV$ and $\mX \gg \ms$, as in our analysis,
the leading terms of the non-relativistic expansion are
\begin{align}
\sigma^{\rmi{bro}}_{\rmi{ann}}  v_{\textrm{rel}}^{ }(X \bar{X} \to VV)
  &=
  \frac{\pi \alpha_d^2}{\mX^2}
  \left( 1-\frac{\mV^2}{2 \mX^2} \right)
  \, ,
\\[2mm]
  \sigma^{\rmi{bro}}_{\rmi{ann}}  v_{\textrm{rel}}^{ }(X \bar{X} \to V s)
  &=
  \frac{3 \pi \alpha_d^2}{8 \mX^2}
  \left( \frac{\mV}{\mX}\right)^2
  \, ,
  \\[2mm]
 \sigma^{\rmi{bro}}_{\rmi{ann}}  v_{\textrm{rel}}^{ }(X \bar{X} \to \chi s)
  &=
  \frac{\pi \alpha_d^2}{4 \mX^2} \left( 1 - \frac{\mV^2}{4 \mX^2} -  \frac{3 \ms^2}{\mX^2}\right)
  \, ,
\end{align}
shown to $\mathcal{O}(\mV^2/\mX^2)$ and $\mathcal{O}(\ms^2/\mX^2)$.

{\small
%
\bibliographystyle{utphys}
\bibliography{ref}

}
\end{document}